\documentclass[a4paper,oneside,12pt,dvipsnames]{article}

\usepackage[T1]{fontenc}
\usepackage[english]{babel}
\usepackage{lmodern}
\usepackage{slashed}
\usepackage{appendix}
\usepackage{fixltx2e}

\usepackage{amssymb}
\usepackage{cite}
\usepackage[font=small,labelfont=bf,margin=1cm]{caption}

\usepackage{amsmath,amsfonts,amsthm} 
\usepackage{graphicx,xcolor}
\usepackage{bbold}    
\usepackage{blkarray}
\usepackage[normalem]{ulem}
\usepackage{multirow}
\usepackage{makecell}

\newcommand{\be}{\begin{equation}}
\newcommand{\ee}{\end{equation}}
\newcommand{\ba} {\begin{equation}\begin{aligned}}
\newcommand{\ea} {\end{aligned}\end{equation}}
\newcommand{\bea}{\begin{eqnarray}}
\newcommand{\eea}{\end{eqnarray}}
\newcommand{\nn}{\nonumber}

\newcommand{\paren}[1]{\left( #1 \right)}
\newcommand{\ele}{\mathcal{L}}
\newcommand{\dmu}{\partial_\mu}

\newcommand{\lraw}{\leftrightarrow}
\newcommand{\olraw}{\overleftrightarrow}
\newcommand{\olaw}{\overleftarrow}
\newcommand{\oraw}{\overrightarrow}

\newcommand{\vH}{\langle H \rangle}
\newcommand{\vs}{\langle \sigma \rangle}

\newcommand{\Msq} {\mathcal{M}\mathcal{M}^\dag}

\def\Tr{{\rm Tr}}

\def\coma{\,,}

\def\lam{\lambda}
\def\Lam{\Lambda}
\def\sig{\sigma}
\def\d{\delta}
\def\De{\Delta}
\def\Ds{\slashed{D}}

\def\ahs{\gamma}

\def\beq{\begin{equation}}
\def\eeq{\end{equation}}

\def\g{\gamma}
\def\s{\sigma}
\def\wH{ \widetilde{H}}
\def\l{\left(}
\def\r{\right)}

\title{
\flushright
 \small{DFPD-2016/TH/04\\FTUAM-16-11\\IFT-UAM/CSIC-16-026
}
\center
\Large{ {\bf The minimal linear $\sigma$ model for the Goldstone Higgs}}}
\author{F.~Feruglio~$^{a)}$, M.~B.~Gavela~$^{b)}$, K.~Kanshin~$^{a)}$, P.~A.~N.~Machado~$^{b)}$,\\
S.~Rigolin~$^{a)}$ and S.~Saa~$^{b)}$\\
\small $^{a)}$Dipartimento di Fisica e Astronomia `G.~Galilei', Universit\`a di Padova
\\
\small INFN, Sezione di Padova, Via Marzolo~8, I-35131 Padua, Italy
\\
\small $^{b)}$Departamento de F\'isica Te\'orica and Instituto de F\'{\i}sica Te\'orica, IFT-UAM/CSIC,\\
\small Universidad Aut\'onoma de Madrid, Cantoblanco, 28049, Madrid, Spain
}

\begin{document}

\maketitle

\abstract{ In the context of the minimal $SO(5)$ linear $\s$-model, a
  complete renormalizable Lagrangian -including gauge bosons and
  fermions- is considered, with the symmetry softly broken to
  $SO(4)$. The scalar sector describes both the electroweak Higgs
  doublet and the singlet $\sigma$.
 Varying the $\sigma$ mass would allow to sweep from the regime of
 perturbative ultraviolet completion to the non-linear one assumed in 
 models  in
 which the Higgs particle is a low-energy remnant of some strong
 dynamics. We analyze the phenomenological implications and
 constraints from precision observables and LHC data. Furthermore, we derive the $d\le6$ 
     effective Lagrangian in the limit of heavy exotic fermions.}


\newpage

\tableofcontents

\newpage

\section{Introduction}

The discovery of the Higgs boson has confirmed the simplest possible picture of electroweak symmetry breaking and mass generation. As predicted long ago
by the Standard Model of particle physics (SM), the Higgs mechanism splits an elementary doublet of scalar particles into an unphysical sector, providing the longitudinal polarization
to the vector bosons $W$ and $Z$, and a single physical spin zero particle. For a long time this picture has been questioned, specially in connection to the
naturalness problem, also called ``electroweak hierarchy problem'', that is the stability of the electroweak scale against quantum corrections and its smallness compared to other higher scales to which the Higgs field may be sensitive, if such new scales of physics do exist in nature. The tension  would be manifest in   the Higgs mass being lower than those putative scales. 

The challenge raised by the lightness of the
Higgs mass becomes increasingly pressing as long as no firm signal of
beyond the Standard Model (BSM) physics appears in new data.  The
resistance to accept extreme fine-tunings has been historically most
fruitful, prompting the identification of new symmetries to justify
dynamically scales smaller than the characteristic overall scale of a
given theory; a magnificent example is for instance the prediction of
the charm particle~\cite{Glashow:1970gm} and its mass~\cite{Gaillard:1974mw}. 
The observed light  Higgs mass poses
a similar conundrum.

Moreover the Higgs boson of the SM would represent the unique example of elementary spin zero particle in nature, while in other known phenomena of 
spontaneous symmetry breaking its role is played by composite excitations. Supersymmetry would justify elementary scalars, actually one copy for each
known fermion, but no direct or indirect  hints of them has been found so far. 
In fact, the only degrees of freedom found in nature  prior to the Higgs
discovery which may originate from scalar fields were Goldstone bosons: the longitudinal components of the
electroweak gauge bosons.  This suggested 
decades~\cite{Kaplan:1983fs,Georgi:1984af,Dugan:1984hq} 
ago a dynamical nature for the Higgs particle as a pseudo-Nambu-Goldstone
boson (PNGB) which would justify a light Higgs. All components of the Higgs multiplet would then share a common Goldstone origin, providing a beautifully homogeneous picture.   In the initial proposal the Higgs originated from one of the Goldstone
bosons produced in the spontaneous breaking of a high-energy $SU(5)$
invariant strong dynamics. Recent attempts tend to start instead from
a $SO(5)$ symmetry~\cite{Agashe:2004rs,Contino:2006qr} 
spontaneously broken to $SO(4)$ at some
high scale $\Lambda$, producing at this stage an ancestor of the Higgs
particle in the form of one of the resulting Goldstone bosons, with
characteristic scale $f$ and $\Lambda\le 4 \pi f$~\cite{Manohar:1983md}. The coset
$SO(5)/SO(4)$ represents the minimal possibility to interpret the
Higgs as a pseudo-Goldstone boson in the presence of a custodial symmetry.

The fermionic couplings of the $SO(5)$ invariant sector to the SM fermions and gauge bosons give an additional -generally soft- breaking of $SO(5)$ resulting in a potential for the Goldstone Higgs. Its minimum breaks spontaneously  the electroweak symmetry at a scale $v$, which phenomenologically needs to differ from $f$, and gives a mass to the Higgs particle.  Moreover, this type of theories~\cite{Kaplan:1991dc} proposes naturally  
a seesaw-like mechanism for quarks and leptons, whose masses would be inversely proportional to the heavy fermion mass scale. 
 It is most suggestive that the seesaw mechanism would not then be
reduced to the realm of neutrino masses - for which it is the best
candidate theory- but it would be instead the universal pattern behind
all fermion masses.

A Goldstone-boson parenthood for the Higgs is not exclusive of those
models, often called ``composite Higgs'' models, but is also embedded in
other constructions such as ``little Higgs'' models, extra-dimensional
 scenarios and others.  
   In concrete models the
  spectra of exotic fermions are directly related to the light fermion masses 
  -- in particular the top mass-- and the Higgs mass. The values of these masses generally require  a spectrum with some
 exotic  fermion masses below the TeV scale, a fact often  in tension with experimental searches~\cite{Aad:2015kqa, CMS:2015alb}.
    It is interesting
to clarify the degree of fine-tuning that the models require, in 
view of the electroweak hierarchy problem.

Most of the literature on composite Higgs models based on $SO(5)$
assumes from the start a strong dynamics and uses an effective
non-linear formulation of the 
model(s)~\cite{Contino:2006qr,Panico:2012uw,Carena:2014ria,Contino:2011np,
Marzocca:2012zn,Redi:2012ha,Carmona:2014iwa,vonGersdorff:2015fta}.   
This approach has the
advantage of being completely general, offering a parametrization of
all possible ultraviolet completions for the symmetry group chosen. At the same
time one of its limitations is that it is applicable only in a finite
domain of energies.  Here instead we construct a complete
renormalizable model which in its scalar part is a linear sigma
model including a new scalar particle $\sigma$,
singlet under the gauge group.  This will allow to gain intuition on
the dependence on the ultraviolet (UV) completion scale of the model,
by varying the $\sigma$ mass: a light $\sigma$ particle corresponds to
a weakly coupled regime, 
 while in the high mass limit the theory should fall back onto an usual
 effective non-linear construction.  Our complete renormalizable model
 can thus be considered either as an ultimate model made out of
 elementary fields, or as a renormalizable version of a deeper
 dynamics, much as the linear $\sigma$ model~\cite{GellMann:1960np} is to
 QCD. One former attempt in this direction~\cite{Barbieri:2007bh} did not
 fully take into account and computed the impact of the fermionic
 sector on the main phenomenological observables, see also Ref.~\cite{Gertov:2015xma}.

While the choice of the minimal bosonic sector is clear, there is a number of possible 
choices for the fermionic sector. 
The option explored in this paper assumes heavy fermions 
in vectorial representations of $SO(5)$, in contrast to   
 models where the SM left doublets are embedded in  $SO(5)$ 
multiplets~\cite{Barbieri:2007bh}.
Direct couplings between SM fermions and the heavy 
fermions will be the source of  the soft $SO(5)$ breaking, while the Higgs particle has tree-level couplings only with the exotic fermionic sector, 
via $SO(5)$-invariant Yukawa 
couplings.     
It will be discussed how the induced Coleman-Weinberg potential requires  soft 
breaking terms to be included in the scalar potential. 

The usual SM Higgs sector is now substituted by a Higgs-$\sigma$ sector, correcting the 
strength of the SM Higgs-gauge boson couplings and opening new interaction channels. 
The phenomenology of the $\sigma$ production and decay will be also studied, including 
fermionic and bosonic tree-level and one loop decays (e.g. gluon-gluon and photon-photon). Analysis of present Higgs data 
will be used to set a constraint on the fine-tuning 
ratio $v/f$. The contribution of the Higgs, $\sigma$ and the exotic fermions to the oblique 
$S$ and $T$ parameters will be computed. One aim of the phenomenological study is to 
clarify the impact of the size of the ultraviolet scale - here represented by the $\sigma$ 
mass - on the tensions of this type of theories.  Particular emphasis will be dedicated to the impact of the $\sigma$ particle on present and future LHC data, produced either via gluon fusion or vector-boson fusion and decaying into a plethora of channels including diphoton final state.

Furthermore, as this paper focuses on the impact of a dynamical $\sigma$ particle, we identify below some of the leading low-energy bosonic operators stemming from the new physics when the exotic heavy fermion 
sector is integrated out: we determine the dominant effective operators made out of the $\sigma$ field and/or SM fields, as a first step towards the identification of a ``benchmark'' 
electroweak  effective Lagrangian including a light dynamical Higgs.  A non-linear effective Lagrangian should result in the limit of very heavy $\sigma$. An interesting characteristic of the non-linear scenario is that the low-energy physical 
Higgs field turns out not to be an exact electroweak doublet, and may appear in the 
effective Lagrangian as a generic SM scalar singlet with arbitrary couplings. The most 
general effective Lagrangian of this type~\cite{Alonso:2012px,Buchalla:2013rka} 
turns out to depend on a plethora of couplings, though. It would be useful to identify the 
reduced pattern of dominant couplings characteristic of models in which the Higgs is a 
pseudo-Goldstone boson in that regime~\cite{future-paper}.

The structure of the paper can be easily inferred from the Table of Contents.

%

%
\section{The $SO(5)/SO(4)$ scalar sector}
%

The complete Lagrangian can be written as the sum of three  terms describing respectively the pure gauge, scalar and fermionic sectors, 
\begin{equation}
  \ele=\ele_{\rm gauge} +\ele_{\rm scalar} + \ele_{\rm fermion},
\end{equation}
where $\ele_{\rm gauge}$ reduces to the SM gauge kinetic terms. 
This section discusses in detail the scalar sector and its
interactions, while the study of the fermionic sector is deferred to
the next section.

In order to define the linear $\s$ model corresponding to an $SO(5)$
symmetry spontaneously broken to $SO(4)$, let us consider a real
scalar field $\phi$ in the fundamental representation of
$SO(5)$. Three among its five components will be ultimately associated
with the longitudinal components of the SM gauge bosons - denoted
below by $\pi_i$, $i=1, 2, 3$, while the other two will correspond to
the Higgs particle $h$ and to an additional  scalar 
$\s$, respectively. For simplicity the results will be often presented
in the unitary gauge ($u.g.$), in which $\pi_i=0$:
\begin{equation}
\phi= \left(\pi_1,\pi_2,\pi_3,h,\sigma\right)^T \quad\stackrel{u.g.}{\to}\quad \left(0,0,0,h,\sigma\right)^T\, .
\label{emb1}
\end{equation}
The scalar Lagrangian describing the scalar-gauge and the scalar-scalar interactions reads
\bea
\ele_{\rm s} = \frac{1}{2} (D_\mu\phi)^T (D^\mu\phi) - V(\phi)\,,
\eea
where the $SU(2)_L\times U(1)_Y$ covariant derivative is given by
\bea
D_\mu\phi=\paren{\dmu+ i g \Sigma_L^i W_\mu^i+ig'\Sigma_R^3B_\mu}\phi \,
\label{covariant}
\eea and $\Sigma_L^i$ and $\Sigma_R^i$ denote respectively the
generators of the $SU(2)_L$ and $SU(2)_R$ subgroups of the custodial
$SO(4)$ group contained in $SO(5)$. The embedding of the gauge group
$SU(2)_L\times U(1)_Y$ inside $SO(5)$, implicitly assumed in
Eqs. (\ref{emb1}-\ref{covariant}), is purely conventional.  As we will
see in section 2.1, both  $h$ and $\sigma$ acquire a vacuum expectation value (vev),
leaving unbroken an $SO(4)'$ subgroup which is rotated with respect to
the group $SO(4)\approx SU(2)_L\times SU(2)_R$ containing
$SU(2)_L\times U(1)_Y$.

For later convenience it is pertinent to introduce the complex notation for the scalar field $\phi$. 
Denoting by $H$ ($\tilde{H}$) the SM Higgs doublet transforming as $(2,1/2)$ ($(2,-1/2)$) under the $SU(2)_L 
\times U(1)_Y$ gauge group,  a complex scalar field in the fundamental representation of $SO(5)$ can be defined as
\bea
 \hat \phi = \left(H^T, \tilde{H}^T,\s\right)^T\,,
\eea
with the convention that the first two entries of this $SO(5)$ multiplet are $SU(2)_L$ doublets with $+1/2$ and 
$-1/2$ eigenvalue of the diagonal $SU(2)_R$ generator, namely 
\begin{equation}
H=
\begin{pmatrix}
H^u \\
H^d
\end{pmatrix} 
\quad \stackrel{\rm u.g.}{\to} \quad \frac{1}{\sqrt{2}}
\begin{pmatrix}
0 \\ h
\end{pmatrix} 
\qquad , \qquad 
\tilde{H}\equiv i\sigma_2 H^* = 
\begin{pmatrix}
\tilde{H}^u \\
\tilde{H}^d
\end{pmatrix} \,
\quad \stackrel{\rm u.g.}{\to} \quad \frac{1}{\sqrt{2}}
\begin{pmatrix}
h \\ 0
\end{pmatrix}\,,
\end{equation}
while the last component, $\s$, is an $SU(2)_L$ and $SU(2)_R$ singlet. The relation between the real and the 
complex notation is given by 
\begin{equation}
\phi= \frac{1}{\sqrt{2}}
\begin{pmatrix}
-i(H^u+\tilde{H}^d) \,,\, H^u-\tilde{H}^d \,,\, i(H^d-\tilde{H}^u) \,,\, H^d +\tilde{H}^u \,,\, \sqrt{2}\sigma
\end{pmatrix}^T \,.
\end{equation}

\subsection{The scalar potential}

The most general $SO(4)$ preserving while $SO(5)$ breaking
renormalizable potential depends a priori on ten parameters. Two of
them can be reabsorbed via a redefinition of parameters~\footnote{Here
  we choose to get rid of the $\sigma^2$ and $\sigma^4$ terms.},
resulting on a Lagrangian dependent on one $SO(5)$ preserving
coupling, $\lambda$, one scale $f$ heralding spontaneous $SO(5)/SO(4)$
breaking, and six $SO(5)$ soft-breaking terms (denoted below
$\alpha,\beta, a_{1,2,3,4}$).  The Lagrangian in the unitary gauge
reads:
\begin{equation}
V(h,\s) = \lam\paren{\sigma^2+h^2-f^2}^2 +\alpha f^3\,\sigma - f^2 \beta \,h^2 +  a_1~f\, \sigma h^2 +  a_2\,\sigma^2 h^2 + a_3~f\,\sigma^3 
  + a_4\,h^4\,.
  \label{potential4}
\end{equation}
 In
order to retrieve the formulae in a general gauge it suffices
to replace $h^2$ by the $SO(4)$ invariant combination
$h^2+\vec{\pi}^2$.

The only strictly necessary soft breaking terms are $\alpha$ and
  $\beta$ as they need to be present to absorb divergences generated
  by one-loop Coleman-Weinberg contributions to the Lagrangian, as
  shown in Appendix~\ref{sec-CW}; only those terms will be considered in what follows~\footnote{Full renormalizability of the theory requires, in general, the presence of all gauge invariant operators of dimension equal to or smaller than four. At two or more loops,  the renormalization procedure may thus require  to include further symmetry breaking terms beyond those considered; we will assume that their finite contributions will be weighted by comparatively negligible coefficients and can be safely omitted in our analysys.}, a procedure already previously adopted in Ref.~\cite{Barbieri:2007bh}.
   The  potential then reads
\begin{equation}
V(h,\s) = \lambda\paren{h^2+\sigma^2-f^2}^2+\alpha f^3\,\sigma-\beta f^2\,h^2\,,
 \label{Laghs}\,
\end{equation}
resulting on a system depending on four parameters.  The scalar
quartic coupling $\lam$ can be conventionally traded by the $\sigma$
mass, given by $m_\sigma^2\simeq 8 \lambda f^2$ for negligible $\alpha$ and
$\beta$; the non-linear model would be recovered in the limit
$m_\sigma\gg f$, that is $\lambda\gg 1$.
 
A consistent electroweak (EW) symmetry breaking requires both scalars $h,\s$ to acquire a non-vanishing vev, 
respectively dubbed as $v$ and $v_\s$ below, as for $v\neq 0$ the 
$SO(4)$ global group and the EW group are spontaneously broken. Note that the  vev of $h$ is identified with the electroweak scale since it can be related to the Fermi constant precisely as in the SM, see Sect.~\ref{sec-scalar-gauge} below. For $\alpha, \beta \neq 0$ and assuming $v\ne0$,  it results 
\begin{equation}
v_\sig^2=f^2\frac{\alpha^2}{4\beta^2} \quad , \qquad v^2 = f^2\left(1-\frac{\alpha^2}{4\beta^2}+\frac{\beta}{2\lam}\right)\,, 
\label{vevs}
\end{equation}
satisfying the condition 
\begin{equation}
v^2+v_\sig^2=f^2\paren{1+\beta/2\lambda}\,, 
\label{vevs2}
\end{equation}
which indicates that the $SO(5)$ vev is ``renormalized'' by the
$\beta$ term in the potential. From Eqs.~(\ref{vevs}) and (\ref{vevs2}) it
follows that both $f^2>0$ and $f^2<0$ are in principle
allowed~\footnote{For $f^2<0$, $\alpha$ would have to be purely imaginary because of hermiticity.}, in appropriate regions of the
parameters ($\alpha$, $\beta$, $\lam$). However, in the
$SO(5)$-invariant limit, for negative $f^2$ the minimum of the
potential is at the origin and in consequence the symmetry is unbroken and there are no
Goldstone bosons.  The focus of this paper is instead set on the interpretation of the Higgs particle as a PNGB, which requires $f^2>0$ as well as $|v| < |v_\sigma|$, the latter condition defining the region in parameter space  continuously connected with the limiting
case $v=0$ in which the Higgs particle becomes a true Goldstone
boson. For $f^2>0$, the positivity of $v^2$ in Eq.~(\ref{vevs}) and the $|v| < |v_\sigma|$ constraint lead respectively to the conditions~\footnote{For $f^2<0$, the inequality Eq.~(\ref{cond}) is reverted.}
\begin{equation}
  \alpha^2 < 4\beta^2\left(1+\frac{\beta}{2\lambda}\right)\,,
\label{cond}
\end{equation}
\be 2\beta^2
\paren{1+\frac{\beta}{2\lam}}<\alpha^2\,,
 \label{condp}
\ee 
 which for $|\beta| \ll \lambda$ would indicate $2 \beta^2 \lesssim \alpha^2 \lesssim 4 \beta^2$. Moreover, in order to get $v^2\ll f^2$, Eq.~(\ref{vevs}) requires a fine-tuning such that $\alpha/2\beta$ is very close to unity.
 
Expanding the $\sigma$ and $h$ fields around their minima, $h \equiv \hat h + v$ and $\sigma \equiv \hat \sigma + v_\sigma$, and diagonalizing the scalar mass matrix, the mass eigenstates  are given by
\be
h_{\text{phys}}=\hat{h}\cos\ahs - \hat{\s}\sin\ahs \qquad , \qquad \sigma_{\text{phys}}= \hat{\s}\cos\ahs + \hat{h}\sin\ahs \,. 
\label{phys-scalars}
\ee 
For simplicity, from now on the notation $h_{\text{phys}}$ and $\sigma_{\text{phys}}$ will be traded by $h$ and $\sigma$, respectively. The mixing angle in Eq.~(\ref{phys-scalars}) is given  by
\begin{equation}
\tan 2\gamma=\frac{4 v v_\sig}{3v_\sig^2-v^2-f^2}
\label{mixing}
\end{equation}
and should remain in the interval $\gamma\in [-\pi/4, \pi/4] $ in
order not to interchange the roles of  
the heavy and light mass eigenstates.  
The mass
eigenvalues are given by
    \begin{equation} 
m^2\textsubscript{heavy, light}=
	4\lambda f^2\left\{ \left(1+\frac{3}{4}\frac{\beta}{\lambda} \right) \pm 
	\left[1+ \frac{\beta}{2 \lambda}\left(1 +\frac{\alpha^2}{2 \beta^2} 
	+\frac{\beta}{8 \lambda}\right) \right]^{1/2} \right\}\,, 
\label{mhsigma}
\end{equation} 
where the plus sign refers to the heavier eigenstate. For $f^2>0$, the squared masses are positive if  the following two conditions
are satisfied~\footnote{For $f^2<0$, both  inequalities in Eq.~(\ref{pedro}) are reverted and as a consequence $\beta<0$.}  
\be
3\beta+4 \lam> 0~~~,~~~~~~~~~~~~~~%
~2\beta^2+4\beta\lam-\alpha^2\lam/\beta> 0\,,
\label{pedro}
\ee
with the second constraint  coinciding with that in Eq.~(\ref{cond}) multiplying it by $1/(4\beta\lambda)$;  
 it follows  that
$\beta>0$. If the  soft
  mass term  proportional to $\beta$ in the scalar potential Eq.~(\ref{Laghs})  would be overall positive (as for instance for $f^2$<0 and $\beta>0$), the minimum would always correspond to an undesired symmetric EW vacuum  
  $v=0$. Assuming the
$SO(5)$ explicit breaking to be small, ${|\beta|}/{4\lam}\ll1$ which may only happen for positive $f^2$, the
masses of the heavy and light eigenstates read
\beq
\begin{split}
m_\textsubscript{heavy}^2=&~ 8\lam f^2 + 2\beta (3 f^2 - v^2) +O\paren{\frac{\beta}{4\lam}}\,,\\
m_\textsubscript{light}^2=&~ 2\beta v^2+O\paren{\frac{\beta}{4\lam}}\,.\\
\end{split}
\eeq

The physical scalars thus correspond to a ``light''   state  with mass
 $O(\sqrt{\beta}v)$ and a ``heavy''  state  with mass $O(\sqrt{\lambda} f)$. 
  It will be later shown that, for a PNGB Higgs particle (that is $v<v_\sigma$ and $f^2>0$), 
  the less fine-tuned regions in parameter space correspond to the case $m\textsubscript{light}=m_h$ and $m\textsubscript{heavy}=m_\sigma$ in the equations above.  In fact, would the $\sigma$ particle be lighter than the Higgs, the roles of the lighter and heavier
    eigenstates would be flipped and the mixing angle $\gamma$ will
    be necessarily outside the region quoted above. The lighter $\sigma$
    scenario is quite different from the typical Higgs PNGB scenarios
    considered in the literature.

 Notice that for $m_h < m_\sigma$ and at variance with the SM case, in the regime of small soft $SO(5)$ breaking the mass of the Higgs and its
 quartic self-coupling are controlled by two different parameters, $\beta$ and $\lam$, respectively. This is consistent with the PNGB nature of the Higgs boson whose mass should now appear protected from growing in the strong interacting regime of the theory --corresponding to large $\lambda$-- in which instead the $\sigma$ mass would increase. In other words,  we have replaced the hierarchy problem for the Higgs particle mass by a sensitivity of the $\sigma$ particle to heavier scales: the $\sigma$ mass represents generically the heavy UV completion. 
The expression for $m_h$ shows that the value of the $\beta$ parameter for small $\beta/4\lambda$ is expected to be 
$\beta\sim {m_h^2}/{2v^2}\sim 0.13$.

\subsection{Scalar-gauge boson couplings}
\label{sec-scalar-gauge}

In the unitary gauge, the kinetic  scalar Lagrangian written in terms or the unrotated fields reads
\begin{equation}
\begin{split}
\ele_{s,kin} =
\frac{1}{2}(\dmu \hat{\sig})^2+\frac{1}{2}(\dmu \hat{h})^2 + \frac{g^2}{4}\paren{\hat{h}+v}^2 W_\mu^+{W^\mu}^-
	+\frac{\paren{g^2+g'^2}}{8}\paren{\hat{h}+v}^2 Z_\mu Z^\mu \,,\nn
\end{split}
\end{equation}
 and justifies the previous identification of the Higgs vev $v$ with the electroweak scale defined from the $W$ mass
 \begin{equation}
  v=246\, \text{GeV}\,.
  \label{vh}
  \end{equation} 
The rotation to the physical  $h,\sigma$ fields results in the following  Lagrangian for the scalar and  scalar-gauge 
interactions, for $m_h<m_\sigma$, 
\bea
\ele_{s} &=& \frac{1}{2}(\dmu\sig)^2+\frac{1}{2}(\dmu h)^2 - \frac{1}{2} m_\sig^2 \sig^2- \frac{1}{2}m_h^2 h^2 
             - \lam\paren{h^2+2\,h\,\sig+\sig^2}^2 \, -\nn \\
  &-&  4\lam (v \cos\ahs - v_\s \sin\ahs) \paren{h^3 + h\,\sig^2} - 
             4 \lam ( v \sin\ahs + v_\s \cos\ahs) \paren{\sig^3 + h^2\, \sig} \, + \nn \\
  &+&  (1+\frac{h}{v}\cos\ahs + \frac{\s}{v}\sin\ahs)^2 
             \left(M^2_W W_\mu^+{W^\mu}^- + \frac{1}{2} M^2_Z Z_\mu Z^\mu\right) \, . 
\label{Lag-scalar-vector-boson}
\eea
The  physical Higgs couplings are thus seen to be weighted down by a $\cos\ahs$  factor with respect to the SM 
Higgs ones, while the $\s$ field acquires the same interactions than $h$ albeit weighted down by $\sin\ahs$, a fact rich in phenomenological consequences to be discussed further below. The SM limit is recovered when the $\s$ field is decoupled from the spectrum  and $\cos\gamma=1$ follows from Eqs.~(\ref{vevs}) and (\ref{mixing}). Conversely, for $m_h>m_\sigma$ the mixing dependence would correspond to the interchange $\cos\ahs\leftrightarrow \sin\ahs$   in Eq.~(\ref{Lag-scalar-vector-boson}).

\subsection{Renormalization and scalar tree-level decays}
\label{renormalization}
The four independent parameters of the scalar Lagrangian can be expressed in terms of four observables, which we choose to be:
\beq
G_F\equiv(\sqrt{2} v^2)^{-1},\qquad m_h,\qquad m_\s,\qquad \sin\gamma\, , 
\label{renorm-scalar}
\eeq with the Fermi constant $G_F$ as measured from muon decay and
$m_h$ from the Higgs pole mass, while $m_\sigma$ could be  determined from future measurements of the
$\sigma$ mass, and $\sin\gamma$
 from either deviations of the Higgs couplings or from the $\sigma$ line shape obtained from its decay into four
  leptons for $m_\sigma \ge 300$ GeV, analogous to the case of a heavy SM Higgs boson~\footnote{For a lighter $\sigma$, the decay width
  becomes too narrow --possibly even below the experimental resolution-- and
  more ingenious procedures would be required to determine the scalar mixing strength, such as for instance
  on-shell to off-shell cross section
  measurements~\cite{Dixon:2013haa}.}.
  
Using Eqs.~(\ref{vevs}), (\ref{mixing}), (\ref{mhsigma}) and
(\ref{vh}), the exact expressions for the $h$ and $\sigma$ vevs in
terms of those physical parameters can be obtained
\beq
\begin{split}
v=&\paren{\sqrt{2}G_F}^{-1/2}\,,\\ 
v_\sig=&\frac{v\sin(2\g) (m_\s^2-m_h^2)}{m_\s^2+m_h^2-(m_\s^2-m_h^2)\cos(2\g)}\,.
\end{split}
\eeq

These expressions in turn allow to express in terms of measured
quantities the four independent parameters of the scalar potential
Eq.~(\ref{Laghs}), which can be written as
\beq
\begin{split}
\lam&=\frac{\sin^2\g m_\sig^2 }{8 v^2}\paren{1 +\cot^2\g \frac{m_h^2}{m_\sig^2} }\,, \\\\ 
\frac{\beta}{4\lam}&=\frac{m_h^2m_\s^2}%
	{\sin^2\g m_\s^4+\cos^2\g m_h^4-2 m_h^2m_\s^2}\,,\\ \\
\frac{\alpha^2}{4\beta^2}&=
	\frac{\sin^2(2\g) (m_\s^2-m_h^2)^2}{4(\sin^2\g m_\s^4+\cos^2\g m_h^4
	-2 m_h^2m_\s^2)}\,,\\ \\
f^2&=
\frac{v^2(\sin^2\g m_\s^4+\cos^2\g m_h^4-2 m_h^2m_\s^2)}%
	{\paren{\sin^2\g m_\s^2+\cos^2\g m_h^2}^2}\,. \\
\end{split}
\label{exact}
\eeq
The above exact formulae show that the mixing angle $\gamma$ does not
coincide with the parameter $v/f\equiv\sqrt{\xi}$ commonly used in the
literature about composite Higgs models, except in the 
\begin{figure}[h]
\centering
\includegraphics[width=0.8\textwidth,keepaspectratio]{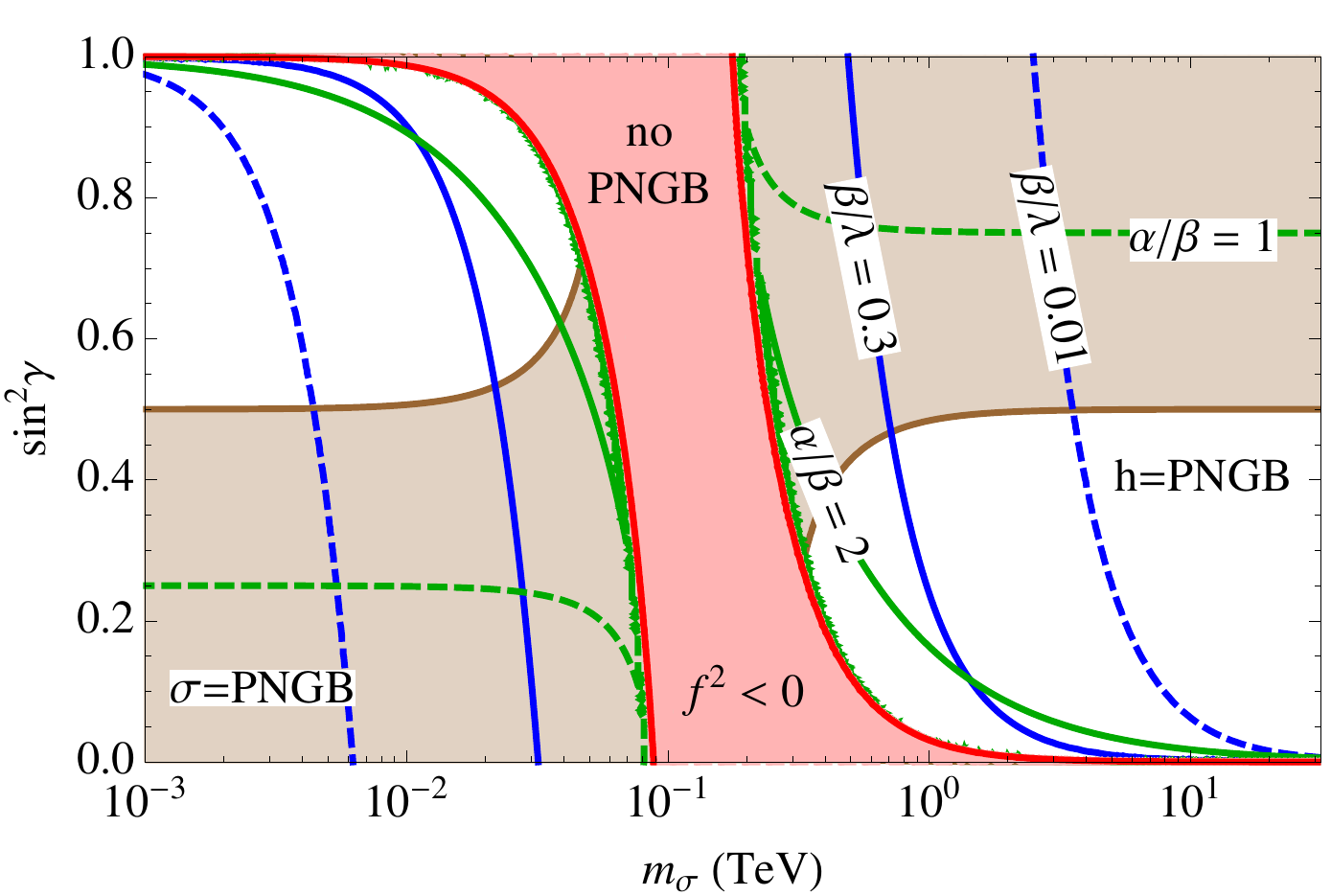}
\caption{ $m_\sigma$ versus $\sin^2 \gamma$ parameter space of the
  scalar sector. The Higgs mass $m_h$ and the Higgs vev $v$ have been
  fixed to their physical values.  The red region corresponds to  $f^2<0$, for which the $SO(5)$-invariant
  part of the potential is unbroken and there are no Goldstone bosons
  in the symmetric limit. The region where
  $|v|>|v_\sigma|$ is shown in brown (these regions are excluded by Higgs
  data, see text).  The Higgs is a pseudo-Goldstone boson within the white regions at
  the bottom-right and the top-left part of the plane.}
\label{Fig-theory-bounds}
\end{figure}
limit $m_\sigma\gg
m_h$ (or more precisely $\beta/4\lam\ll 1$ and $v^2\ll f^2$), where for
sizeable $\sin\gamma$  the last equation above leads to 
\be 
 \sin^2\gamma\underset
{m_\sigma/m_h\gg1}{\longrightarrow} \frac{v^2}{f^2} - 4
\frac{m_h^2}{m_\sigma^2}\,. 
\label{xi-aprox}
\ee

A few comments regarding the parameter space and the scalar
  spectrum are in order, as arbitrary values of $m_\sigma$ and
$\sin^2\g$ are not allowed if we insist on interpreting the Higgs
boson as the pseudogoldstone boson of a spontaneous $SO(5)$
breaking.  Fig.~\ref{Fig-theory-bounds} displays 
  the $(m_\sigma,\sin^2\g)$ plane: at each point the scalar sector is completely defined  as 
   $m_h$ and
  $v$ are fixed to their physical values. The differently colored regions correspond to
  \begin{itemize}
  \item No $SO(5)$ breaking in the light red region, where $f^2<0$; its red borders depict the $f^2=0$ frontier; 
  
  \item The $\sigma$ particle being the PNGB of the spontaneous breaking of $SO(5)$ in the light brown region, where $v_\sigma < v$;
  
    \item The Higgs as the PNGB of the $SO(5) \to SO(4)$ breaking in the white areas, where $v<v_\sigma$ and the Higgs would became a true goldstone boson in the absence of EW breaking ($v\rightarrow 0$).
  
    \end{itemize}
 
\noindent
 A complementary divide is provided by the value of the Higgs mass:
 
  \begin{itemize}
  \item On the  $m_h<m_\sigma$ region to the right of the figure, the  physical Higgs couplings to SM particles are weighted down by $\cos\gamma$ with respect to SM values, see for instance Eq.~(\ref{Lag-scalar-vector-boson}).  It will be shown in the next sections that present LHC Higgs data only allow for values $\sin^2\g< 0.18$  at $2\sigma$ CL, though, leaving as allowed parameter space a fraction of the lower white (Higgs PNGB) section of the figure. The analysis in the next sections will thus focus  in this regime, for which Fig.~\ref{Fig-theory-bounds} already suggests a lower bound on $m_\sigma$ of a few hundreds of GeV.  The relative importance of the soft
  breaking terms is also illustrated through the curves depicted for fixed
  $\alpha/\beta$ and $\beta/\lambda$; 

    \item On the $m_\sigma<m_h$ area to the right of the figure, the  physical Higgs couplings to SM particles are instead weighted down by $\sin\gamma$, whose value will thus be bounded by $\sin^2\gamma > 0.82$ at $2\sigma$ CL. It thus remains as available zone the upper part of the upper white (Higgs PNGB) region. Nevertheless, the quartic coupling $\lambda$ is there very small,  typically $\lambda <10^{-3}$, making the $SO(5)$ invariant potential very flat and potentially unstable against radiative corrections; furthermore, if the soft breaking parameters are required to be small  compared to the symmetric term, $\alpha, \beta < \lambda$,  their values  may require extra fine-tuning with respect to radiative corrections from the fermionic sector to be discussed further below. For these reasons we will not dwell further below on the case $m_\sigma<m_h$ even if phenomenologically of some interest.
  \end{itemize}
  
 Extending the renormalization scheme to the gauge sector, we choose the two extra
observables needed to be the mass of
the $Z$ boson and the fine structure constant,
\beq 
M_Z, \qquad
\alpha_{em}=\frac{e^2}{4\pi}\,,
\label{renorm-scalar}
\eeq
with $M_Z$ and $ \alpha_{em}$ as determined from Z-pole mass
measurements and from Thompson scattering, 
respectively~\cite{Agashe:2014kda}.   
In our model, the relation between the gauge boson masses is the same than that for the SM, 
\beq
M_W=\cos\theta_W M_Z \coma 
\eeq
where the weak angle is given at tree-level by
\beq
\sin^2\theta_W=\frac{1}{2}\paren{1-\sqrt{1-\frac{4\pi \alpha_{em}}{\sqrt{2}G_FM_Z^2}}}\,.
\eeq

Using all the above, it is straightforward to compute the relevant
tree-level branching ratios for the heavy and light scalar boson
decays into SM bosons: 
\bea
\Gamma(h\to W W^*) &=& \Gamma_{\rm SM}(h\to WW^*)\cos^2\gamma\,,  \nn \\
\Gamma(h\to ZZ^*) &=&\Gamma_{\rm SM}(h\to ZZ^*)\cos^2\gamma\,, \nn \\
\Gamma(\sig\to W^+W^-) &=&\frac{\sqrt{2} G_F}{16\pi}m_\sig^3 \sin^2\!\ahs\left[1+\mathcal{O}\left(\frac{M_W^2}{m_\sig^2}\right)\,\right]\,,\nn \\
\Gamma(\sig\to ZZ)&=&\frac{\sqrt{2} G_F}{32\pi} m_\sig^3 \sin^2\!\ahs\left[1+\mathcal{O}\left(\frac{M_Z^2}{m_\sig^2}\right)\right]\,, \nn \\
\Gamma(\sig\to h\,h)&=&\frac{\sqrt{2} G_F}{32\pi} m_\sig^3 \sin^2\!\ahs\left[1+\mathcal{O}\left(\frac{m_h^2}{m_\sig^2}\right)\right]\,, 
\label{tree-scalar-decays}
\eea 
where the SM widths can be found for instance in
  Ref.~\cite{Gunion:1989we}. The $\sigma$ partial widths above will dominate the total $\sigma$ width unless the mixing is unnaturally tiny, and thus measuring the branching ratios is not enough to infer the value of the scalar mixing. It is easy to see that
  \begin{equation}
    \frac{\Gamma_\sigma}{m_\sigma} \simeq \frac{m_\sigma^2\sin^2\gamma}{8v^2},
  \end{equation}
  and thus the measurement of the line shape of the $\sigma$ seems  feasible only for $m_\sigma$ above the EW breaking scale
  (assuming non-negligible mixing). In that regime, the value of $\sin\gamma$ can be inferred from the line shape and all other
observables in Eq.~(\ref{tree-scalar-decays}) can then be predicted in
terms of the physical parameters defining our
  renormalization scheme. Other bosonic decay channels requiring
  one-loop amplitudes will be discussed later on.
 

%

%
\section{Fermionic sector}
\begin{figure}
\centering
\includegraphics[width=0.7\textwidth,keepaspectratio]{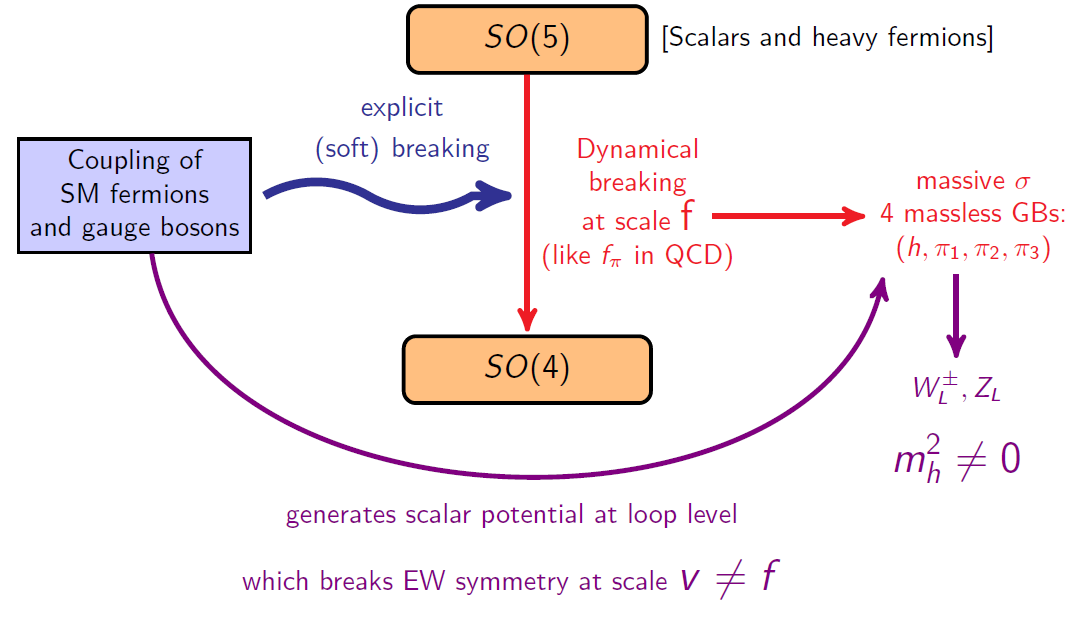}
\caption{Schematics of the $SO(5)\to SO(4)$ model}
\label{Fig-schema}
\end{figure}

\begin{figure}
\centering
\includegraphics[width=0.7\textwidth,keepaspectratio]{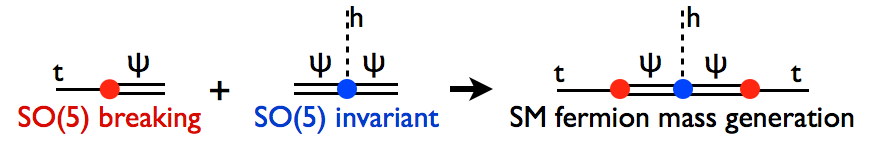}
\caption{Schematics of light fermion mass generation. The light SM fermions -here the top quark- couple to the 
 heavy partners breaking  explicitly $SO(5)$. The middle image depicts the $SO(5)$ invariant Yukawa interactions between the Higgs and the heavy partners.
 The combination of both couplings induces and effective top Yukawa coupling and thus a massive top quark.}
\label{mass-cartoon}
\end{figure}

%
The fermionic sector is unavoidably an important source of model dependency as diverse choices of $SO(5)$ fermionic multiplets are possible.
Moreover the achievement of the desired symmetry breaking pattern in ``composite Higgs'' models relies on the fermionic sector.
A schematic picture is given in Fig.~(\ref{Fig-schema}) considering a high energy global symmetry group - SO(5) in the case under discussion:

 \begin{itemize}
 \item Heavy scalar and fermion representations of the high-energy
   global symmetry are considered.  $SO(5)$ breaks spontaneously to
   $SO(4)$ at a scale $f$, resulting in four massless Goldstone
   bosons: the three longitudinal components of the electroweak gauge
   bosons and a ``Higgs Goldstone boson''. The fifth component $\sigma$
   remains massive.
 \item Furthermore, $SO(5)$ is explicitly broken by the
   coupling of the exotic heavy representations to the SM fermions (soft breaking) and
   to the gauge bosons (hard breaking). This induces at one-loop a potential for the $h$
   field with a non-trivial minimum, providing a mass for $h$ and
   breaking the SM electroweak symmetry at a scale $v\ne f$.
\end{itemize}
 Several ``minimal'' possibilities have been explored in the
literature for the exotic fermionic representations (see for instance
Refs.~\cite{Barbieri:2007bh, Carena:2014ria}). The setup considered in
this paper contains: 
\begin{enumerate} 
\item Heavy (exotic) vector-like fermions in complete representations
  of $SO(5)$, either in the fundamental representation, denoted below
  by $\psi$ - or singlets denoted by $\chi$.
\item A scalar field $\phi$ in the fundamental representation of
  $SO(5)$, which contains the $h$ and $\sigma$ particles. Its vev
  breaks $SO(5)$ spontaneously to $SO(4)'$. By construction {\em only}
  the heavy exotic fermions couple directly to the scalar $\phi$.
\item The Higgs field couples to the
  exotic fermions only via $SO(5)$ invariant Yukawa couplings. The sources of $SO(5)$ breaking lie instead in the electroweak gauge
  interactions and in mixing terms between the heavy exotic fermions
  and the SM fermions.  
  Such a breaking is fed via loop corrections to the scalar
  potential, where it is modeled by two $SO(5)$ soft breaking terms
  which are custodial preserving.
\end{enumerate}
This choice of fermionic representation respects an approximate custodial
  symmetry which  protects the $Zbb$
  coupling~\cite{Agashe:2006at}.  Fig.~(\ref{mass-cartoon})
  illustrates an intriguing characteristic of the fermionic sector in
  this class of models --which are often denominated by the generic
  name of ``partial compositeness''~\cite{Kaplan:1991dc}: {\it a
    seesaw-like mechanism is at work in the generation of all
    low-energy fermion masses}. The heavier the exotic fermions the
  lighter the light fermions.  

 To ensure correct hypercharge assignments for the SM fermions coupled directly to  heavy exotic fields,    
  the global symmetry  is customarily enlarged by (at least) an extra $U(1)_X$ sector,  leading finally to a pattern of   spontaneous global symmetry breaking  given by  
\bea
SO(5) \times U(1)_X \to SO(4) \times U(1)_X \approx SU(2)_L \times SU(2)_R \times U(1)_X \,, 
\eea
with   the hypercharge
corresponding now to a combination of the new generator and that of
$SU(2)_R$ generator, see Eq.~(\ref{covariant}), 
\bea
Y = \Sigma^{(3)}_R + X\,.
\label{hypercharge}
\eea
As the global $U(1)_X$ symmetry remains unbroken, no additional
Goldstone bosons are generated. 
 Two different $U(1)_X$ charges are compatible with SM hypercharge
 assignments: $2/3$ and $-1/3$.  We will indeed consider two
 different copies of heavy fermions for each representation,
 differentiated by the $U(1)_X$, as they are necessary to induce mass terms for
 both the SM up and the down quark sectors. 
 Schematically, the fundamental and singlet representations can be
 decomposed under $SU(2)_L$ quantum numbers as follows, 
\bea
\psi^{(2/3)} &\sim& (X,Q,T^{(5)}) \quad \quad, \quad\quad \psi^{(-1/3)} \sim(Q',X',B^{(5)})\,, \nn \\
\chi^{(2/3)} &\sim& T^{(1)}  \hspace{1.25cm} \quad\quad ,\quad\quad \chi^{(-1/3)} \sim B^{(1)} \, ,\nn 
\eea
where $X^{(')},Q^{(')}$ denote the two different $SU(2)_L$ doublets
contained in the fundamental representation of $SO(5)$. In each
multiplet, the first doublet has $\Sigma^{(3)}_R=1/2$ while the second
one has $\Sigma^{(3)}_R=-1/2$.  $T_{(1,5)},B_{(1,5)}$ denote instead
$SU(2)_L \times SU(2)_R$ singlets, respectively in the {\bf 5} and
{\bf 1} representation of $SO(5)$. Table~\ref{tabQN} summarizes the relevant
quantum numbers for all heavy fermions.

\begin{table}[htb]
\centering 
\renewcommand{\arraystretch}{1.5}
\footnotesize
\begin{tabular}{|c|c|c|c|c|c|c| }
\hline
Charge/Field & $X$ & $Q$ & $T_{(1,5)}$ & $Q'$ & $X'$ & $B_{(1,5)}$ \\[0.5ex] 
\hline
$\Sigma^{(3)}_R$ & $+1/2$ & $-1/2$ & 0 & $+1/2$ & $-1/2$ & 0 \\
\hline
$SU(2)_L \times U(1)_Y$ & $(2,+7/6)$ & $(2,+1/6)$ & $(1,+2/3)$ & $(2,+1/6)$ & $(2,-5/6)$ & $(1,-1/3)$ \\
\hline 
$x$ & $+2/3$ & $+2/3$ & $+2/3$ & $-1/3$ & $-1/3$ & $-1/3$ \\
\hline
$q_{EM}$ & $\begin{matrix} X^u = +5/3 \\ X^d = +2/3 \end{matrix}$ & $\begin{matrix} Q^u = +2/3 \\ Q^d = -1/3 \end{matrix}$ 
         & $+2/3$ & $\begin{matrix} Q'^u = +2/3 \\ Q'^d = -1/3 \end{matrix}$  
         & $\begin{matrix} X'^u = -1/3 \\ X'^d = -4/3 \end{matrix}$ & $-1/3$ \\
\hline 
\end{tabular}
\caption{Heavy fermion charges assignments.}\label{tabQN}
\end{table}

%
\subsection*{The fermionic Lagrangian}
\label{Sec-fermionic-Lag}
%

For the SM fermions, the analysis below will be restricted to the third generation of SM quarks for simplicity,  denoting by  $q_L$  and  $t_R$ and $b_R$  the doublet and singlets, respectively. 
   It would be straightforward to extend the results to the other two generations, for instance introducing heavier replica of the exotic sector, leading to very minor additional phenomenological impact.

Assuming the ``minimal'' content specified in the previous sections, the fermionic Lagrangian is given by
\bea
\mathcal{L}_F
   &=& \bar{q}_{L} i\Ds  q_L + \bar{t}_R i\Ds  t_R + \bar{b}_R i\Ds  b_R\nn\\ 
  & +& \bar{\psi}^{(2/3)}\left(i\Ds-M_5\right)\psi^{(2/3)}+\bar{\psi}^{(-1/3)}\left(i\Ds-M'_5\right)\psi^{(-1/3)}\nn \\
   &+&\hspace{-0.25cm} \bar{\chi}^{(2/3)}\left(i\Ds-M_1\right)\chi^{(2/3)}+\bar{\chi}^{(-1/3)}\left(i\Ds-M'_1\right)\chi^{(-1/3)}\nn \\
   &-\Big[&\hspace{-0.25cm} y_1\,\bar{\psi}_{L}^{(2/3)} \phi \,\chi_{R}^{(2/3)}+y_2\,\bar{\psi}_{R}^{(2/3)} \phi \,\chi_{L}^{(2/3)} 
    +y'_1\,\bar{\psi}_{L}^{(-1/3)} \phi \,\chi_{R}^{(-1/3)}+y'_2\,\bar{\psi}_{R}^{(-1/3)} \phi \,\chi_{L}^{(-1/3)}\nn \\
   &+&\hspace{-0.25cm} \Lam_1\paren{\bar{q}_L{\De_{2\times5}^{(2/3)}}}\psi_R^{(2/3)}		          						      
    + \Lam_2 \,\bar{\psi}_L^{(2/3)} \paren{\De_{5\times1}^{(2/3)} t_R} + \Lam_3 \,\bar{\chi}_L^{(2/3)} t_R \nn \\
   &+&\hspace{-0.25cm} \Lam_1'\paren{\bar{q}_L{\De_{2\times5}^{(-1/3)}}}\psi_R^{(-1/3)}   
     + \Lam'_2 \,\bar{\psi}_L^{(-1/3)} \paren{\De_{5\times1}^{(-1/3)} b_R} +\Lam'_3 \,\bar{\chi}_L^{(-1/3)} b_R+h.c.\Big] \,.
\label{SO5Lag}
     \eea
The first lines contain the kinetic terms for the SM fermions. The second and third lines include  the kinetic and mass terms for the exotic fermions. The kinetic terms become $SO(5)$-invariant
in the gaugeless limit. The fourth line contains the $SO(5)$ invariant Yukawa couplings of the exotic sector to the Higgs field.
     Finally, the last two lines of the Lagrangian contain the $SO(5)$ soft-breaking interactions of SM fermions with exotic fermions.  $\Delta_{2\times5}$ and $\Delta_{5\times 1}$ denote suitable spurions connecting $SO(5)$ and 
$SU(2)\times U(1)$ representations. If the primed parameters were set to zero no bottom mass would be generated through 
this mechanism. All parameters in Eq.~(\ref{SO5Lag}) are assumed real for simplicity, that is, we will assume CP invariance in what follows.

It is useful to rewrite the Lagrangian in Eq.(\ref{SO5Lag}) in terms of  $SU(2)_L$ components. For this purpose, from this point and until Eq.~(\ref{matrix-K2}) below, $h$ and $\sigma$ will denote again the unshifted and unrotated original scalar fields in Eq.~(\ref{Laghs}):
\bea
\mathcal{L}_F
    &=& \bar{q}_{L} i\Ds  q_L + \bar{t}_R i\Ds t_R + \bar{b}_R i\Ds  b_R 
       + \bar{Q} \left(i\Ds-M_5\right) Q + \bar{X} \left(i\Ds-M_5\right) X  \nn \\
 &+& \hspace{-0.25cm} \bar{T}^{(5)} \left(i\Ds-M_5\right) T^{(5)} + \bar{T}^{(1)} \left(i\Ds-M_1\right) T^{(1)} 
       + \bar{Q'} \left(i\Ds-M'_5\right) Q' + \bar{X'} \left(i\Ds-M'_5\right) X'  \nn \\
    &+& \hspace{-0.25cm}\bar{B}^{(5)} \left(i\Ds-M'_5\right) {B}^{(5)} + \bar{B}^{(1)} \left(i\Ds-M'_1\right) {B}^{(1)} \nn \\
    &-\Big[& \hspace{-0.25cm}y_1 \paren{\bar{X}_L H{}  T_R^{(1)} +\bar{Q}_L \widetilde{H}{}  T_R^{(1)}
        +\bar{T}_L^{(5)}\sig T_R^{(1)}}+y_2\,\paren{\bar{T}_L^{(1)} H{}^\dagger X_R 
        +\bar{T}_L^{(1)}\widetilde{H}{}^\dagger Q_R+\bar{T}_L^{(1)}\sig T_R^{(5)}} \nn \\
    &+& \hspace{-0.25cm}y'_1 \paren{\bar{X'}_L \widetilde{H} B_R^{(1)} +\bar{Q'}_L H B_R^{(1)}
        +\bar{B}_L^{(5)}\sig B_R^{(1)}} + y'_2\,\paren{\bar{B}_L^{(1)} \widetilde{H}^\dagger X'_R 
        +\bar{B}_L^{(1)} H^\dagger Q'_R+\bar{B}_L^{(1)}\sig B_R^{(5)}} \nn \\
    &+& \hspace{-0.25cm}\Lam_1\bar{q}_L Q_R+ \Lam_1'\bar{q}_L Q_R'+\Lam_2\bar{T}_L^{(5)} t_R+\Lam_3\bar{T}_L^{(1)} t_R 
      +\Lam'_2\bar{B}_L^{(5)} b_R+\Lam'_3\bar{B}_L^{(1)} b_R + h.c.\Big]\,.
\label{SO5LagD}
\eea
Eq.~(\ref{SO5LagD}) shows that the light fermion masses must be proportional to the $SO(5)$ invariant Yukawa couplings of heavy fermions and to the explicitly $SO(5)$ breaking light-heavy fermionic interactions.  
  The generation of light quark masses requires a vev for the scalar doublet $H$.
For instance,  a $\bar{t}_Lt_R$ mass term is seen to result from the following chain of couplings, 
\begin{equation}
q_l \underset {\Lambda_1}{\longrightarrow}  Q_R \underset {M_5}{\longrightarrow}  Q_L \underset {y_1\langle\tilde{H}\rangle}{\longrightarrow} T_R^{(1)} \underset {M_1}{\longrightarrow} T_L^{(1)}\underset {\Lambda_3}{\longrightarrow} t_R\,, 
\end{equation}
suggesting 
\begin{equation}
m_t \propto   y_1\,\frac{\Lambda_1\, \Lambda_3}{M_1\, M_5}\,v\,,
\label{topmasscartoon}
\end{equation}
 see also Fig.~\ref{mass-cartoon} and Sect.~\ref{EffectiveLag}. 
Furthermore, both the  $+2/3$ and $-1/3$ electrically charged sectors acquire off-diagonal mixing terms. 

The expression for the fermionic Lagrangian  Eq.~(\ref{SO5LagD}) can be rewritten in a compact form defining a fermionic vector whose components are ordered by their  electrical charges  $q_{EM} = (+5/3,+2/3,-1/3,-4/3)$,
\bea
\Psi = 
\l
X^u , \
 \mathcal{T} , \
\mathcal{B} ,\
X'^d
\r,
\eea
where  $\mathcal{T}$ and $\mathcal{B}$ include the top and bottom quarks together with their heavy fermionic partners
\bea
 \mathcal{T} = \l t,Q^u,X^d,T^{(5)},T^{(1)},Q'^{u} \r , \quad 
 \mathcal{B} = \l b,Q'^d,X'^u,B^{(5)},B^{(1)},Q^d \r .
\eea
The fermion mass terms in the weak basis can then be written as
\bea
\mathcal{L}_\mathcal{M} = -\bar{\Psi}_{L} \ \mathcal{M}(h,\sigma)  \  \Psi_R,
\label{Lag-K}
\eea
where here and in what follows the sum over all components of the fermionic vector is left implicit and the block diagonal $14\times14$ fermion mass matrix $\mathcal{M}$ reads 
\bea
\mathcal{M}(h,\sigma) &=& \text{diag}\l 
M_5,\mathcal{M}^{\mathcal{T}}(h,\sigma),\mathcal{M}^{\mathcal{B}}(h,\sigma), M'_5 \r, \\[3mm]
\mathcal{M}^{\mathcal{T}}(h,\sigma) &=&
\begin{blockarray}{cccccc}
    \begin{block}{(cccccc)}
    0  &\Lambda_1  &0 &0  &0   &\Lambda_1'\\
   0  & M_5&  0& 0  & y_1\frac{h}{\sqrt{2}}&0\\
  0    &0 &M_5  & 0   & y_1\frac{h}{\sqrt{2}} &0\\
    \Lambda_2 &  0 &0 &  M_5   & y_1\sigma &0\\
   \Lambda_3 & y_2 \frac{h}{\sqrt{2}}  &y_2 \frac{h}{\sqrt{2}} & y_2\sigma & M_1&0\\
    0  &0 & 0 &   0& 0& M_5'\\
    \end{block}
  \end{blockarray}\,\,,  \\
\mathcal{M}^{\mathcal{B}}(h,\sigma) &=& 
\mathcal{M}^{\mathcal{T}}(h,\sigma)
 \ \text{with} \ \{ y_i, \Lambda_i, M_i\}\leftrightarrow \{ y'_i, \Lambda'_i, M'_i\} \,.
\label{matrix-K2}
\eea
The mass matrices can be diagonalised by bi-unitary (or for the case of the real parameters bi-orthogonal)  transformations, 
\bea
\label{FermRot}
\Psi^{\text{phys}}_L &=& L  \Psi_L \,,  \quad 
\Psi_R^{\text{phys}} = R  \Psi_R\,,  \quad \mathcal{M}^{\text{diag}} = L^\dag \mathcal{M} R\,.
\eea
 These matrices can be diagonalized analytically 
in some interesting limits; in general they will be diagonalized numerically.

 The physical light eigenstates are admixtures of the light and heavy fermion fields appearing in the Lagrangian.   
The scalar fields vevs induce in addition heavy fermion mass
splittings. Notice however that, even in the limit of vanishing Yukava
couplings, the exotic fermions get mixed via the $SO(5)$ breaking
couplings. Moreover, although the various dimensional couplings
$\Lambda_i$ and $\Lambda_i'$ in Eqs.~(\ref{SO5Lag}) and
$(\ref{SO5LagD})$ may be of the same order,  the top and bottom components of the heavy doublets are 
splitted by $SO(4)$ breaking terms, generically of $O(y_iv)$.

\section{ Phenomenology}

In this section,  bounds are derived first on the model parameters resulting from present LHC Higgs data and from electroweak precision tests - namely S,T 
and $g_L^b$. Future signals are discussed next, focusing in particular on $\sigma$ physics.

\subsection{Bounds from Higgs measurements}
\label{sec-h-bounds}

The tree-level mixing of the scalar singlet $\sigma$ with the Higgs resonance $h$ can be strongly bounded from present data and in particular from $h$ to $ZZ$ and $W^+W^-$ decays, and from $h$-gluon-gluon transitions: the Higgs coupling strength to SM fields is weighted down simply by a $\cos\g$ factor with respect  the SM value, as previously explained and shown in Eq.~(\ref{Lag-scalar-vector-boson}).   
We use the latest ATLAS and CMS combined results for the
gluon-gluon and vector boson mediated Higgs production 
processes~\cite{CMS-ATLAS-comb}.

A $\chi^2$ fit taking into account the correlation between the corresponding
coupling modifiers in the combined fit 
of the 7 and 8 TeV LHC data -given by figure 23.B of Ref.~\cite{CMS-ATLAS-comb}-
constrains directly $\cos\g$, translating into the following bound 
\beq
\sin^2\g \lesssim 0.18~\text{(at $2\sigma$)}\,,
\label{xilimit}
\eeq
which in the $m_\sigma\gg m_h$ limit would point to a value for  the non-linearity parameter of composite Higgs models, $\xi\equiv {v^2}/{f^2} \sim\sin^2\g$, consistent with the limits found in the literature~\cite{Panico:2015jxa}, see Eqs.~(\ref{exact}) and (\ref{xi-aprox}) and the discussion below.
\subsection*{Comparison with literature on non-linear realizations}
Ref.~\cite{Panico:2015jxa} shows that in non-linear realizations of the composite Higgs scenario the behaviour of the  Higgs
couplings modifications varies  depending on the $SO(5)$ fermionic representations chosen. In particular they compare the so called  
$MCHM_4$ and $MCHM_5$ scenarios:
\begin{itemize}
\item In $MCHM_4$, the fermions (both the embedded light ones and the heavy partners) are in the {\bf 4} (spinorial) representation of $SO(5)$; the coupling modifiers then obey $k_V^{(4)}=k_F^{(4)}=(1-\xi)^{1/2}$, leading to a bound from Higgs data $\xi^{(4)}<0.18$ at $2\sigma$ CL. $MCHM_4$  is actually ruled out by its impact on the $Zbb$ coupling and thus for instance disregarded  in Ref.~\cite{Carena:2014ria}.
\item In $MCHM_5$, the fermions are instead in the {\bf 5} (fundamental) of $SO(5)$, and in this case $k_V^{(5)}=(1-\xi)^{1/2}$ differs from $k_F^{(5)}=({1-2\xi}){(1-\xi)^{-1/2}}$, that is, ${k_F^{(5)}}/{k_V^{(5)}}
		\approx 1 -\xi $ for small $\xi$ values. LHC Higgs data set then a bound $\xi^{(5)}<0.12$ at $2\sigma$ CL.
\end{itemize}
Now, the heavy fermion configuration of our model seems alike to that in $MCHM_5$ if the $\sigma$ particle was disregarded ($\sigma$ does not intervene in the tree-level Higgs data analysis of SM couplings),  and in spite of including one heavy fermion in a {\bf 1} (singlet) of SO(5) as the latter could be integrated out to mimic the $MCHM_5$ spectrum discussed for instance in Refs.~ \cite{Anastasiou:2009rv} or  \cite{Carena:2014ria}. Indeed, according to the notation  in Ref.~\cite{Carena:2014ria}, the fermion representation in our model would be given by  $MCHM_{Q-T-B}\to MCHM_{5-5,1-5,1}$; nevertheless, 
the behaviour for the
modifiers --and thus the resulting $\xi$ bound-- is the same than in $MCHM_4$, see Eq.~(\ref{xilimit}).  Plausibly, this apparent paradox may be resolved  by taking into account that the limit of very heavy $\sigma$ corresponds to the $\lambda$ coupling reaching the non-perturbative regime, and a resumation of the strong interacting effects may  be needed to  fully reach the non-linear regime.

\subsection {Precision electroweak constraints}
\label{precision}

Analyses available on precision tests for composite Higgs
models, such as that in Ref.~\cite{Ghosh:2015wiz}, tend to consider non-linear versions of the theory
where the only scalar present is the Higgs particle, but for 
Ref.~\cite{Barbieri:2007bh},
which discusses qualitatively the interplay of scalar and exotic
fermion contributions. See also Ref.~\cite{Gertov:2015xma}.  We present here an
explicit computation of the scalar ($h$ and $\sigma$) and exotic
fermion contributions, discussing the impact of varying the $\sigma$
mass. $S,T$ and $U$ parameters are considered together with $g_L^b$ and
parameter correlations.
%
\subsubsection{$S$, $T $ and $g^b_L$ }
%
%
Consider the parameter definitions in Ref.~\cite{Peskin:1991sw}, 
\bea
\alpha\, S&=& 4 s_W c_W \frac{d \Pi_{30}(q^2)}{dq^2}|_{q^2=0} = 4 s_W
c_W \, F_{30}\,, \nn \\ 
\alpha\, T&= &\frac{1}{M_W^2} [\Pi_{11}(0) -
  \Pi_{33}(0)]= \frac{1}{M_W^2} [A_{11} - A_{33}] \,,\nn \\ 
\alpha\,
U&= & -4 s_W^2 \frac{d }{dq^2}[\Pi_{33}(q^2)-\Pi_{11}(q^2)]|_{q^2=0}=
4 s_W^2 (F_{11}-F_{33})\,,
\eea 
where $c_W$ ($s_W$) denotes the cosinus (sinus) of the Weinberg angle
$c_W= M_W/M_Z$, and the electroweak vacuum polarization functions are
given by
\begin{equation}
\Pi_{ij}^{\mu\nu}(q)=-i[\Pi_{ij}(q^2) g^{\mu\nu} + (q^\mu q^\nu -terms)] \,; \qquad \Pi_{ij}(q^2)\equiv A_{ij} (0)+ q^2 F_{ij } +...\\
\end{equation}
with $i,j= W,Z$ or $i,j= 0,3$ for the $B$ or the $W_3$ bosons,
respectively, and the dots indicating an expansion in powers of
$q^2$. We will not consider further $U$ as it typically corresponds to
higher order (mass dimension eight) couplings while only low-energy
data (e.g. LEP) will be used here~\footnote{Some other handful
  parameter definitions are: $\epsilon_1\equiv \alpha T\,;
  \epsilon_3=\alpha S \,/(4 s_W^2)$, see Ref.~\cite{Altarelli:1990zd}.}. On the
contrary, relevant constraints could stem from deviations induced in the
$Z\bar{b}_Lb_L$ coupling, parametrized by $g^b_L$ in the decay
amplitude
\be
\mathcal{M}_{Z\to{\bar{b}_L} b_L} = - \frac{e\, g^b_L}{s_W c_W}
\bar{b} (p_2) \,\slashed{\epsilon}(q) \frac{1- \gamma_5}{2}\, b(p_1),
\label{gLb}
\ee
where $\epsilon(q)$ denotes the $Z$ boson polarization and $p_i$ the $b$ quark and antiquark momenta.

The values of $S$, $T$ and $g^b_L$  are allowed to deviate from the SM prediction within the constraints~\cite{Ciuchini:2014dea,Ghosh:2015wiz} 
\bea
\Delta S&\equiv& S-S_{SM}=0.0079 \pm 0.095\,\,,\nn\\
\Delta T&\equiv& T - T_{SM}= 0.084\pm0.062\,, \nn\\
\Delta g^b_L& \equiv &{g^b_L} - {g^b_L}_{SM}= (-0.13\pm0.61)\times10^{-3}\,,
\eea
with the ($S,T,g^b_L $) correlation matrix  given by
\be
\begin{pmatrix}
1 & 0.864 & 0.06 \\
0.864 & 1 & 0.123 \\
0.06 & 0.123 & 1 \\
\end{pmatrix}\,.
\ee

\subsubsection*{Scalar  contributions in the linear $SO(5)$ model: $h$ and $\sigma$}

Given the scalar couplings in Eq.~(\ref{Lag-scalar-vector-boson}),  their contributions  to $S$ and $T$ can be formulated as
\begin{equation}
\Delta T^{(h\,\textrm{and}\,\sigma)}= - \Delta {T_{SM}^{h}}(m_h) + c_\gamma^2 {\Delta T_{SM}^h} (m_h)+ \Delta T^{(\sigma)} =  s_\gamma^2\left[ -\Delta T_{SM}^{h}(m_h) +\Delta T_{SM}^{h}(m_\sigma)\right] \,,
\label{DeltaT_h+sigma}
\end{equation}
\begin{equation}
\Delta S^{(h \,\textrm{and}\,\sigma)}= - \Delta S_{SM}^{h}(m_h) + c_\gamma^2 \Delta S_{SM}^h (m_h) +\Delta S^{(\sigma)} =  s_\gamma^2\left[
-\Delta S_{SM}^h (m_h) +\Delta S_{SM}^h (m_\sigma) \right] \,,\label{DeltaS_h+sigma}\,
\end{equation}
where the $\sigma$ contributions   $\Delta T^{(\sigma)}$ and $\Delta S^{(\sigma)}$ have been simply written in terms of the usual SM formulae  for the Higgs contribution $\Delta T_{SM} ^{h}$ and $\Delta S_{SM}^{h}$  with the replacement $m_h \rightarrow m_\sigma$ and using the formulae valid for masses much above the electroweak scale.  The scalar contribution to $\Delta T$ is then given by
\begin{equation}
\begin{split}
\Delta T^{(h\,\textrm{and}\,\sigma)}=& \frac{3 G_F M_W^2}{8 \pi^2 \sqrt{2}} s_\gamma^2\Bigg( - m_h^2 \frac{\log(m_h^2/M_W^2)}{M_W^2-m_h^2} +  m_\sigma^2 \frac{\log(m_\sigma^2/M_W^2)}{M_W^2-m_\sigma^2}\\
 & + \frac{M_Z^2}{M_W^2} \left\{  m_h^2 \frac{\log(m_h^2/M_Z^2)}{M_Z^2-m_h^2} -  m_\sigma^2 \frac{\log(m_\sigma^2/M_Z^2)}{M_Z^2-m_\sigma^2}\right\} \Bigg)\,,
\end{split}
\end{equation}
which in the limit ${m_\sigma \gg m_h, M_W, M_Z}$ reduces to 
\begin{equation}
\Delta T^{(h\,\textrm{and}\,\sigma)}\sim \, s_\gamma^2\,  \frac{3 G_F M_W^2}{8 \pi^2 \sqrt{2}} \,  \frac{s_W^2}{c_W^2} \log(m_\sigma^2/M_W^2)\,.
\end{equation}

For the $\Delta S$ corrections, the formulation in Refs.~\cite{Novikov:1992rj,Orgogozo:2012ct} is used, leading to
 \begin{equation}
\begin{split}
\alpha \Delta S_{SM}^{h}(m)
=  s_W^2  \frac{2G_F}{\sqrt{2}\pi^2}M_W^2\Bigg(  
\frac{x}{12(x-1)}\log(x)+ \left(-\frac{x}{6} +\frac{x^2}{12}\right)F(x)-\left(1-\frac{x}{3}+\frac{x^2}{12}\right)F'(x) \Bigg)\,,
\end{split}
\end{equation}
where $x\equiv m^2/M_Z^2$ and for $x<4$:
 \begin{equation}
 \begin{split}
 &F(x)= 1+ \left(\frac{x}{x-1} - \frac{1}{2}x\right)\log{x} - x\sqrt{\frac{4}{x}-1}\arctan\sqrt{\frac{4}{x}-1}\,,\\
 &F'(x)= -1+ \frac{x-1}{2}\log{x} + (3-x) \sqrt{\frac{x}{4-x}}\arctan\sqrt{\frac{4}{x}-1}\,,
 \end{split}
 \end{equation}
 while for $x>4$: 
 \begin{equation}
 \begin{split}
 &F(x)= 1+ \left(\frac{x}{x-1} - \frac{1}{2}x\right)\log{x} - x\sqrt{1-\frac{4}{x}}\,\log\left(\sqrt{\frac{x}{4}-1} +\sqrt{\frac{x}{4}}\right) \,,\\
 &F'(x)= -1+ \frac{x-1}{2}\log{x} + (3-x) \sqrt{\frac{x}{x-4}}\,\log\left(\sqrt{\frac{x}{4}-1} +\sqrt{\frac{x}{4}}\right) \,.
 \end{split}
 \end{equation} 
  In the limit of very large $m_\sigma$, the $\sigma$ contribution to
  $S$ can be approximated by:
\begin{equation}
\alpha\, \Delta S^{(\sigma)}\underset{\sigma\rightarrow \infty}{\longrightarrow} s_\gamma^2 s_W^2\frac{2G_F}{\sqrt{2}\pi^2}M_W^2\left[ \frac{1}{12}\log\left(\frac{m_\sigma^2}{M_W^2}\right)\right]\,,
\end{equation}
consistent with the statements in the literature for a very heavy
Higgs particle~\cite{Peskin:1991sw}.
\begin{figure}
\centering
\includegraphics[width=0.5\textwidth,keepaspectratio]{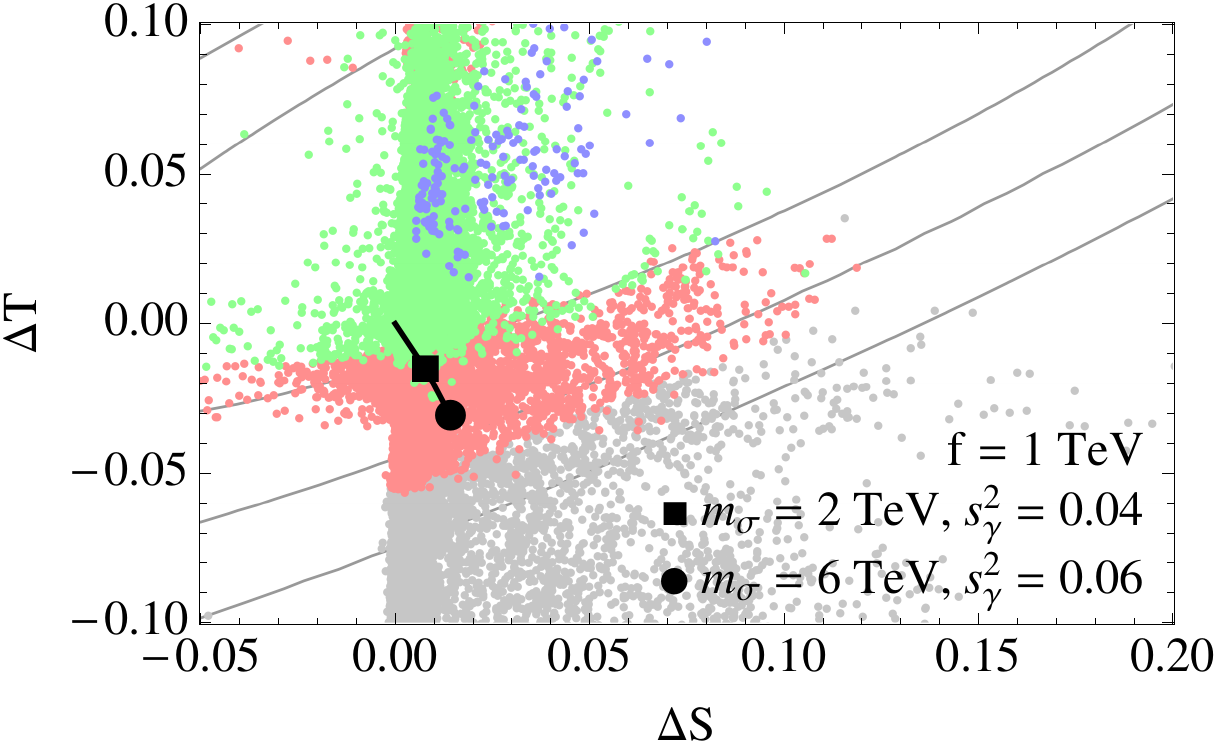}
\caption{Uncombined contributions of the scalar sector (black curve)
  and the exotic fermionic sector to the parameters $S$ and $T$.}
\label{Fig-ST-sigma}
\end{figure}

The black curve in Fig.~\ref{Fig-ST-sigma} displays examples of the
$\Delta S$ and $\Delta T$ corrections induced by the $\sigma$ scalar
as it follows from the formulae shown above.  As earlier explained,
the set of parameters in the scalar potential
$(f\,,\lambda\,,\alpha\,,\beta)$ has been traded by four observables:
$G_F$, $m_h$, $m_\sigma$ and the scalar mixing $\g$ (with the latter two  yet to be experimentally measured). It is nevertheless theoretically
illuminating to indicate the corresponding values for $f$ and the
scalar quartic self-coupling $\lambda$ for each example
analyzed, and their values are shown in all figures to follow.
 We present numerical results for two typical parameter regimes:
\begin{itemize}
\item $m_\sigma= 2$ TeV, $s_\gamma^2=0.04$, which corresponds to $f=
  1$ TeV and to scalar potential couplings $\lambda= 0.38$,
  $\alpha=0.35$ and $\beta=0.16$, which clearly lie within the
  perturbative regime of the linear $SO(5)$ sigma model.
\item $m_\sigma= 6$ TeV, $s_\gamma^2=0.06$, which also correspond to 
  $f=1$ TeV, while $\lambda=4.3$ -closer to the limit of validity of
  the perturbative expansion- and $\alpha= 0.25$, $\beta=0.13$; this
  pattern corresponds then to a mainly $SO(5)$ symmetric scenario with
  small soft symmetry breaking.
\end{itemize}
Fig.~\ref{Fig-ST-sigma} shows a sizeable negative contribution of the
$\sigma$ particle to $\Delta T$ which increases with $m_\sigma$, and
positive contribution to $\Delta S$; the result is consistent with the
pattern expected in Ref.~\cite{Barbieri:2007bh}, and similar to that for the
heavy Higgs case (see e.g.~Ref.~\cite{Haber:2010bw}). In the limit
$m_\sigma \rightarrow m_h$ the total scalar contribution matches that
in the SM due to the Higgs particle. It is easy to extrapolate the $S$
and $T$ scalar contributions to other mixing regimes as they scale
with $s_\gamma^2$: for instance the effect would be amplified by a
factor of $\sim 3$ when raising the mixing towards the maximal value
allowed, see Eq~(\ref{xilimit}).

 For $g_L^b$ instead we will not analyze the one-loop $\sigma$ contributions, as they would be proportional to the bottom Yukawa couplings and thus negligible compared to the top and top-partner contributions to be discussed next.  


\subsubsection*{Fermionic contributions}

The heavy fermion sector may have an impact  on the oblique parameters and on $g^b_L$.
 This sector adds additional parameter dependence on top of the four renormalization parameters already discussed for the scalar sector of the linear $SO(5)$ sigma model. The fermionic parameter space is quite large and adjustable, and thus in practice  $m_\sig$ and $\g$  will be treated here as independent from them. It will also be assumed that the inclusion of  quarks and
leptons from the first two generations does not alter significantly the analysis of electroweak precision tests, as lighter
fermions tend to have very small mixing with their heavy partners. 

\begin{figure}
\centering
\includegraphics[width=\textwidth,keepaspectratio]{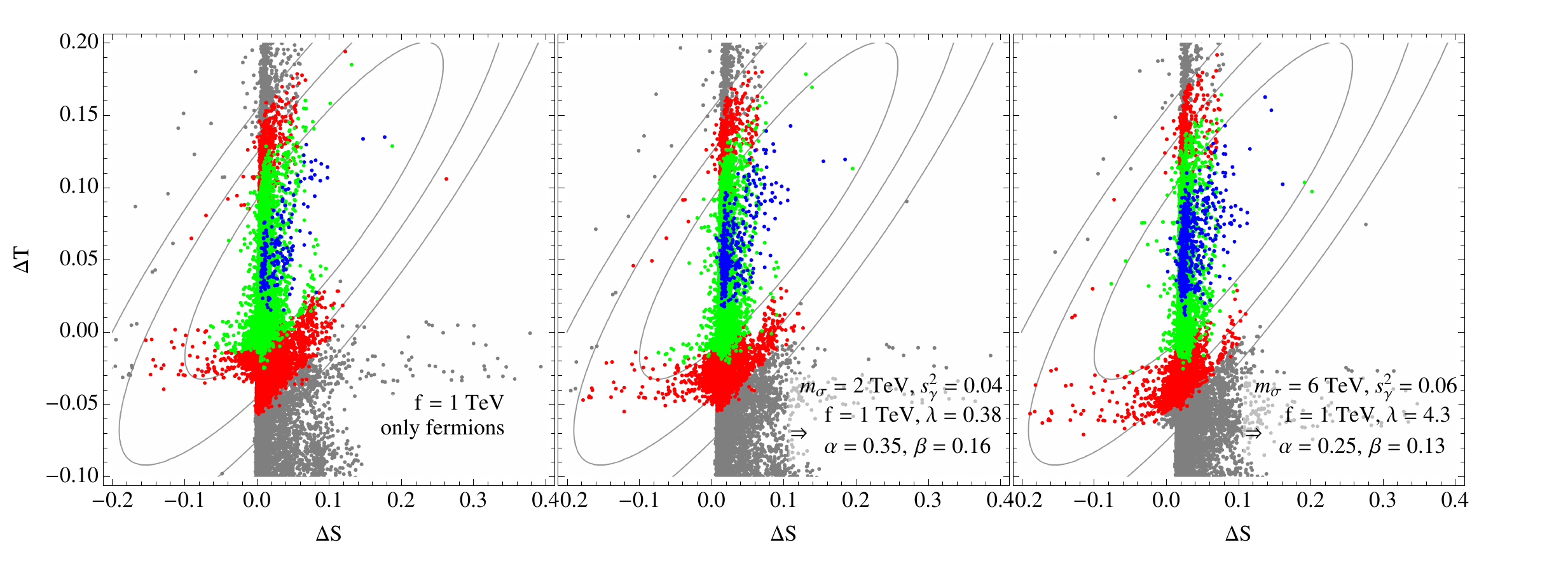}
\caption{Combined contributions to $S$ and $T$ from the scalar
    sector and the exotic fermionic sector. The blue, green and red
    points are allowed at $1,\,2,\,3\sigma$ by the combined
    $(S,T,g_L^b)$ fit, while gray points are outside the $3\sigma$
    region.}
\label{Fig-ST-all}
\end{figure}

The gauge boson couplings  to  neutral (NC) and charged (CC) fermionic currents in the weak basis can be read from  Table~\ref{tabQN}. After rotation to the mass basis, the corresponding Lagrangians can be written as~\cite{Anastasiou:2009rv}:
\bea
\label{couplingMatrices}
\nn
\mathcal{L}_\text{NC} &=& 
 \bar{\Psi}^{\text{phys}}\gamma^\mu 
  \left[ \frac{g}{2}
\left(  C_{L} P_L + C_{R} P_R \right)   W_\mu^3 
 - g'  (Y_{L} P_L + Y_{R} P_R ) B_\mu \right] \Psi^{\text{phys}} \\ \nn
&=&  \bar{\Psi}^{\text{phys}} \gamma^\mu
\left[
\frac{g}{2 c_W} \left( C_{L} P_L +C_{R} P_R -  2 s_W^2 \mathcal{Q} \right)  Z_\mu
-e  \mathcal{Q} A_\mu  \right]  \Psi ^{\text{phys}}\,, 
\\ 
\mathcal{L}_{CC} &=& 
\bar{\Psi}^{\text{phys}} \gamma^\mu \left[ \frac{g}{\sqrt{2}} 
\left( V_{L} P_L + V_{R} P_R \right) W_\mu^+ \right] \Psi^{\text{phys}}  + h.c.\,,
\label{GaugeLag}
\eea
where  
$P_L$ and $P_R$ are chirality projectors, $\Psi ^{\text{phys}}$ denotes the generic fermionic vector in the physical mass basis and $e$ is the absolute value of the electric charge unit. In the model under discussion, the matrices $C$ and $Y$ are related via the electric charge matrix --see also Eq.~(\ref{hypercharge}): 
\be
Y_{\alpha}=\mathcal{Q}-\frac{1}{2} C_{\alpha}\,\quad  \alpha = L \text{ or } R\,,
\ee
 with 
 \bea
\mathcal{Q} = \l +\frac{5}{3}, +\frac{2}{3}\  \bold{1}_{6\times 6},-\frac{1}{3}\  \bold{1}_{6\times 6}, -\frac{4}{3}\r\,. 
\label{Q}
\eea
The relation between the NC coupling matrices in the mass basis, $C_{L,R}$ and $Y_{L,R}$,  and their counterparts in the interaction basis (same symbols in curly characters below) is given by
\bea
&&C_L = L \mathcal{C}_L L^\dagger , \ C_R = R\, \mathcal{C}_R R^\dagger \,, \quad \mathcal{C}_{L\text{\bf ;}R} = \text{diag}(+1,\mathcal{C}^\mathcal{T}_{L\text{\bf ;}R},\mathcal{C}^\mathcal{B}_{L\text{\bf ;}R},-1) \,,\label{XLR}\\
&& \mathcal{C}^{\mathcal{T}}_{L\text{\bf ;}R} =-\mathcal{C}^{\mathcal{B}}_{L\text{\bf ;}R} =  \text{diag} ( +1;0 , +1,-1, 0, 0, +1) \,\label{chiLR}; 
\eea
\bea
&&Y_L = L \mathcal{Y}_L L^\dagger , \ Y_R = R \mathcal{Y}_R R^\dagger\,,\quad \mathcal{Y}_{L\text{\bf ;}R} = \text{diag}\l+\frac{7}{6},\mathcal{Y}^\mathcal{T}_{L\text{\bf ;}R}\,,\mathcal{Y}^\mathcal{B}_{L\text{\bf ;}R},-\frac{5}{6}\r\,,  \\
&&\mathcal{Y}^\mathcal{T}_{L\text{\bf ;}R} =\text{diag}\l\frac{1}{6}\text{\bf ;}\,\, \frac{2}{3},\frac{1}{6},\frac{7}{6},\frac{2}{3},\frac{2}{3},\frac{1}{6}\r\,, \,\mathcal{Y}^\mathcal{B}_{L\text{\bf;}R} = \text{diag}\l\frac{1}{6}\text{\bf;}-\frac{1}{3},\frac{1}{6},-\frac{5}{6},-\frac{1}{3},-\frac{1}{3},\frac{1}{6}\r .
\eea
Analogously, for the CC coupling matrices  $V_{L,R}$: 
\bea
&&V_L = L  \mathcal{V} L^\dagger\, ,  V_R = R  \mathcal{V} R^\dagger \,;
\mathcal{V}_{L\text{\bf ;}R} =
\left(
\begin{array}{cccc}
 0 & \mathcal{V}^{X^u\mathcal{T}} & {\mathbf 0_{1\times6}} & 0 \\
{\mathbf 0_{6\times1}} & {\mathbf 0_{6\times6}} & \mathcal{V}^{\mathcal{T}\mathcal{B}}_{L\text{\bf ;}R}  & {\mathbf 0_{6\times1}} \\
 {\mathbf 0_{6\times1}} & {\mathbf 0_{6\times6}}& {\mathbf 0_{6\times6}} & \mathcal{V}^{\mathcal{B}X'^d}  \\
  0 & {\mathbf 0_{1\times6}} & {\mathbf 0_{1\times6}}& 0 \\
\end{array}
\right)\,, \\
 && \qquad \qquad  \mathcal{V}^{X^u\mathcal{T}}=\l \mathcal{V}^{\mathcal{B}X'^d} \r ^\dagger =(0,0,1,0,0,0)\,,\label{calV2}
 \eea
while $\mathcal{V}^{\mathcal{T}\mathcal{B}}_{L}$ is a $6\times6$ matrix whose elements are null but for its $(1,1)$, $(2,6)$ and $(6,2)$ entries with value $1$, and $\mathcal{V}^{\mathcal{T}\mathcal{B}}_{R}$ is a $6\times6$ matrix with null elements but for its $(2,6)$ and $(6,2)$ entries with value $1$.

\paragraph{T parameter.} The contribution of the fermionic sector to the $T$ parameter,  $\Delta T^f$ , is given by~\cite{Lavoura:1992np}
\bea \nn
\Delta T^f=& \dfrac{3}{16 \pi s_W^2 c_W^2} &  \Big \{
\sum_{ij} \left[ \l (V^{ij}_{L})^2 + (V^{ij}_{R})^2 \r  \theta_+ (\eta_i,\eta_j) + 
2  V^{ij}_{L} V^{ij}_{R} \theta_- (\eta_i,\eta_j)\right] - \Big. \\ \nn
&&-\frac{1}{2 }\sum_{ij}  \left[ \l (C^{ij}_{L})^2 + (C^{ij}_{R})^2 \r  \theta_+ (\eta_i,\eta_j) + 
2  C^{ij}_{L} C^{ij}_{R} \theta_- (\eta_i,\eta_j)\right]
\Big. \Big \}   \\
&&- \frac{3}{16\pi s_W^2 c_W^2 M_Z^2} \l m_t^2 + m_b^2 - 2 \frac{m_t^2 m_b^2}{m_t^2-m_b^2} \ln\frac{m_t^2}{m_b^2} \r\,,
\eea
where $m_i$ denotes the fermion masses, $m_i \equiv \mathcal{M}^{\text{diag}}_{ii} $, and  $\eta_i \equiv m_i^2/M_Z^2$. The last line in this equation corresponds to the subtraction of the SM contribution from the light fermions (top and bottom). The  $\theta_\pm$ functions are defined as~\cite{Lavoura:1992np}:
\bea
\theta_+(\eta_1,\eta_2) &=& \eta_1 + \eta_2 - \frac{2 \eta_1 \eta_2}{\eta_1-\eta_2} \ln\frac{\eta_1}{\eta_2} 
- 2 (\eta_1 \ln \eta_1 + \eta_2 \ln \eta_2) +\text{div}  \  \frac{\eta_1 + \eta_2}{2} \,, \\
\theta_-(\eta_1,\eta_2) &=& 2 \sqrt{\eta_1 \eta_2} \l \frac{\eta_1+\eta_2}{\eta_1-\eta_2}\ln \frac{\eta_1}{\eta_2} 
-2+ \ln(\eta_1\eta_2)-\frac{\text{div}}{2}
\r \,. 
\eea
\paragraph{S parameter.}The  fermionic contribution to  $S$, $\Delta S^f$,  can be computed following Ref.~\cite{Lavoura:1992np},
\bea
\Delta S^f  = - 
\frac{1}{\pi} \sum_{ij} \Big\{ (C^{ij}_LY^{ij}_L+C^{ij}_RY^{ij}_R) \left[-\frac{\text{div}}{12} -\frac{5}{9}+ \frac{\eta_i+\eta_j}{3} + \frac{\ln(\eta_i\eta_j)}{6} 
\right.\nn\\[2mm] \left.+ \frac{\eta_i-1}{12} f(\eta_i,\eta_i)  + \frac{\eta_j-1}{12} f(\eta_j,\eta_j) - \frac{\chi_+(\eta_i,\eta_j)}{2} \right]  \nn\\[2mm]
- (C^{ij}_LY^{ij}_R+C^{ij}_RY^{ij}_L)\left[2\sqrt{ \eta_i\eta_j}+\sqrt{\eta_i\eta_j}\ \frac{f(\eta_i,\eta_i)+f(\eta_j,\eta_j)}{4}+\frac{\chi_-(\eta_i,\eta_j)}{2}\right] \Big\} -\Delta S^f_{SM}  \,,
\eea
with the functions $f(\eta_1,\eta_2)$ and $\chi_\pm(\eta_i,\eta_j)$ as defined in Ref.~\cite{Lavoura:1992np}, ``div'' standing for the divergent contributions typically appearing in dimensional regularisation, and the last term corresponding to the subtraction of the SM  light (top and bottom) fermionic contributions.~\footnote{The SM fermionic contributions to $S$ and $T$ with only one generation of quarks follow from Eq.~(\ref{GaugeLag}) considering a two-component fermion field $\Psi^{SM}=(t,b)$, with 
$\mathcal{M}^{SM}=\text{diag}(m_t,m_b)$ and coupling matrices $\mathcal{Q}^{SM}= Y^{SM}_R=\text{diag}\l+2/3,-1/3\r$, $C^{SM}_L=\text{diag}\l+1,-1\r$,
$Y^{SM}_L = + \frac{1}{6} \  \mathbb{1}_{2\times 2} $ , 
$ V^{SM}_L =\text{antidiag}\l 0,1\r$, 
$C^{SM}_R =V^{SM}_R = \mathbb{0}_{2\times 2}$.} 

\paragraph{Anomalous $Zbb$ coupling.}
We follow Ref.~\cite{Anastasiou:2009rv} for the computation of the
corrections to the $ g^b_L$ parameter defined in Eq.~(\ref{gLb}),
$\delta g^b_L$. Only the top and bottom sectors will be taken
into account as the mass generation mechanism for the lighter
  fermions are expected to have a lesser impact on EW precision tests
  since either the exotic fermions involved are much heavier or the Yukawa 
  couplings connecting them to the SM fermions are much smaller.   Moreover, the bottom
  quark mass will be neglected ($y_1'=y_2'=0$)~\footnote{The
  cancellation of divergences in the computation of $\delta g^b_L$ has
  been verified in this approximation.}.
  The fermion-gauge couplings relevant to this case are the Z couplings for the charge $2/3$ and 
$-1/3$ sectors which can be read from Eqs.~\eqref{Q} and \eqref{chiLR}, 
 and the couplings to the $W^\pm$ boson between the $(2/3,R)$ and
the $(-1/3,L)$ sectors (see  the matrix $\mathcal{V}^{\mathcal{T}\mathcal{B}}_{L}$  defined after Eq.~\eqref{calV2}). In addition to the NC and CC couplings in Eq.~(\ref{GaugeLag}), the interactions of the charged longitudinal gauge boson components ``$\pi_i$'' are needed, 
\beq
\begin{split}
\mathcal{L}_{\pi^\pm} =
\bar{\Psi}^{\text{phys}} \frac{g}{\sqrt{2}} 
\left( W_{L} P_L + W_{R} P_R \right) \Psi^{\text{phys}}\pi^+  + h.c.
\end{split}
\eeq
where 
\beq
W_L = R  \mathcal{W}_L  L^\dagger , \qquad  W_R = L  \mathcal{W}_R R^\dagger \,,
\eeq
with $\mathcal{W}_L$  and $\mathcal{W}_R$ being the mixing matrices in the interaction basis, given in the present model  by
\bea
&&\mathcal{W}_{L\text{\bf ;}R} =
\left(
\begin{array}{cccc}
 0 & \mathcal{W}^{X^u\mathcal{T}}_{L\text{\bf ;}R} & {\mathbf 0_{1\times6}} & 0 \\
 {\mathbf 0_{6\times1}} & {\mathbf 0_{6\times6}} & \mathcal{W}^{\mathcal{T}\mathcal{B}}_{L\text{\bf ;}R} & {\mathbf 0_{6\times1}} \\
 {\mathbf 0_{6\times1}} & {\mathbf 0_{6\times6}}& {\mathbf 0_{6\times6}} & \mathcal{W}^{\mathcal{B}X'^d}_{L\text{\bf ;}R}  \\
 0 & {\mathbf 0_{1\times6}} & {\mathbf 0_{1\times6}} & 0 \\ 
\end{array}
\right)\coma \\ \nonumber \\
 && \mathcal{W}^{X^u\mathcal{T}}_{L\text{\bf ;}R}=\frac{\sqrt{2}}{g}(0,0,0,0,-y_2\text{\bf ;}-y_1,0)\,,\,\l \mathcal{W}^{\mathcal{B}X'^d}_{L\text{\bf ;}R} \r ^\dagger=\frac{\sqrt{2}}{g}(0,0,0,0, y'_1\text{\bf ;}~y'_2 ,0)\,. \nonumber 
\eea
The $6\times6$ matrix $\mathcal{W}^{\mathcal{T}\mathcal{B}}_{L}$  in this equation has all elements  null  but for its $(5,6)$ and $(6,5)$ entries which take values $y_1$ and $-y'_2$, respectively, while $\mathcal{W}^{\mathcal{T}\mathcal{B}}_{R}$ is a $6\times6$ matrix of null elements but for its $(5,6)$ and $(6,5)$ entries which take values $y_2$ and $-y'_1$, respectively. 
In practice, only the entries connecting --after rotation-- the charge
$2/3$ fermions to $b_L$ enter the computation.
\begin{figure}
\centering
\includegraphics[width=\textwidth,keepaspectratio]{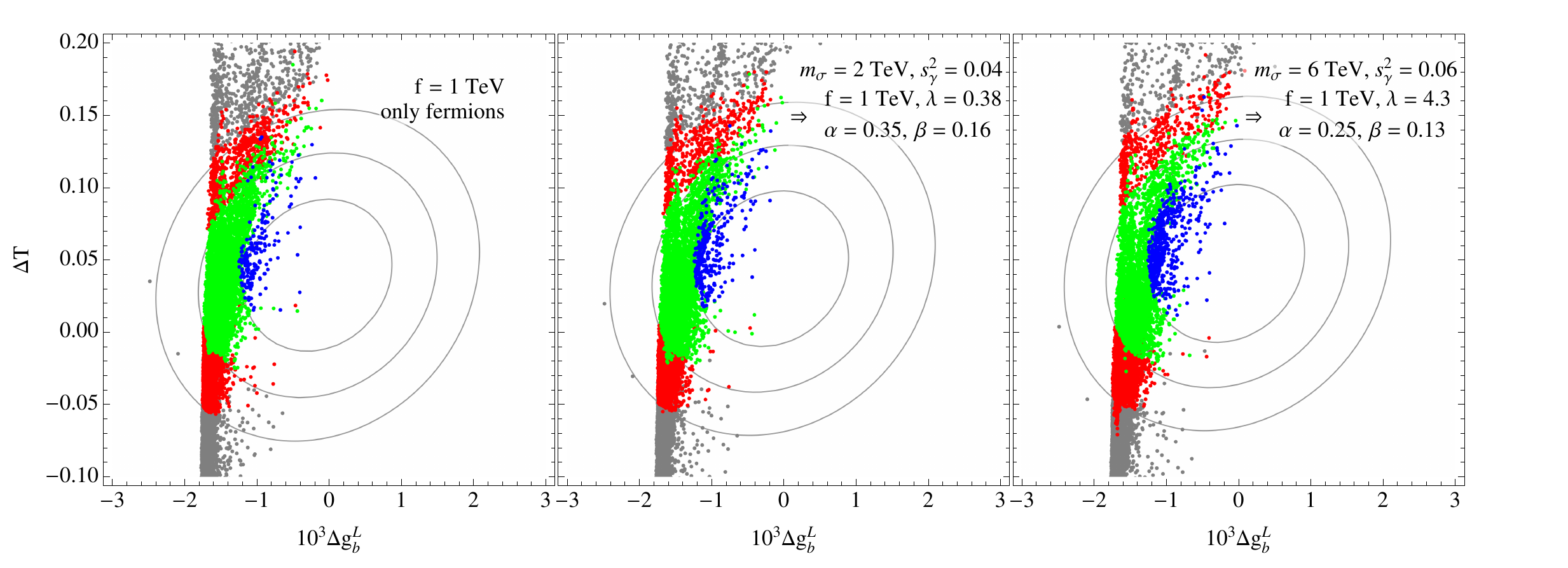}
\caption{Scalar and fermionic  impact on the $T$ parameter and
    on the $Z$-$b_L$-$b_L$  coupling $g_b^L$.}
\label{Fig-STRb-all}
\end{figure}

In the numerical analysis, the two sets of values considered earlier on for the numerical analysis of the pure scalar contributions will be retained:  ($m_\sigma= 2$ TeV,  $\sin^2_\gamma= 0.04$) and  ($m_\sigma= 6$ TeV,  $\sin^2_\gamma= 0.06$), both corresponding to $f= 1$ TeV and within the soft breaking regime $\alpha, \beta <\lambda$, with the latter being kept within its perturbative range~\footnote{$m_\sigma=4\pi f$ is
  roughly where perturbativity is lost in chiral perturbation
  theory~\cite{Manohar:1983md}.}. Note that, for a $\sigma$ particle much heavier than the Higgs, values of $f$ below 700~GeV would be difficult to
accommodate experimentally as $\sin\g^2\simeq
v^2/f^2$, see Eq.~(\ref{xilimit}).   The exotic fermionic masses will be allowed to vary randomly between 800~GeV and $\mathcal{O}(10~{\rm
  TeV})$, as the heavy top partners with electric charges $+5/3$ and $+2/3$ are bounded to be above
$800-1000$~GeV~\cite{Aad:2015kqa, CMS:2015alb}, depending on the
dominant decay mode. In the light fermion sector, the top and bottom masses will be allowed to vary within the intervals  $m_t=173\pm5$~GeV and $m_b=4.6\pm2$~GeV, respectively, for illustrative purposes.

 Figs.~\ref{Fig-ST-sigma} to 6 depict the points that satisfy a $\chi^2$ global fit
to the precision pseudo-observables  $S$, $T$ and $\d g_L^b$, where the blue, green, and red
points are the allowed $1\sigma, 2\sigma$ and $3\sigma$ regions, respectively, while
gray points lie above the $3\sigma$ limit.   The central values,
uncertainties and correlation matrix are taken from
Ref.~\cite{Ghosh:2015wiz}.  The ellipses drawn in the $\Delta
S-\Delta T$ plane in Figs.~\ref{Fig-ST-sigma}  and \ref{Fig-ST-all} are the projection for $\Delta g_b^L=0$, while those in the 
$\Delta T-\Delta g_b^L$ plane in Fig.~\ref{Fig-STRb-all} use the $\Delta S$ value coming from
the scalar sector. The latter is a good approximation since $S$ gets in practice a
very small correction from the heavy fermions, as seen in Fig.~(\ref{Fig-ST-all}).

{ \bf S versus T}.  The fermion sector can lead to large
deviations in the value of the $T$ parameter.
  In Fig.~(\ref{Fig-ST-sigma}) and in the first panel of Fig.~(\ref{Fig-ST-all}) only the fermionic contributions are depicted. 
 The last two panels in Fig~(\ref{Fig-ST-all})
 show the fermion plus scalar combined results: the lighter the
 $\sigma$ particle, the less tension follows with respect to
 electroweak precision data, in particular due to the impact on
 $\Delta T$, although it is to be noted that even for large $m_\sigma$ fermionic
   contributions can bring $(S,T)$ within the experimentally allowed
   region.

 The sign of the fermionic contributions to $S$ and $T$
  can be largely understood in terms of the light-heavy fermion
  mixings and the mass hierarchy between the heavy eigenstates.  For
  instance, large mixing values with a heavy singlet are known to
  induce large positive contributions to $\Delta T$, as pointed out in
  Ref.~\cite{Anastasiou:2009rv}, as a result from the custodial
  symmetry being broken by the singlet-doublet mixing.  It is possible
  to illustrate the analysis more in detail following
  Ref.~\cite{Dawson:2012di}, which uses a different fermionic but
  nonetheless illuminating embedding. They consider heavy vector
  fermions which couple directly to both the light doublet $q$ and the
  light singlet $q_R$. When only a heavy singlet is present, the
  expected contributions to $\Delta T$ and $\Delta S$ are both
  significant (though the first are more important) and
  positive for the regime we consider, see their Eq.~(32). Instead,
  when only a heavy vector doublet was taken into account, the sign of
  the correction to the oblique parameters was proportional to the
  sign of the mass splitting between the heavy eigenstates with charge
  $2/3$ and $-1/3$, resulting in sizeable contributions to $\Delta T$
  and very small to $\Delta S$.

\begin{figure}
\centering
\includegraphics[width=\textwidth,keepaspectratio]{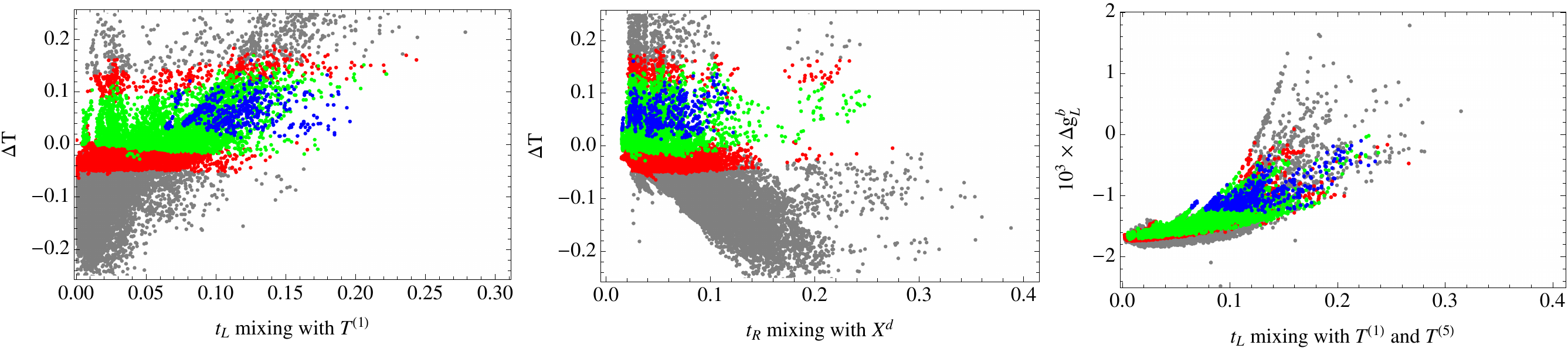}
\caption{Examples of correlations between the fermion mixing strength and
    electroweak precision measurements. The label ``$t_{L,R}$ mixing
    with $\Psi_1$, $\Psi_2$, \dots'' indicates
    $\left(|U^{L,R}_{1}|^2+|U^{L,R}_{2}|^2+\dots\right)^{1/2}$, where
    $U^{L,R}$ indicates the left or right rotation that diagonalizes
    the mass matrix.} 
\label{Fig-correlations-all}
\end{figure}

 It is not possible to apply those conclusions in
 Ref.~\cite{Dawson:2012di} directly here, though, as in our setup the
 light fermion mass generation involves necessarily and simultaneously
 {\it both} a heavy doublet and a heavy singlet, see
 Eq.~(\ref{topmasscartoon}): the light doublet $q$ mixes
 directly only with the heavy doublet $Q$, while $q_R$ mixes with
 $T_1$. 
 Nevertheless, the mainly positive fermionic corrections to $\Delta S$ found
 are consistent with being dominated by the participation of a heavy
 singlet. The results, in Fig.~\ref{Fig-correlations-all} show
   indeed that a large mixing between $t_L$ and the singlet $T^{(1)}$
   leads to a positive $\Delta T$ (left panel) while the 
 negative corrections to $\Delta T$ obtained are consistent instead
 with a large mixing between $t_R$ and the doublet component
   $X^d$ (middle panel).

{ \bf T versus $g_L^b$}. The deviations induced in the $Zbb$ coupling
provide additional bounds: even if the model parameters do not impact on
$\Delta g_b^L$ at tree level, the top partners may induce at loop
level deviations from the SM value.  Fig.~\ref{Fig-STRb-all} depicts
the purely fermionic and the scalar plus fermion combined
contributions in the $T-g_L^b$ parameter space. 
 Finally, the right panel of
   Fig.~\ref{Fig-correlations-all} shows a sizeable and positive impact
   on $g_L^b$ of the mixing between $t_L$ and  the
charge $2/3$ heavy singlets $T^{(1)}$
   and $T^{(5)}$.

As a final remark, there are considerable mixings in the fermion
  sector for which the dominant effects go schematically as 
  $\tan\theta_{ij}\sim\Lambda_i/M_j$.  It could  be thus suspected that large
  deviations in the $Wtb$ coupling should occur. However, these
  rotations are mainly driven by the $SO(5)$ breaking couplings $\Lambda_i$ and $\Lambda'_i$,
  which are custodial symmetry preserving.  Therefore, a large
  rotation in the top sector is mostly compensated by a corresponding
  one in the bottom sector, leading to practically no deviation in
  $V_{tb}$.

%
\subsection{Higgs and $\sigma$ coupling to gluons }
%

This section and the next one deal with the scalar to photons and to gluons effective couplings, arising at one-loop level. Define the scalar-gluon-gluon amplitudes $hgg$ ($\sigma gg$)  as
\be
{\cal A}_{h(\sigma)}\equiv{\cal A}_{h(\sigma)\leftrightarrow gg} (m^2_{h(\sigma)}) =
-i \frac{\alpha_s}{\pi}~g_{h(\sigma)}~(p\cdot k~ g^{\mu\nu}-p^\mu k^\nu)\delta^{ab}\,,
\label{amplgg}
\ee
where $g_{h}$ and $g_\sigma$ are scale dependent functions 
that parametrize the amplitude strength, $\alpha_s=g_s^2/4\pi$ with
$g_s$ denoting the QCD coupling constant,  $p$ and $k$ stand for the
gluon four-momenta, and $a,b$ are color indices. In the case of the SM, the $hgg$ coupling is induced only
at one loop level and the amplitude is dominated by the top quark, 
\be g_h^{SM} =
\left(\frac{y_t}{\sqrt{2}}\right)~\frac{1}{m_t}I\left(\frac{m_h^2}{m_t^2}\right)
\,,
\label{ghSM}
\ee
where $y_t$ is the top Yukawa coupling ($m_t\equiv y_t \,v/\sqrt{2}$)
and $I(m_h^2/m_t^2)/m_t$ is the loop factor with
\be
I\left(\frac{q^2}{m^2}\right)=\int_0^1dx~\int_0^{1-x}dz~\frac{1-4xz}{1-xz \frac{q^2}{m^2}}\quad \approx \quad 
\left\{
\begin{array}{c}
1/3~~~~~{\rm for}~~~~m^2\gg q^2\nonumber\\
\\
0~~~~~~~~{\rm for}~~~~m^2\ll q^2\nonumber
\end{array}
\right\} \,.
\label{SMloop}
\ee
The SM bottom contribution corresponds to $I(m^2_h/m^2_b)\approx
10^{-2}$ and is thus usually neglected~\footnote{The large mass limit
  in the integral is customarily applied for $m_h < 2 m_i$, which
  includes the top case.}.

There are no direct $hgg$ or $\sigma gg$ couplings in the Lagrangian discussed here, but effective $h gg$ and $\sigma gg$ interactions arise  
via fermion loops. Expanding the global field-dependent mass matrix $\mathcal{M}(h,\s)$ in Eqs.~(\ref{Lag-K})-(\ref{matrix-K2}) around the scalar field vevs, $v$ and $v_\s$,  and defining the following constant matrices
\bea
\mathcal{\overline{M}} \equiv \mathcal{M}(v,v_\sigma) \quad, \quad 
\frac{\partial\mathcal{\overline{M}}}{\partial h} \equiv \left.\frac{\partial\mathcal{M}(h,\s)}{\partial h}
\right|_{\scriptsize \renewcommand{\arraystretch}{0.75} \!\!\begin{array}{l} h=v \\ \s=v_\s \end{array}}
 \quad, \quad 
\frac{\partial\mathcal{\overline{M}}}{\partial \s} \equiv \left.\frac{\partial\mathcal{M}(h,\s)}{\partial \s}
\right|_{\scriptsize \renewcommand{\arraystretch}{0.75} \!\!\begin{array}{l} h=v \\ \s=v_\s \end{array}} \nn \,,
\eea
 the fermionic mass Lagrangian Eq.~(\ref{Lag-K}) can be written as 
 \begin{eqnarray}
- \mathcal{L}_Y&=&\bar \Psi_L\,\mathcal{\overline{M}}\,\Psi_R \,+\, 
       \hat h\,\bar \Psi_L\frac{\partial \mathcal{\overline{M}}}{\partial h}\,\Psi_R \,+\,
       \hat\s\,\bar \Psi_L \frac{\partial \mathcal{\overline{M}}}{\partial \s}\, \Psi_R \,+\, h.c.
\end{eqnarray}
where $\hat h$ and $\hat\s$ are the unrotated scalar fluctuations, see Eq.~(\ref{phys-scalars}). Performing the rotation to the fermionic mass eigenstate 
basis $\{\Psi_i \to \Psi^{phys}_i\}$, 
\be
\begin{array}{lcl}
\l \mathcal{\overline M} \r_{ij} \to m_i\, \delta_{ij} \quad , \quad
\l \frac{\partial\mathcal{\overline{M}}}{\partial h}\r_{ij}  \to  (Y_h)_{ij} \quad , \quad
\l \frac{\partial\mathcal{\overline{M}}}{\partial \s}\r_{ij}  \to  (Y_\sigma)_{ij} \, , 
\end{array}
\ee
where $m_i$, $Y_h$ and $Y_\s$ are respectively the masses and the couplings to the unrotated scalars fields $\hat h, \hat \s$ 
of the physical fermionic states~\footnote{ For instance, in this notation $(Y_{h})_{tt}=y_t/\sqrt{2}$.}. For simplicity, CP invariance will be assumed in what follows.

It is straightforward to obtain the physical $h \leftrightarrow gg$ and $\sigma \leftrightarrow  gg$ amplitudes 
 combining those involving the unrotated $\hat h$ and $\hat \s $ fields. The latter will require the substitution of the SM loop factor in Eq.~(\ref{SMloop}) as follows, 
 \be
\frac{y_t}{\sqrt{2}}~\frac{1}{m_t}I\left(\frac{m^2_h}{m_t^2}\right) 
\qquad \to \qquad \left\{
\begin{array}{l}
\displaystyle\sum_i (Y_h)_{ii}\frac{1}{m_i}~I\left(\frac{q^2}{m_i^2}\right) \qquad \text{for } \hat h \leftrightarrow gg\\ \\
\displaystyle\sum_i (Y_\sigma)_{ii}\frac{1}{m_i}~I\left(\frac{q^2}{m_i^2}\right)\qquad \text{for } \hat \sigma \leftrightarrow gg
\end{array} \right \}\,,
\label{hsigmaloop}
\ee
 where $q^2=m_h^2$ for $h\leftrightarrow gg$ on-shell transitions, while  $q^2=m_\sigma^2$ for $\sigma \leftrightarrow gg$ on-shell transitions, and where the sum runs over all colored fermion species present in the model.
%
\subsubsection*{\boldmath${h \leftrightarrow gg}$ transitions}
%
If all fermion masses were much larger 
than $m_h$, it would be possible to simply factorize the constant integral outside the sum as follows:
\bea
\sum_i \frac{\left(Y_h\right)_{ii}}{m_i}~I\left(\frac{m^2_h}{m_i^2}\right) 
&\approx&  \frac{1}{3}\sum_i \frac{\left(Y_h\right)_{ii}}{m_i} 
= \frac{1}{6}\frac{d}{dh}\log\det(\overline{\mathcal{M}} \, \overline{\mathcal{M}}^\dagger) \,,
\label{allheavy}
\eea
where the last term is written in the original unrotated fermionic basis since trace and determinant are 
invariant under a change of basis. All fermions in the model under consideration are indeed much heavier than the Higgs particle but for the bottom, whose loop contribution $I(m^2_h/m^2_b)$ is negligible. Therefore the false "heavy" bottom contribution included in Eq.~(\ref{allheavy}) should be 
removed at energies $q^2 \approx m^2_h$, resulting in the following effective couplings at the scale $m_h$:

\bea
g_{\hat{h}}(m^2_h) &\approx & 
      \frac{1}{6}\frac{d}{dh}\left.\left(\log\det(\mathcal{M}\,\mathcal{M}^\dagger)-\log(m_b(h,\sigma)\,m^*_b(h,\sigma))\right) 
      \right|_{\scriptsize \renewcommand{\arraystretch}{0.75} \!\!\begin{array}{l} h=v \\ \s=v_\s \end{array}} \nn \\ 
   &=& \frac{1}{3v} + O\l\frac{v}{M'_1 M'_5}\r \,,\\
g_{\hat\s}(m^2_h) &\approx & 
      \frac{1}{6}\frac{d}{d\s} \left.\left(\log\det(\mathcal{M}\,\mathcal{M}^\dagger)-\log(m_b(h,\sigma)\,m^*_b(h,\sigma))\right) 
      \right|_{\scriptsize \renewcommand{\arraystretch}{0.75} \!\!\begin{array}{l} h=v \\ \s=v_\s \end{array}} \nn \\
   &=& -\frac{y_2}{3 M_5}\frac{\Lam_2}{\Lam_3} + O\l\frac{v_\s}{M'_1 M'_5},\frac{v_\s}{M^2_5}\r\,,
\eea
where the eigenvalue of the field dependent mass matrix corresponding to the bottom quark reads: 
\be
m_b (h,\sigma) = \frac{y'_1 \Lambda'_1\Lambda'_3-y'_1 y'_2 \Lambda'_1\Lambda'_2~\sigma/M'_5}
                 {M'_1 M'_5-y'_1 y'_2~(h^2+\sigma^2)}\, \frac{h}{\sqrt{2}} \,.
\label{m0b}
\ee
The $h\leftrightarrow gg$ amplitude is then given by Eq.~(\ref{amplgg}), with 
\be
g_{h} \equiv {g}_{\hat{h}}(m^2_h)\cos\gamma - {g}_{\hat\s}(m^2_h)\sin\gamma \,. \nn \\
\ee
In the limit $m_t\gg m_h$, the $hgg$ effective coupling is exactly as
in the SM. The contribution from the heavy vector-like quarks
tends to cancel out for bare vector-like masses
  substantially larger than $v$, a result well-known in the
literature.

%
\subsubsection*{\boldmath${\sigma \leftrightarrow gg}$ transitions}
%

With analogous procedure, the $\sigma gg$ amplitude can be obtained
using Eq.~(\ref{hsigmaloop}) for $q^2=m_\sigma^2$. 
The difference with the previous case is that now the top
  quark is lighter or comparable in mass to $\sigma$ and it cannot be
  integrated out, that is $m_b\ll m_t,m_\s \ll m_i$, where here $m_i$
  denotes the heavy fermion masses, and in consequence it is necessary
  to subtract the bottom contribution and to take into account the
  $q^2 $ dependence in the top loop. In the approximation
  $I(m^2_\s/m^2_b) \approx 0$, it results
\bea
g_{\hat{h}}(m^2_\s) & \approx & 
      \frac{1}{6}\frac{d}{dh} \left.\left(\log\det(\mathcal{M}\,\mathcal{M}^\dagger)
      -\log(m_t(h,\sigma)\,m^*_t(h,\sigma)) 
      -\log(m_b(h,\sigma)\,m^*_b(h,\sigma))
\right) 
      \right|_{\scriptsize \renewcommand{\arraystretch}{0.75} \!\!\begin{array}{l} h=v \\ \s=v_\s \end{array}} \nn\\
&\, & + \frac{1}{v}I\left(\frac{m^2_\sigma}{m_t^2}\right) \nn \\ 
 & = &  \frac{1}{v}I\left(\frac{m^2_\sigma}{m_t^2}\right)-\frac{2}{3} v \l \frac{y_1 y_2}{M_1 M_5} + \frac{y'_1 y'_2}{M'_1 M'_5}\r + 
      O\l\frac{v v_\sigma^2}{M^2_1 M^2_5}, \frac{v v_\sigma^2}{{M'}^2_1 {M'}^2_5}\r \,,\label{ghath}
      \eea
      \bea
g_{\hat\s}(m^2_\s) & \approx & 
      \frac{1}{6}\frac{d}{d\s}\left.\left(\log\det(\mathcal{M}\,\mathcal{M}^\dagger)
      -\log(m_t(h,\sigma)\,m^*_t(h,\sigma))
      -\log(m_b(h,\sigma)\,m^*_b(h,\sigma))\right)
      \right|_{\scriptsize \renewcommand{\arraystretch}{0.75} \!\!\begin{array}{l} h=v \\ \s=v_\s \end{array}}  \nn \\
    &\,& -\frac{y_2}{M_5}\frac{\Lam_2}{\Lam_3}I\left(\frac{m_\sigma^2}{m_t^2}\right)\nn\\
    & = & - \frac{y_2}{M_5}\frac{\Lam_2}{\Lam_3}I\left(\frac{m_\sigma^2}{m_t^2}\right) -\frac{2}{3} v_\s \l \frac{y_1 y_2}{M_1 M_5} + \frac{y'_1 y'_2}{M'_1 M'_5}\r + 
      O\l\frac{v^3_\s }{M^2_1 M^2_5}, \frac{v^3_\s}{{M'}^2_1 {M'}^2_5}\r \,,\label{ghatsigma}
\eea
where the eigenvalue of the field dependent mass matrix corresponding to the top quark reads 
\be
m_t (h,\sigma) = \frac{y_1 \Lambda_1\Lambda_3 - y_1 y_2 \Lambda_1\Lambda_2~\sigma/M_5}
                 {M_1 M_5 - y_1 y_2~(h^2+\sigma^2)}\, \frac{h}{\sqrt{2}} \,.
\label{m0t}
\ee
Note that  the dominant contribution to  the $\sigma gg$ effective coupling requires both $y_1$ and $y_2$ 
to be non-vanishing. In contrast,  the dominant contribution to the top quark mass is proportional to $y_1$ 
 but independent of $y_2$.

The $\sigma\leftrightarrow gg$ amplitude is finally given by Eqs.~(\ref{amplgg}), (\ref{ghath}) and (\ref{ghatsigma}), with 
\be
g_{\s} \equiv {g}_{\hat{h}}(m^2_\s)\sin\gamma + {g}_{\hat\s}(m^2_\s)\cos\gamma\,.  \nn \\
\ee
The matrix elements modulus square for $gg\to h$ and $gg\to\s$, averaged over the polarisations of the initial state,  are then given by 
\bea
\overline{|{\cal A}_{h}|^2}&=&\frac{\alpha_S^2 m_h^4}{64\pi^2}\,g_h^2 \,,\nn\\
\overline{|{\cal A}_{\s}|^2}&=&\frac{\alpha_S^2 m_\s^4}{64\pi^2}\,g_\s^2\,\label{Ahs}.
\eea
In terms of those amplitudes, the cross section at the parton level can be expressed as
\bea
\sigma_{part}(gg\to h)&=&\overline{|{\cal A}_{h}|^2}\frac{\pi}{s_{part}}~\delta(s_{part}-m_h^2)\nonumber\\
&=&\overline{|{\cal A}_{h}|^2}\frac{\pi}{\tau s^2}~\delta(\tau-\frac{m_h^2}{s})\,,
\eea
where as usual $s_{part}$ denotes the center-of-mass energy at the parton level $s_{part}=\tau s$. A similar expression holds for $\sigma_{part}(gg\to\s)$.
By convoluting the cross-section with the gluon densities $G(x)$ we finally obtain
\bea
\sigma(pp\to h)&=&\overline{|{\cal A}_{h}|^2}~\frac{\pi}{m_h^2 s}\times \int_{m_h^2/s}^1~\frac{dx}{x}~G(x)\,\,G
      \left(\frac{m_h^2}{sx}\right)\,,\nn\\
\sigma(pp\to \s)&=&\overline{|{\cal A}_{\s}|^2}~\frac{\pi}{m_\s^2 s}\times \int_{m_\sigma^2/s}^1~\frac{dx}{x}~G(x)\,\,G
      \left(\frac{m_\sigma^2}{sx}\right)\,.
      \label{xsectpp}
\eea

\vspace{0.5cm}
In resume, for very heavy fermion partners the $h$-gluon-gluon transitions are dominated by the top quark contribution, while they have a more significant impact on $\sigma$-gluon-gluon transitions.

\subsection{Higgs and $\sigma$ decay into $\gamma\gamma$}
There are no direct $h\gamma\gamma$ or $\sigma \gamma\gamma$ couplings in our Lagrangian. They arise instead  as effective interactions  from loops of fermions and, in the case of $h\gamma\gamma$, also of massive vector bosons. 
As in the case of $h$ and $\sigma$ production via $gg$ fusion, we distinguish between mass eigenstates $h,\s$ and the unrotated (interaction )
eigenstates $\hat h,\hat \s$.

Let the scalar-photon-photon amplitudes $h\gamma\gamma$ and $\sigma \gamma \gamma$  be defined as
\begin{equation}
{\cal A}_{h(\sigma)\leftrightarrow \gamma \gamma} (m^2_{h(\sigma)}) =
i \frac{\alpha}{\pi}~\Omega_{h(\sigma)}~(p\cdot k~ g^{\mu\nu}-p^\mu k^\nu)\delta^{ab}\,,
\label{amplgammagamma}
\end{equation}
where again $\Omega_h$ and $\Omega_\sigma$ are scale dependent functions. The decay amplitudes are then given by
\be
\Gamma(h\to\gamma\gamma)=\frac{\alpha^2 m_h^3}{64\pi^3} |\Omega_{h}|^2\,, \qquad
\Gamma(\s\to\gamma\gamma)=\frac{\alpha^2 m_\sigma^3}{64\pi^3} |\Omega_{\s}|^2\,.
\ee
In the model under study, the contributions can again be decomposed as
\bea
\Omega_{h}&=&\cos\gamma~ \Omega_{\hat h}(m_h^2)-\sin\gamma~ \Omega_{\hat \s}(m_h^2)\,,\nn\\
\Omega_{\s}&=&\sin\gamma~ \Omega_{\hat h}(m_\sigma^2)+\cos\gamma~ \Omega_{\hat \s}(m_\sigma^2)\,.\nn
\eea
 While both unrotated scalar fields $\hat h$ and $\hat \sigma$ couple to fermions,  only $\hat h$ couples to the $W$ boson, 
\be
\nn
\Omega_{\hat h}(q^2)=\Omega_{\hat h}^F(q^2)+\Omega_{\hat h}^W(q^2)~~~~,~~~~~~~~~~~~\Omega_{\hat \sigma}(q^2)=\Omega_{\hat \sigma}^F(q^2)\,,
\ee
where the superscripts $F$ and $W$ stand for fermionic and gauge contributions, respectively.
The latter is akin to the SM
one, that is,
\be
\Omega_{\hat h}^W(q^2)= \frac{g^2 v}{8 m_W^2} ~I_W\left(\frac{4\,m_W^2}{q^2}\right),
\label{OmegahathW}
\ee
where the factor $g^2 v$ results from the Higgs$-WW$ vertex, and the remaining part $I_W/8 m_W^2$ results from the kinematics of the loop integral 
\be
I_W(x)=2+3 x+3 x(2-x)f(x) \, , \qquad
f(x)=\left\{
\begin{array}{ll}
\arcsin^2(1/\sqrt{x})&x\ge 1\\
-\frac{1}{4}\left[\log\frac{1+\sqrt{1-x}}{1-\sqrt{1-x}}-i\pi
\right]^2&x<1
\end{array}
\right.\,.
\ee
\subsection*{\boldmath$h\leftrightarrow \gamma \gamma$ transitions}
At the Higgs mass scale, the SM $\hat h WW$ coupling in Eq.~(\ref{OmegahathW})  is given by
\be
\Omega_{\hat h}^W(m_h^2)\approx \frac{4.2}{v}\,.
\ee
The SM quark contributions are in turn given by
\be
\Omega_h^{SM,F}=-6\sum_f  Q_f^2 \left(\frac{y_f}{\sqrt{2}}\right)~\frac{1}{m_f} I\left(\frac{m_h^2}{m_f^2}\right)\, \approx\, -\frac{8}{9}\frac{1}{v}\,,
\label{OmegahSM}
\ee
where  $y_f$ is the fermion Yukawa coupling, $m_f=y_f\,v/\sqrt{2}$, and the remaining factor $I/m_f$ 
results from the loop integral. The last expression in Eq.~(\ref{OmegahSM}) corresponds to the top contribution, which dominates the SM fermionic contribution. The SM decay $h\to \gamma \gamma$ decay rate is as given in Eq.~(\ref{amplgammagamma})   with $\Omega_h= \Omega_h^{SM, W}+\Omega_h^{SM, F}$. In the model under consideration these expressions for the quark contributions to the $h\to\gamma\gamma$ transitions are generalized as follows, in analogy with the $gg$ fusion analysis above, 
\bea
\Omega_{\hat h}^F(m_h^2)&=&-2 \sum_f N_C^f Q_f^2~ \omega_f^h(m_h^2) =-2 \left[3\left(\frac{2}{3}\right)^2~\omega_{2/3}^h(m_h^2)+
                3\left(-\frac{1}{3}\right)^2~ \omega_{-1/3}^h(m_h^2) \right]\,,\nn\\
\Omega_{\hat \sigma}^F(m_h^2)&=&-2 \sum_f N_C^f Q_f^2~ \omega_f^\sigma(m_h^2) =-2 \left[3\left(\frac{2}{3}\right)^2~\omega_{2/3}^\sigma(m_h^2)+
                3\left(-\frac{1}{3}\right)^2~ \omega_{-1/3}^\sigma(m_h^2) \right]\,,\nn\\
\eea
where $N_C^f$ is the number of colors of a given quark species $f$, and  $\omega_f^h$ are scale-dependent functions, which for charge $2/3$ and $-1/3$ fermions read
\bea
\omega_{2/3}^h(m_h^2) & \equiv & \left. 
   \frac{1}{6}\frac{d}{dh}\left(\log\det(\mathcal{M}_\mathcal{T}\,\mathcal{M}^\dagger_\mathcal{T})\right) 
   \right|_{\scriptsize \renewcommand{\arraystretch}{0.75} \!\!\begin{array}{l} h=v \\ \s=v_\s \end{array}}
     = \frac{1}{3v}\,, \nn \\ 
\omega_{-1/3}^h(m_h^2) & \equiv  & \left.
   \frac{1}{6}\frac{d}{dh}\left(\log\det(\mathcal{M}_\mathcal{B}\,\mathcal{M}^\dagger_\mathcal{B}) 
   -\log(m_b(h,\sigma)\,m^*_b(h,\sigma))\right) 
   \right|_{\scriptsize \renewcommand{\arraystretch}{0.75} \!\!\begin{array}{l} h=v \\ \s=v_\s \end{array}}   \\
    & = & -\frac{2}{3} v \frac{y'_1 y'_2}{M'_1 M'_5} + \mathcal{O}\l\frac{v\, v^2_\s}{{M'}^2_1 {M'}^2_5} \r\, \nn.
\eea
For the $\omega^\sigma$ functions, it holds instead 
\bea
\omega_{2/3}^\sigma(m_h^2) & \equiv & \left. 
   \frac{1}{6}\frac{d}{d\sigma}\left(\log\det(\mathcal{M}_\mathcal{T}\,\mathcal{M}^\dagger_\mathcal{T})\right)
   \right|_{\scriptsize \renewcommand{\arraystretch}{0.75} \!\!\begin{array}{l} h=v \\ \s=v_\s \end{array}}
     = -\frac{y_2}{3M_5}\frac{\Lambda_2}{\Lambda_3}+ \mathcal{O}\l\frac{v_\s}{M_5^2} \r\,, \nn \\ 
\omega_{-1/3}^\sigma(m_h^2) & \equiv  & \left.
   \frac{1}{6}\frac{d}{d\sigma}\left(\log\det(\mathcal{M}_\mathcal{B}\,\mathcal{M}^\dagger_\mathcal{B}) 
   -\log(m_b(h,\sigma)\,m^*_b(h,\sigma))\right) 
   \right|_{\scriptsize \renewcommand{\arraystretch}{0.75} \!\!\begin{array}{l} h=v \\ \s=v_\s \end{array}}   \\
    & = & -\frac{2}{3}\frac{v_\sigma y'_1 y'_2}{M'_1 M'_5}+ \mathcal{O}\l\frac{v_\sigma^3}{{M'}^2_1 {M'}^2_5} \r \,.\nn
\eea
In these expressions the bottom contribution was neglected,
  while it has been assumed $m_h \ll m_i$
for the top mass and all other exotic fermion masses $m_i$.
\subsection*{\boldmath$\sigma\leftrightarrow \gamma \gamma$ transitions}
Similarly, for $\s$ decaying into two photons the contributions for the unrotated field $\hat \sigma$ are given by
\bea
\Omega_{\hat \s}^F(m_\sigma^2) &=&-2 \sum_f N_C^f Q_f^2~ \omega_f^\s(m_\sigma^2) = 
   -2 \left[3\left(\frac{2}{3}\right)^2~ \omega_{2/3}^\s(m_\sigma^2)+3\left(-\frac{1}{3}\right)^2~\omega_{-1/3}^\s(m_\sigma^2)\right],\nonumber\\
\Omega_{\hat h}^F(m_\sigma^2) &=&-2 \sum_f N_C^f Q_f^2~ \omega_f^h(m_\sigma^2) = 
   -2 \left[3\left(\frac{2}{3}\right)^2~ \omega_{2/3}^h(m_\sigma^2)+3\left(-\frac{1}{3}\right)^2~\omega_{-1/3}^h(m_\sigma^2)\right],\nonumber
\eea
where 
\bea
\omega_{2/3}^\sigma(m_\sigma^2) & \equiv & \left. 
   \frac{1}{6}\frac{d}{d\s}\left(\log\det(\mathcal{M}_\mathcal{T}\,\mathcal{M}^\dagger_\mathcal{T})
   -\log(m_t(h,\sigma)\,m^*_t(h,\sigma))\right) 
   \right|_{\scriptsize \renewcommand{\arraystretch}{0.75} \!\!\begin{array}{l} h=v \\ \s=v_\s \end{array}}     \!\!\!- \frac{y_2}{M_5}\frac{\Lambda_2}{\Lambda_3}I\left(\frac{m^2_\sigma}{m_t^2}\right) \nn \\ 
     & = &  - \frac{y_2}{M_5}\frac{\Lambda_2}{\Lambda_3}I\left(\frac{m^2_\sigma}{m_t^2}\right) -\frac{2}{3} v_\s \frac{y_1 y_2}{M_1 M_5} 
     + \mathcal{O}\l\frac{v^3_\s}{{M}^2_1 {M}^2_5} \r \,,\nn \\ 
\omega_{-1/3}^\s(m_\sigma^2) & \equiv  & \left.
   \frac{1}{6}\frac{d}{d\s}\left(\log\det(\mathcal{M}_\mathcal{B}\,\mathcal{M}^\dagger_\mathcal{B}) 
   -\log(m_b(h,\sigma)\,m^*_b(h,\sigma))\right) 
   \right|_{\scriptsize \renewcommand{\arraystretch}{0.75} \!\!\begin{array}{l} h=v \\ \s=v_\s \end{array}}  \nn \\
    & = & -\frac{2}{3} v_\s \frac{y'_1 y'_2}{M'_1 M'_5}
    + \mathcal{O}\l\frac{v^3_\s}{{M'}^2_1 {M'}^2_5} \r\,,
\eea
while for $\omega^h(m_\sigma^2)$ it results
\bea
\omega_{2/3}^h(m_\sigma^2) & \equiv & \left. 
   \frac{1}{6}\frac{d}{dh}\left(\log\det(\mathcal{M}_\mathcal{T}\,\mathcal{M}^\dagger_\mathcal{T})
   -\log(m_t(h,\sigma)\,m^*_t(h,\sigma))\right) 
   \right|_{\scriptsize \renewcommand{\arraystretch}{0.75} \!\!\begin{array}{l} h=v \\ \s=v_\s \end{array}}     \!\!\!+ \frac{1}{v}I\left(\frac{m^2_\sigma}{m_t^2}\right) \nn \\ 
        & = &  \frac{1}{v}I\left(\frac{m^2_\sigma}{m_t^2}\right) -\frac{2}{3} v \frac{y_1 y_2}{M_1 M_5} 
     + \mathcal{O}\l\frac{v v^2_\s}{{M}^2_1 {M}^2_5} \r \,,\nn \\ 
\omega_{-1/3}^h(m_\sigma^2) & \equiv  & \left.
   \frac{1}{6}\frac{d}{dh}\left(\log\det(\mathcal{M}_\mathcal{B}\,\mathcal{M}^\dagger_\mathcal{B}) 
   -\log(m_b(h,\sigma)\,m^*_b(h,\sigma))\right) 
   \right|_{\scriptsize \renewcommand{\arraystretch}{0.75} \!\!\begin{array}{l} h=v \\ \s=v_\s \end{array}}  \nn \\
    & = & -\frac{2}{3} v \frac{y'_1 y'_2}{M'_1 M'_5}
    + \mathcal{O}\l\frac{v v^2_\s}{{M'}^2_1 {M'}^2_5} \r\,,
\eea
where it has been assumed that $m_b\ll  m_t, m_\s$ while $m_\s \ll m_i$ for all the other heavy quarks.

Finally, the physical $h$ and $\s$ decay widths into two photons are given by
\bea
\Gamma({ h}\to\gamma\gamma) &=& 
       \frac{\alpha^2 m_h^3}{64\pi^3}~\left|\cos\gamma~\left[\Omega_{\hat h}^{W}(m_h^2) + \Omega_{\hat h}^{F}(m_h^2)\right] - \sin\gamma~\Omega_{\hat \s}^F(m_h^2)\right|^2 \,,\nn \\
\Gamma({ \sigma}\to\gamma\gamma) &=&
       \frac{\alpha^2 m_\sigma^3}{64\pi^3}~\left|\sin\gamma~ \left[\Omega_{\hat h}^{W}(m_\sigma^2) +\Omega_{\hat h}^{F}(m_\sigma^2)\right] + \cos\gamma~\Omega_{\hat \s}^F(m_\sigma^2)\right|^2\,. 
\eea

Quantitatively,  $\sigma\to\gamma\gamma$ transitions are dominated by the $W^\pm$  loop contributions unless the scalar mixing is small enough for the heavy partner loop contribution to be significant.

\section{ The $\sigma$ resonance at the LHC}
\label{sec:pheno-sigma}

May a light $\sigma$ resonance be lurking in LHC data? In that case,
what distinguishes the phenomenology expected for an approximate
$SO(5)$ invariant scenario and that for a generic singlet scalar
freely added to the SM Lagrangian? What is the parameter space allowed
at present and the discovery reach of the next LHC run? In this
section we address these questions.

The constraints from electroweak precision tests explored in
Sect.~\ref{precision} showed that a scenario with a light $\sigma$
particle tends to diminish the tension with data. On the
other side, from the theoretical viewpoint the assumption of a PNGB
nature for the Higgs boson within an approximate global $SO(5)$
symmetry mildly broken by soft terms prefers a sizeable mass for the
$\sigma$ particle, see Fig.~(\ref{Fig-theory-bounds}). The PNGB
interpretation implies the existence of a non-zero mixing between
$\sigma$ and $h$, specially when considering naturalness as a
guideline since $\sin^2\gamma\sim\xi\ll 1$ would require a strong
fine-tuning of the theory -see the discussion after Eq.~(\ref{condp}) and Eqs.~(\ref{exact}), (\ref{xi-aprox}) and (\ref{xilimit}).

 As argued in Sect.~\ref{renormalization},
the scalar potential is completely determined by the masses $m_h$ and
$m_\sigma$, the constant $G_F$, and the scalar mixing $\sin\gamma$. 
The conclusions obtained for the linear $\sigma$ model  together with generic soft breaking terms are of general validity. 
The extra ingredients needed to  determine the phenomenology
of the $\sigma$ particle are its couplings to the vector-like fermions of the
theory, which introduce instead significant model-dependence and may have important consequences particularly in the
production of this scalar. 

In order to estimate the LHC constraints
on the model, we recast many LHC searches for
scalar resonances into the $\sigma$ parameter space,  calculating the
production cross section and decays of the $\sigma$ particle. The production of the $\sigma$ particle at the LHC  may proceed mainly via two
processes, gluon fusion and vector boson fusion (VBF). 
Gluon fusion usually dominates the  production due to the
large gluon pdfs. Nevertheless, this conclusion is somewhat model-dependent as the
heavy fermion couplings to $\sigma$ may a priori enhance or diminish the cross
section.  VBF depends essentially on the mixing angle $\gamma$, but it
typically yields a lower production cross section than gluon fusion for
$m_\sigma<\mathcal{O}(1~{\rm TeV})$, for which it will have unnoticeable impact in what follows.   

Consider then the cross section for $\sigma$ production via gluon fusion $\sigma (gg \to \sigma)$. To account
for higher order corrections to $\Gamma(\sigma\to gg)$, we will profit from the results in the literature for a  heavy SM-like Higgs boson $H'$, using the 
following approximation
\begin{equation}
  \sigma(gg\to\sigma) \simeq \frac{|A(\sigma\to gg)|^2}{|A_{SM}(H'\to gg)|^2}
                       \,\, \sigma_{SM}(gg \to H')\,,
\end{equation}
where  $A(\sigma,H'\to gg)$  refers to
leading order (one loop) amplitudes and $\sigma_{SM}(gg \to H')$ is the
NNLO standard gluon fusion production cross section given in
Ref.~\cite{Heinemeyer:2013tqa}. For illustrative purposes we discuss next the LHC impact of the $\sigma$ particle in two steps: first an ``only scalars'' analysis  will be considered, to add next to it the effect of the rather model-dependent fermionic  sector.

\begin{figure}
\centering
\includegraphics[width=0.48\textwidth,keepaspectratio]{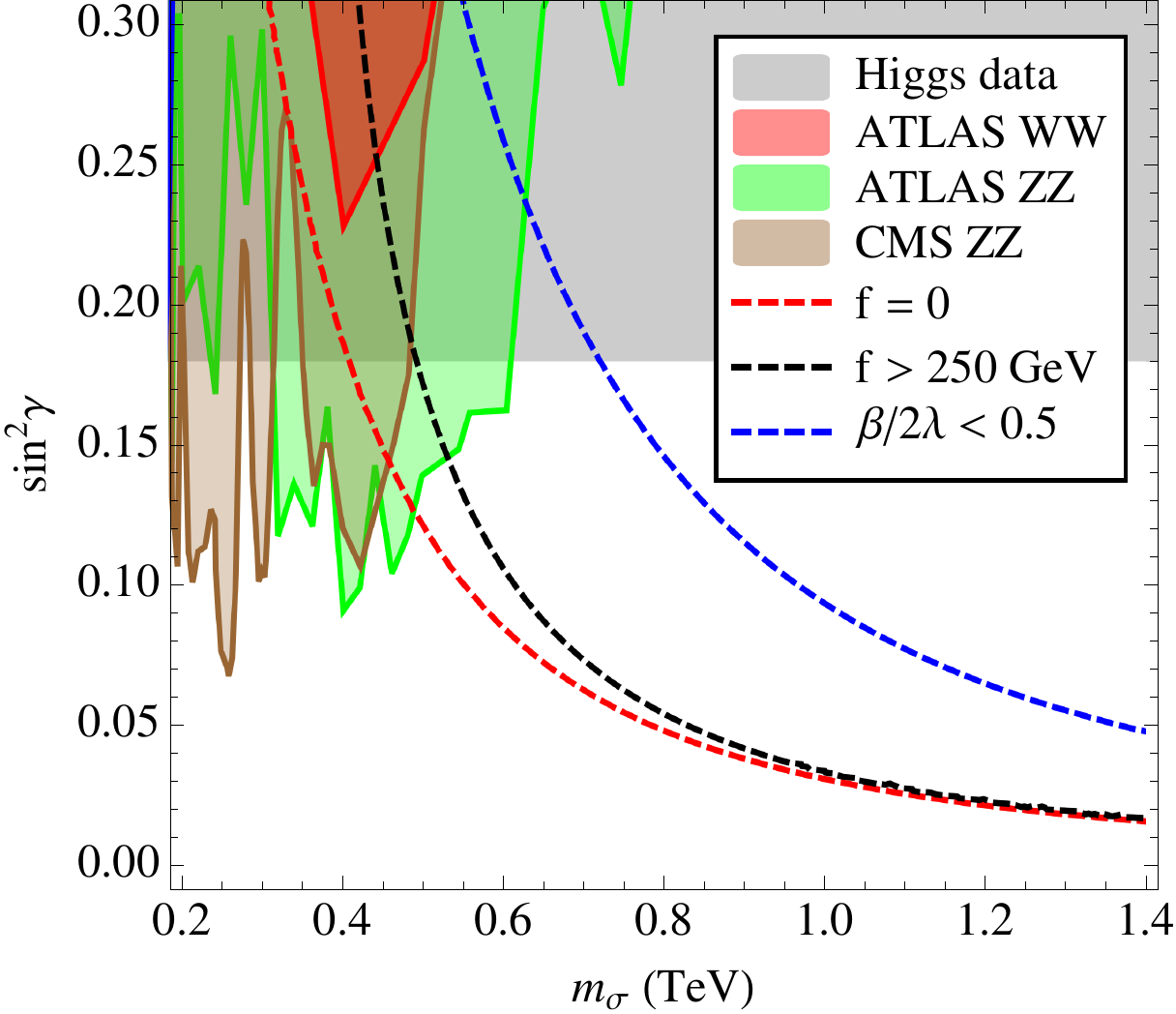}
\includegraphics[width=0.48\textwidth,keepaspectratio]{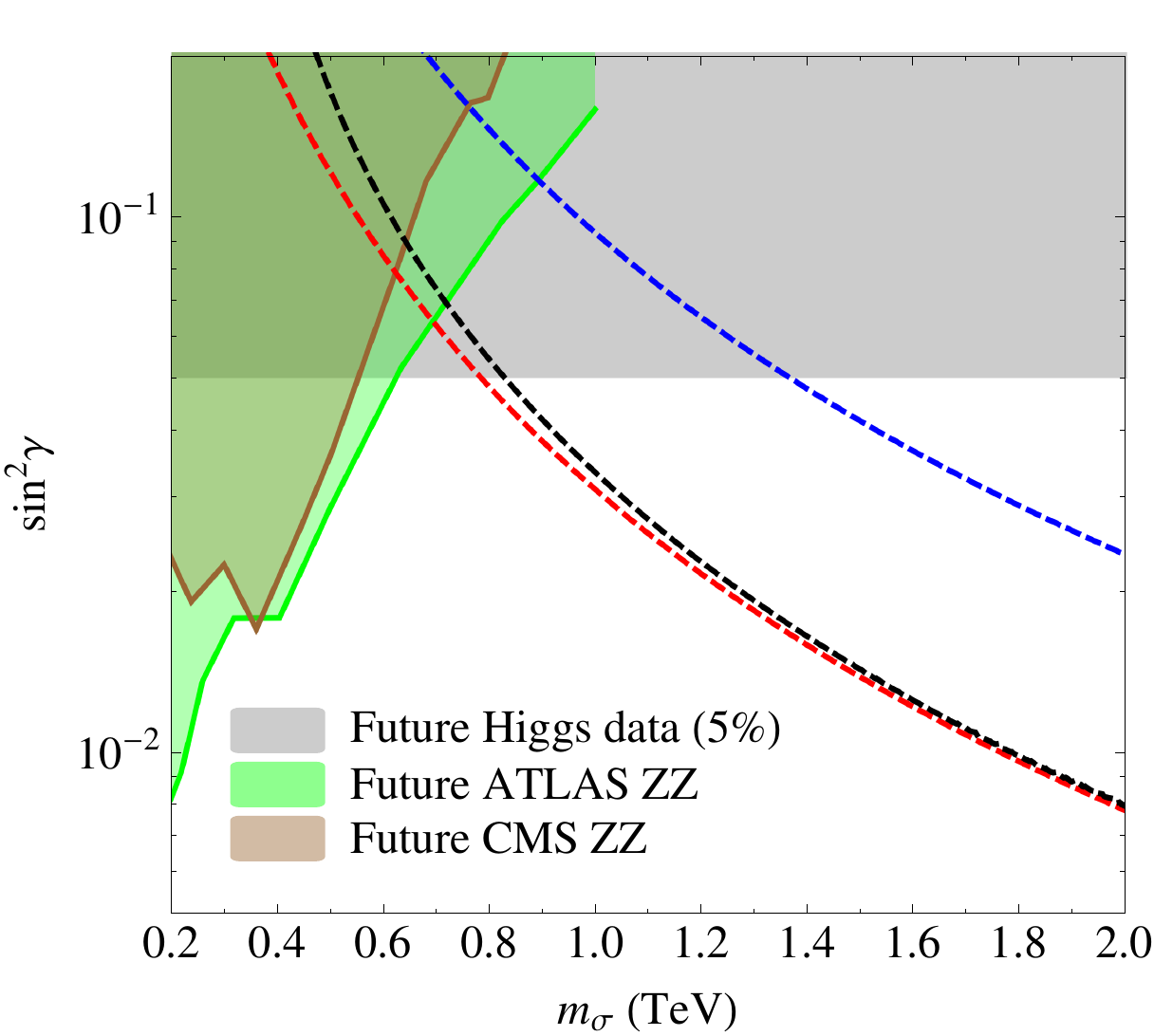}
\caption{Present LHC (left panel) and future LHC run-2 (right panel)
  constraints on $\sin^2\gamma$ versus $\sigma$ mass parameter space
  in the case where gluon fusion is dominated by the top loop. For the
  latter, a total luminosity of 3ab$^{-1}$ was assumed.}
  \label{Fig-sigma-LHC}
\end{figure}

In the only scalars scenario, that is, a case in which the impact of
the heavy fermions on gluon fusion is negligible compared to the top
contribution, the production amplitude can be approximated by the top
loop contribution for a heavy SM Higgs weighted down by $\sin\gamma$.
  Under this
assumption, we have recasted the LHC searches for a heavy Higgs-like
particle into constraints in the $\{m_\sigma, \sin^2\gamma\}$ plane, and the results are shown
on the left panel of Fig.~\ref{Fig-sigma-LHC}. The searches taken into
account here include  diphoton~\cite{Aad:2014ioa,Khachatryan:2015qba}, diboson~\cite{Aad:2015agg, Aad:2015kna,CMS:xwa, CMS:bxa} and dihiggs\cite{Khachatryan:2015yea, Aad:2015xja} decays.
The figure shows that present LHC data are sensitive to $\sin^2\gamma \simeq
0.1$ for $m_\sigma< 600~{\rm GeV}$, otherwise Higgs measurements put
a bound on $\sin^2\gamma<0.18$ independently of $m_\sigma$.  It is worth noting that these bounds apply well beyond the model discussed in this paper: they are valid for any physics scenario in which the role of the Higgs particle is substituted by a Higgs-scalar system with a generic mixing angle $\gamma$, independently of the details of the theory. 
In addition, by 
combining the LHC data with theoretically motivated constraints as
those mentioned above, interesting bounds can be derived: a PNGB
nature for the Higgs boson corresponds to the area to the right of the
 red curve depicted, see also Fig.~\ref{Fig-theory-bounds}, corresponding to the minimal theoretical
  requirement $f^2>0$  for $SO(5)$ to be spontaneously broken,
  resulting in the bound $m_\sigma> 500$~GeV in particular from the impact of ATLAS $H_{\rm heavy}\to ZZ$ searches. If $f^2$ values above the electroweak scale are instead required (black curve) $m_\sigma> 550$~GeV follows.
 The future prospects for
this ``only scalars'' scenario are depicted on the right panel of
Fig.~\ref{Fig-sigma-LHC}. It shows the future LHC sensitivity in the
$ZZ$ decay channel of the 14~TeV LHC run with an integrated luminosity
of 3~ab$^{-1}$, for both ATLAS and CMS~\cite{Holzner:2014qqs}, as well
as the mixing disfavoured by Higgs data assuming a 5\% precision on
the Higgs couplings to SM particles. In the absence of any beyond the
SM signal, future LHC data together with the aforementioned theory
constraints could push the limit on the $\sigma$ mass above
900~GeV--1.4~TeV.

The difference between the LHC predictions of the model discussed in
this paper and those stemming from extending the SM by a generic
scalar singlet (see e.g. Ref.~\cite{Martin-Lozano:2015dja}) is the
underlying $SO(5)$ structure of the former, which prescribes a
specific relation between the quartic terms in the potential as well
as specific soft breaking terms. In the  generic
extra singlet scenarios, the allowed parameter space is given by
the entire white area in Fig.~\ref{Fig-sigma-LHC}, while a PNGB nature
for the Higgs restricts the allowed region to the area to the right of
the curves depicted in the figure.

\begin{figure}
\centering
\includegraphics[width=0.48\textwidth,keepaspectratio]{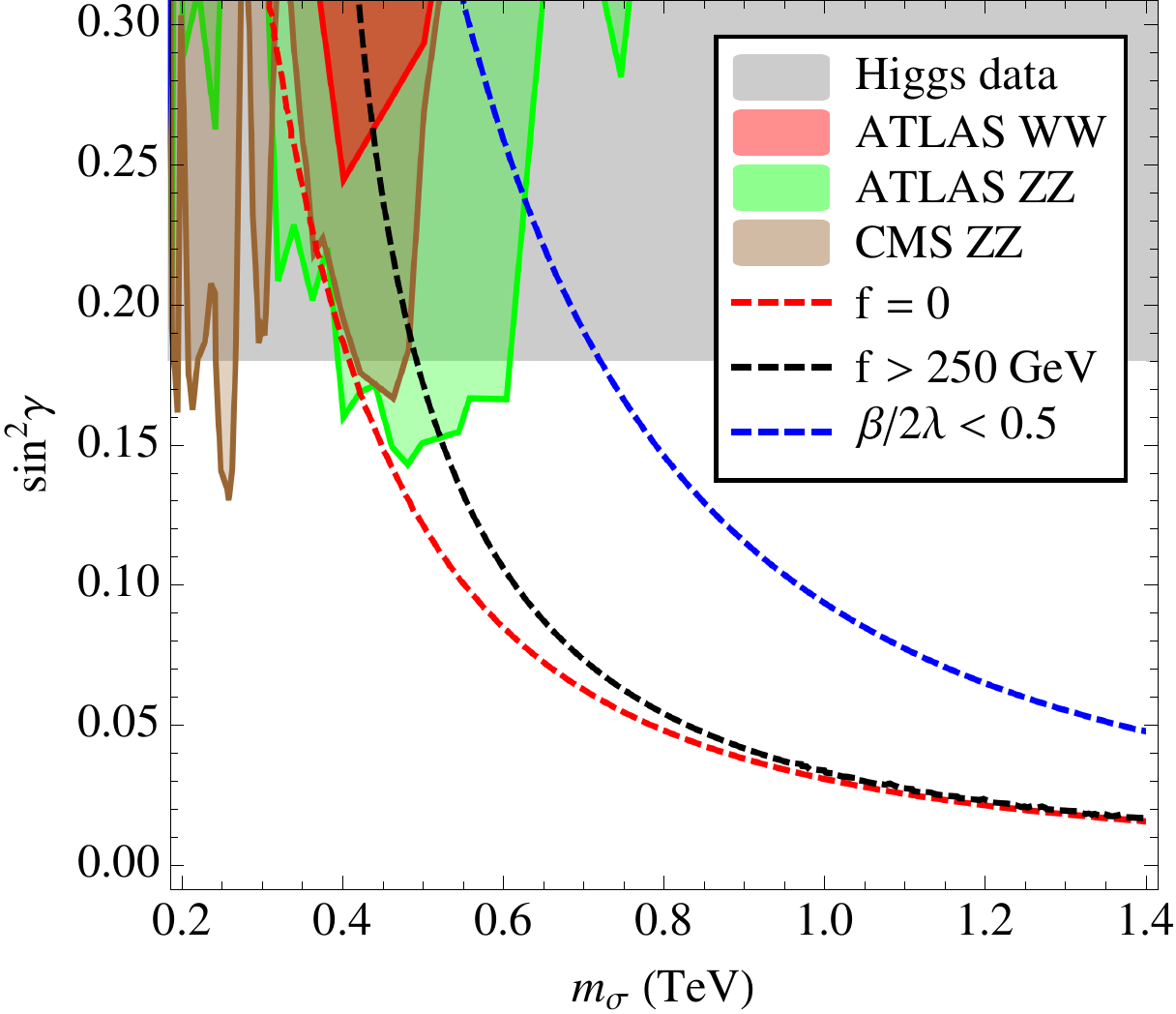}
\includegraphics[width=0.48\textwidth,keepaspectratio]{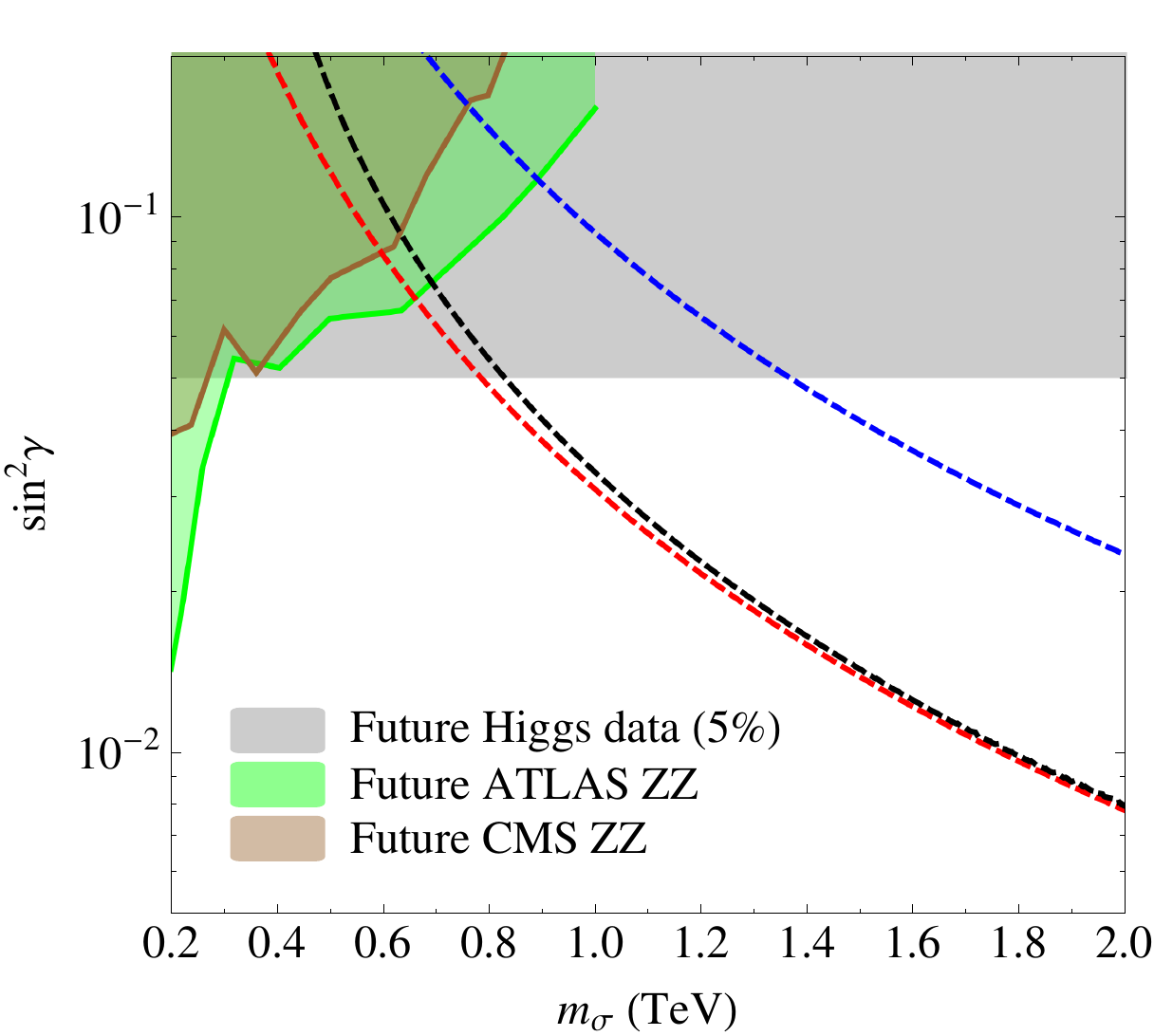}
\caption{Same as fig.~(\ref{Fig-sigma-LHC}), but considering a
  sizeable contribution of the heavy fermion sector to gluon
  fusion. See text for details.}
  \label{Fig-sigma-LHC-22}
\end{figure}

Finally, the impact of the heavy fermions of the model on the gluon
fusion cross section may be significant. Using the approximate expressions
in Eqs.~(\ref{Ahs}) and (\ref{xsectpp}), assuming that the factor $y_1
y_2/M_1M_5$ in Eqs.~(\ref{ghath}) and (\ref{ghatsigma}) is the largest
contribution between $1/(4\pi f)^2$ and 1~TeV$^{-2}$ (the latter
  will also be assumed when $f^2<0$), the results obtained are
depicted in Fig.~\ref{Fig-sigma-LHC-22}.  It shows that the present
LHC bounds on $\sin^2\gamma$ can be weakened by $\mathcal{O}(30-50\%)$
with respect to the ``only scalars'' bounds in
Fig.~\ref{Fig-sigma-LHC}. This is due to a destructive interference
between the heavy fermions and the top loops, for the set of parameters considered.
 Moreover, future searches will
be much more sensitive to the heavy fermion sector as they probe
smaller mixing angles, and therefore they enter regions in parameter
space where the top quark is relatively less important to the $\sigma$
phenomenology.

\subsubsection*{ The 750 GeV di-photon excess}
It is a natural question whether the mild $750$~GeV di-photon
excess  
observed by ATLAS~\cite{ATLAS-exo} and CMS~\cite{CMS:2015dxe} could be explained
by the $\sigma$ resonance under discussion. This is highly unlikely
because of the constraints imposed on the scalar couplings by the
approximate $SO(5)$ symmetry of the scalar potential, as well as the
uniqueness of the signal, as explained next.

Since the decay $\sigma\to\gamma\gamma$ is loop induced, the
corresponding branching ratio tends to be very small. In order to be
able to account for the excess observed, the $h-\sigma$ mixing needs
to be tiny, so that for instance the $WW$ and $ZZ$ channels are
suppressed and the loop-induced processes may dominate the decay.
 This requires $\sin^2\gamma\ll (m_h/m_\sigma)^2\simeq 0.03$, which Eq.~(\ref{exact}) shows to require on one side $f^2<0$ --for which the Higgs cannot be interpreted  as a
PNGB, and on the other a very small and fine-tuned $\alpha$ value as  $\alpha^2\propto\beta^2\sin^2(2\gamma)$; overall a very unnatural scenario.

Furthermore, since the mixing is so small, the
top loop essentially does not contribute to the production and decay
anymore. Hence, to obtain a large enough
$\Gamma(\sigma\to\gamma\gamma)$ and $\Gamma(\sigma\to gg)$
 the fermion content needs a higher multiplicity than what is assumed
in this paper, as well as large Yukawa couplings, since $\sigma$
mainly couples to $T_1$ and $T_5$ with electric charges $2/3$. Even
assuming such an extreme and extended configuration, the stability of
the potential could become a very serious issue as the fermion
contribution to the beta function of the quartic couplings is
negative. Additional field content would then be possibly required to
compensate for this effect, making the model extremely
\emph{ad hoc}. Therefore, we find no compelling argument to interpret
the 750~GeV excess as corresponding to the $\sigma$ scalar studied
here.


%
\section{\boldmath$d\le6$ Fermionic Effective Lagrangian}
\label{EffectiveLag}
%
\begin{table}
\centering
\renewcommand{\arraystretch}{1.75}
\begin{tabular}{| c| c|}
\hline
Coefficient & Leading Order in $f/M$ \\[0.5ex]
\hline \hline
${\mathcal Z}_{q_L}$ & $\l 1 + \frac{\Lam_1^2}{M_5^2} + \frac{{\Lam'}_1^2}{{M'}_5^2} \r$ \\
\hline
${\mathcal Z}_{t_R}$ & $\l 1 + \frac{\Lam_2^2}{M_5^2} + \frac{\Lam_3^2}{M_1^2} \r$ \\
\hline
${\mathcal Z}_{b_R}$ & $\l 1 + \frac{{\Lam'}_2^2}{{M'}_5^2} + \frac{{\Lam'}_3^2}{{M'}_1^2} \r$ \\
\hline
\end{tabular}
\caption{Table with the definitions for the renormalization factors.} \label{tabrenfactors}
\label{Zvalues}
\end{table}

\begin{table}
\centering
\renewcommand{\arraystretch}{2.75}
\begin{tabular}{|c | c| c| c|}
\hline
& Operator & $c_i$ & Leading Order in $f/M$ \\[0.5ex]
\hline \hline
\multirow{2}{*}{dim 4}
&$\bar q_L~\wH~t_R$ & $-y_t$ & $- \l  \frac{y_1 \Lam_1\Lam_3}{M_1 M_5} \r {\mathcal Z}^{-1/2}_{q_L} {\mathcal Z}^{-1/2}_{t_R}$ \\
\cline{2-4}
&$\bar q_L~H~b_R$ & $-y_b$ & $- \l\frac{y'_1 \Lam'_1\Lam'_3}{M'_1 M'_5} \r {\mathcal Z}^{-1/2}_{q_L} {\mathcal Z}^{-1/2}_{b_R}$ \\
\hline \hline
\multirow{2}{*}{dim 5}
&$\s\,(\bar q_L \wH t_R)$ &
$c^{t}_{\s 1}$ & $\frac{y_t}{M_5}\paren{y_2\frac{\Lam_2}{\Lam_3}-\l y_1\frac{\Lam_2\Lam_3}{M_1 M_5}
+ y_2\frac{\Lam_2\Lam_3}{M^2_1} \r\mathcal Z^{-1}_{t_R} }$\\
\cline{2-4}
&$\s\,(\bar q_L H b_R)$ &
$c^{b}_{\s 1}$ & $\frac{y_b}{M'_5}\paren{y'_2\frac{\Lam'_2}{\Lam'_3}- \l y'_1\frac{\Lam'_2\Lam'_3}{M'_1 M'_5}
+ y'_2\frac{\Lam'_2\Lam'_3}{{M'}^2_1} \r \mathcal Z^{-1}_{b_R}}$\\
\hline\hline
\multirow{9}{*}{dim 6}
&$\s^2\,(\bar q_L \wH t_R)$ &
$c^{t}_{\s 2}$ &\makecell{$-\frac{y_t}{M_1 M_5}\Bigl(y_1 y_2 - \l
y_1 y_2\l 2\frac{\Lam_2^2}{M_5^2} + \frac{\Lam_3^2}{M_1^2} \r +
\frac{3 y_2^2 \Lam_2^2 + y_1^2\Lam_3^2}{2M_1 M_5} \r \mathcal Z^{-1}_{t_R}$ \\
\hspace{2cm} $+2\frac{\Lam_2^2\Lam_3^2}{M_1 M_5}
		\paren{\frac{y_1^2}{M_5^2}+\frac{2 y_1 y_2}{M_5 M_1}
	+\frac{y_2^2}{M_1^2}}\mathcal Z^{-2}_{t_R}\Bigr) $}\\
\cline{2-4}
&$\s^2\,(\bar q_L H b_R)$ &
$c^{b}_{\s 2}$ &\makecell{$-\frac{y_b}{M'_1 M'_5}\Bigl(y'_1 y'_2  -\l
y'_1 y'_2\l 2\frac{\Lam'{}_2^2}{M'{}_5^2} + \frac{\Lam'{}_3^2}{M'{}_1^2} \r +
\frac{3 y'{}_2^2 \Lam'{}_2^2 + y'{}_1^2\Lam'{}_3^2}{2M'_1 M'_5} \r \mathcal Z^{-1}_{b_R}$ \\
\hspace{2cm} $ +2\frac{\Lam'{}_2^2\Lam'{}_3^2}{M'_1 M'_5}
		\paren{\frac{y'{}_1^2}{M'{}_5^2}+\frac{2 y'_1 y'_2}{M'_5 M'_1}
	+\frac{y'{}_2^2}{M'{}_1^2}}\mathcal Z^{-2}_{b_R}\Bigr) $}\\
\cline{2-4}
&$\,|H|^2\,(\bar q_L \wH t_R)$ &
$c^{t}_{H2}$ &\makecell{ $-\frac{y_t}{M_1 M_5}\Bigl(2y_1 y_2
-\paren{2y_1y_2\frac{\Lam_3^2}{M_1^2}
+y_1^2\frac{\Lam_3^2}{M_1 M_5}}\mathcal Z^{-1}_{t_R}$ \\
\hspace{3.5cm} $-\paren{y_1y_2\frac{\Lam_1^2}{M_5^2}
+\frac{y_1^2}{2}\frac{\Lam_1^2}{M_1 M_5}}\mathcal Z^{-1}_{q_L}\Bigr)$}\\
\cline{2-4}
&$\,|H|^2\,(\bar q_L H b_R)$ &
$c^{b}_{H2}$ &\makecell{$-\frac{y_b}{M'_1 M'_5}\Bigl(2y'_1 y'_2
-\paren{2y'_1y'_2\frac{\Lam'{}_3^2}{M'{}_1^2}
+y'{}_1^2\frac{\Lam'{}_3^2}{M'_1 M'_5}}\mathcal Z^{-1}_{b_R}$ \\
\hspace{3.5cm} $-\paren{y'_1y'_2\frac{\Lam'{}_1^2}{M'{}_5^2}
+\frac{y'{}_1^2}{2}\frac{\Lam'{}_1^2}{M'_1 M'_5}}\mathcal Z^{-1}_{q_L}\Bigr)$}\\
\cline{2-4}
&$(H^\dagger i\overleftrightarrow D{}_\mu\,H)(\bar q_L \g^\mu q_L)$ &
$c^{(1)}_L$ & $\frac{1}{4}\paren{\frac{y_1^2\Lam_1^2}{M_1^2 M_5^2}-\frac{{y'}_1^2{\Lam'}_1^2}{{M'}_1^2 {M'}_5^2}}{\mathcal Z}^{-1}_{q_L}$\\
\cline{2-4}
&$ (H^\dagger i\overleftrightarrow D{}^i_\mu\,H)(\bar q_L \tau^i \g^\mu q_L) $ &
$c^{(3)}_L$ & $-\frac{1}{4}\paren{\frac{y_1^2\Lam_1^2}{M_1^2 M_5^2}+\frac{{y'}_1^2{\Lam'}_1^2}{{M'}_1^2 {M'}_5^2}}{\mathcal Z}^{-1}_{q_L}$\\
\cline{2-4}
&$(H^\dagger i\overleftrightarrow D{}_\mu\,H)(\bar t_R \g^\mu t_R)$&
$c^{t}_R$ & 0 \\
\cline{2-4}
&$(H^\dagger i\overleftrightarrow D{}_\mu\,H)(\bar b_R \g^\mu b_R)$ &
$c^{b}_R$ & 0 \\
\cline{2-4}
& $i(\wH^\dagger D{}_\mu\,H)(\bar t_R \g^\mu b_R)$ &
$c^{tb}_R$ & 0 \\
\hline
\end{tabular}
\label{Effectiveops}
\caption{\small Leading order low-energy effective operators induced  and  their 
coefficients. Note that the Yukawa couplings defined
in the two first rows appear as well in coefficients of some higher order operators.
The renormalization factors present have been defined in Table~\ref{Zvalues}.
Those operators made out exclusively of SM fields have been written in the Warsaw basis~\cite{Grzadkowski:2010es}.}
 \label{tabops}
\end{table}

The linear model described is renormalizable and valid for any mass range of the fermionic and/or scalar exotic fields. Two simplifying limits are specially interesting: 
i) the heavy fermion regime,  
$M \gg m_\s \gg v $, where $M$ generically represents the exotic fermionic scales $M_i$ in Eq.~(\ref{SO5Lag}); ii) the heavy singlet regime,  $m_\s \gg M \gg v$. We have concentrated in this paper 
 on the first scenario, considering a not-so-heavy extra 
singlet in the spectrum and its phenomenological consequences. The second limit is instead interesting to elucidate the connection
between the linear (weak) and non-linear (strong) BSM physics scenarios: the $m_\sigma\rightarrow \infty$ regime should lead to the non-linear scenarios usually explored in the literature about composite Higgs; it will be discussed in detail 
in a subsequent publication \cite{future-paper}. 

When condition i) is satisfied, some important low-energy effects (and model dependencies) induced by the 
exotic fermions are easily inferred by integrating them out. 
The procedure is quite lengthy;  
 here only the resulting  mass-dimension ($d$)  $4$, $5$ and $6$ effective operators and their coefficients are summarized.  For energies $E< M$,
 the effective Lagrangian describing the  $d\le6$ interactions of fermions  with gauge and scalar fields can be written as
\bea
\mathcal{L}_{eff} &=& 
    \bar q_L i\Ds ~ q_L + \bar t_R i\Ds\,t_R + \bar b_R i\Ds\,b_R + \sum_i c_i \mathcal{O}_i\,,
\label{efflagd6}
\eea
where the set $ \{\mathcal{O}_i\}$ includes operators of dimension four (for which the induced coefficients are the leading contributions to the top and bottom Yukawa couplings), five and six. We will use the ``Warsaw basis''~\cite{Grzadkowski:2010es} below.

In most models (for instance in composite Higgs ones) it is reasonable to assume that the goldstone 
boson scale $f$ and the scalar vevs all satisfy $f\,,v\,, v_\s\,  \ll M$.  In what follows we will thus assume  $f /M \ll 1$ for simplicity,
while  $\Lambda \approx M$ will be considered with $\Lambda$ denoting generically the composite $\Lambda_i$ scales in Eq.~(\ref{SO5Lag}).  The light field kinetic energies get contributions which require wave function renormalization in order to recover canonically normalized kinetic energies, 
\bea
q_L & & \rightarrow \quad  {\mathcal Z}^{-1/2}_{q_L} q_L\,, \\
q_R & & \rightarrow \quad 
\begin{pmatrix}{\mathcal Z}^{-1/2}_{t_R}  & 0 \\ 0 & {\mathcal Z}^{-1/2}_{b_R}\end{pmatrix} 
\begin{pmatrix} t_R \\ b_R \end{pmatrix}\,, 
\eea
where ${\mathcal Z}^{-1/2}_{t_R}$ and ${\mathcal Z}^{-1/2}_{b_R}$ are given in Table~\ref{Zvalues}. The operators obtained and their coefficients resulting after those redefinitions, at leading order  in $f/M$, are  shown in Table~\ref{tabops}, where the following definitions have been used, 
\bea
(H^\dagger i\olraw D_\mu\,H) &\equiv& i\paren{H^\dagger (\oraw D_\mu H) - (H^\dagger \olaw D_\mu) H} \, , \nn \\ 
(H^\dagger i\olraw D^i_\mu\,H) &\equiv& i\paren{H^\dagger\tau^i (\oraw D_\mu H) - (H^\dagger \olaw D_\mu)\tau^i H} \,. \nn
\eea

In writing Eq.~(\ref{efflagd6}) and Table~\ref{tabops}  the unshifted scalar fields have been assumed. This fact introduces a potential subtlety that we discuss next. Consider for instance the top and bottom quark masses corresponding to the first two operators in the table, which are their respective Yukawa couplings: when the Higgs gets a vev, mass terms for the light quarks are generated. 
Additional contributions to the light quark masses stem however  from the next six operators in the list, for $\s=\vs$ and $H = \vH$. The corrections induced in the top and bottom mass are of higher order in $f/M$, though, and do not need to be retained when working at leading order. Finally, 
\bea
m_t &=& \frac{v}{\sqrt{2}} \l \frac{y_1 \Lam_1\Lam_3}{M_1 M_5} \r \frac{1}
        {\sqrt{\l 1 + \frac{\Lam_1^2}{M_5^2} + \frac{{\Lam'}_1^2}{{M'}_5^2} \r
         \l 1 + \frac{\Lam_2^2}{M_5^2} + \frac{\Lam_3^2}{M_1^2}\r}} \l 1 + \mathcal{O}\left(\frac{f}{M}\right) \r \,,\nn \\
m_b &=& \frac{v}{\sqrt{2}} \l \frac{y'_1 \Lam'_1\Lam'_3}{M'_1 M'_5} \r \frac{1}
        {\sqrt{ \l 1 + \frac{\Lam_1^2}{M_5^2} + \frac{{\Lam'}_1^2}{{M'}_5^2} \r 
         \l 1 + \frac{{\Lam'}_2^2}{{M'}_5^2} + \frac{{\Lam'}_3^2}{{M'}_1^2} \r}} \l 1 + \mathcal{O}\left(\frac{f}{M}\right) \r \,.\nn    
\eea
The same reasoning applies to other couplings. For example the fermion-$\sigma$ coupling via the $\mathcal{O}^t_{\s1}$ operator~\cite{Grzadkowski:2010es} would get corrections proportional to $c^t_{\s2} \vs$ --see Table~(\ref{tabops})-- which are of higher order in the $f/M$ expansion, and can thus be disregarded when restraining to the leading contributions.


\section{Conclusions}
 A composite Higgs would manifest with deviations of the Higgs couplings to fermions
and gauge vector bosons from the SM predictions. Current data about the Higgs properties are in good agreement with the SM, but the present experimental precision
still allow for deviations at the level of 20\% or more, while many channels predicted by the SM have not yet been tested.

Completely clarifying the mechanism of electroweak breaking is one of the main goals of particle physics today and an important role, on the theory side, is played by
simple and motivated extensions of the SM which could provide guidance in the experimental search. To this purpose we have formulated a 
model where the scalar sector of the SM is minimally extended to include an additional scalar particle $\sigma$. We are motivated by the attractive possibility that the Higgs itself
can be interpreted as a pseudo-Goldstone boson associated to the breaking of an approximate symmetry. The most economic custodial preserving possibility is offered by a global $SO(5)$
spontaneously broken down to $SO(4)$, thus generating the four components of the Higgs doublet as Goldstone bosons.
Indeed this case has already been vastly analyzed in the literature in the strongly interacting regime, either
in the context of four-dimensional models where $SO(5)$ is nonlinearly realized or in
five-dimensional models with a warped space-time metric. The latter, weakly coupled duals to strongly coupled four-dimensional conformal theories, 
are believed to provide a calculable framework for composite Higgs models. 

In our model, where a scalar fiveplet of $SO(5)$ comprises both an electroweak doublet and an extra singlet $\sigma$,
the $SO(5)$ symmetry is linearly realized and the theory is renormalizable. It could be viewed as the simplest UV completion in the class of models based on the coset $SO(5)/SO(4)$.
In this way we lose generality, but we gain in calculability and predictability. We can study accurately the regime where the symmetry breaking sector is in the perturbative regime
and provide a useful interpolation between the weakly and the strongly interacting cases. The $SO(5)$ invariant part of the symmetry breaking sector contains just two parameters, the symmetry breaking scale and the mass of the $\sigma$ particle. In the $SO(5)$ invariant limit the Higgs particle is a true massless Goldston boson.
Other two parameters, arising from the one-loop effective potential when gauge and Yukawa interactions are turned on, break the $SO(5)$ symmetry softly,
fix the relative orientation between the residual $SO(4)$ invariance and the $SU(2)\times U(1)$ electroweak group and provide the Higgs boson a mass.
A mixing angle $\gamma$ defines the two physical mass eigenstates as mixtures of the electroweak doublet and the singlet. 

As the Higgs mass and the value of the electroweak scale are known, the scalar parameter space is thus completely defined in terms of the $\sigma$ mass and $\sin\gamma$. We have identified in it the areas in which the Higgs can be considered a pseudo-Goldstone boson, resulting in two well-differentiated regions corresponding respectively to a $\sigma$ particle lighter and heavier than the Higgs particle. The former case has phenomenological interest, but it turns out to require fine-tuned parameters in the scalar potential and indeed it would call for an explanation of the stability of such a light $\sigma$; in consequence, we have focused most of the analysis on the region in which the $\sigma$ is heavier than the Higgs, to which the remarks that follow apply.

The  UV completion of the theory  would require further explanation as far as the $\sigma$ particle is light enough for the theory to remain in the perturbative realm, as the model would have then replaced the hierarchy problem for the Higgs mass for that of the $\sigma$ mass. It is nevertheless a most useful tool to explore a dynamical origin for the Higgs and the possibility of new degrees of freedom appearing in foreseen experiments. For heavy $\sigma$ the symmetry breaking sector becomes strongly interacting and in the limit of infinite $m_\sigma$ we fall into a nonlinear realization of $SO(5)$. 
The couplings of the physical Higgs $h$ to $W$ and $Z$
are suppressed by $\cos\gamma$, while the heavier $\sigma$ state couples to $W$ and $Z$ with a strength proportional to $\sin\gamma$. 
Present data on the Higgs decay into $WW$ require a relatively small mixing between the Higgs field and the new scalar, leaving however a considerable room to a departure from the SM picture. We have also identified the differences in experimental impact of the case of a generic singlet scalar added to the SM and the case of the approximately $SO(5)$ invariant scalar sector under discussion.

The scalar sector of the non-linear $SO(5)$ $\sigma$ model developed is minimal, simple and of general validity.  
A significant model dependence comes in when considering the fermion sector. We chose to describe fermion masses within a simple realization of 
the partial compositeness idea. We introduced a set of new vector-like heavy fermions, which couple to the full $SO(5)$ scalar fiveplets. Ordinary fermions
do not directly interact with scalars but mix with the heavy fermions. At low energies such a mixing gives rise to the usual Yukawa couplings and fermion masses.
To keep our model as minimal as possible, we assign the heavy fermions to singlets and fiveplets. By focusing on the third generation of quarks, we introduce one singlet and one fiveplet per each charge sector. Even in this minimal setup, the fermion sector of the model brings in 14 parameters: four heavy fermion masses,
four independent Yukawa couplings and six mixing parameters.  For completeness and as first step towards a low-energy benchmark effective Lagrangian, we have separately integrated out the heavy fermions and obtained the ensuing dimension four, five and six effective operators made out of only SM fields or $\sigma$ plus SM fields,  which include  Yukawa and other couplings for the lighter particles. Nevertheless, all the phenomenological analyses have been performed with dynamical heavy fermions.

We have analyzed precision electroweak observables, globally parametrized in terms of the $S$, $T$ and $R_b$ parameters. Concerning the scalar contribution, from an explicit
one-loop computation we recover a well-known result: a positive contribution to the $S$ parameter and a negative contribution to the $T$ parameter. At variance with generic 
models where $SO(5)$ is nonlinearly realized, we do not have any cut-off ambiguity. In our model the extra contributions $\Delta S$ and $\Delta T$ are finite. They vanish
when $\sin\gamma$ goes to zero, since the scalar $\sigma$ does not couple any more to $W$ and $Z$ in this limit. They also vanish when $h$ and $\sigma$ have the same mass,
since in this case the angle $\gamma$ loses any physical meaning. If $h$ and $\sigma$ are close in mass, the contribution to $\Delta S$ and $\Delta T$ is reduced, compared
to nonlinear realizations. The main impact of the heavy fermion contribution is on the $T$ parameter. Letting the heavy fermions to be as light as $800$ GeV, 
the lower limit from direct searches, contributions to $T$ of both signs as large as 0.2 in magnitude are generated by scanning the parameter space. Positive (negative) $\Delta T$ are correlated to a sizable mixing
between $t_L$ ($t_R$) and an electroweak singlet (doublet) heavy fermion component. Fermionic contributions to $S$ are typically positive and smaller, while those to
$R_b$ are mainly negative and of order few per mil. Even the largest scalar contributions to $\Delta S$ and $\Delta T$, obtained when
$\sin^2\gamma$ saturates its experimental bound and the scalar $\sigma$ is very heavy, can always be compensated by the fermionic ones
for an appropriate choice of the parameters, thus keeping $S$ and $T$ within the experimentally allowed region.

The Higgs production at LHC proceeds mainly via gluon fusion as in the SM. The amplitude comprises the usual SM contribution weighted by $\cos\gamma$ 
and an extra contribution proportional to $\sin\gamma$ which decouples in the heavy fermion limit. The interference between the two can be both constructive or destructive.
The effective coupling of the Higgs to a diphoton pair is modified in a similar way. In the large mass limit for the heavy fermions --taking the limit in such a way that the Higgs and light masses remain finite and at their physical values-- their
impact can be neglected and deviations from the SM only depend on $\sin\gamma$. In particular the Higgs decay to $WW$ allows to put the bound $\sin^2\gamma<0.18$, at $2\sigma$. 

The production of $\sigma$ at the LHC can mainly proceed via vector boson fusion or gluon fusion, the latter typically dominating for a $\sigma$ mass not exceeding few TeV.
There are no direct couplings of $\sigma$ to gluons but, as for the Higgs, an effective coupling arises from fermion loops. 
In the regime where the heavy fermions decouple the effective $\sigma$ coupling to gluons is controlled by the top loop weighted by a $\sin\gamma$ factor. 
Present direct LHC searches for a scalar particle are already sensitive to $\sin^2\gamma$ of order 0.1-0.2, for $m_\sigma<600$ GeV. If we combine this result
with the theoretical requirement that the Higgs behaves as a pseudo-Goldstone boson, we can already exclude $\sigma$ masses below about 500 GeV.
This limit can be pushed to $900-1400$ GeV by future LHC data from run 2, in the absence of any signal of new physics.

While in principle our model contains the ingredients to explain the recently observed 750 GeV diphoton excess in terms of $\sigma$ production and decay
into a photon pair, such an interpretation is rather unnatural since it would involve a tuning of the parameters in the scalar sector and, to boost both production and decay, a heavy fermion multiplicity much larger that the one adopted in the present version. Clearly, together with the modifications of the SM Higgs couplings, the prediction of an additional scalar, potentially observable at the next LHC run,
is the distinctive feature of our model. However, due to the mixing with an electroweak doublet, the new particle is not expected to decay exclusively into photons 
but rather into a variety of channels, much as the SM Higgs does.
\vspace{2cm}
\section*{Acknowledgements}
It is a pleasure to thank Marcela Carena, Tony Gherghetta, Oleksii Matsedonskyi  and Luca Merlo for
useful discussions. The authors (each identified by the first letter
of her/his last name) acknowledge partial financial support by the
European Union through the FP7 ITN INVISIBLES (PITN-GA-2011-289442)
(FGKMRS), by  the Horizon2020 RISE InvisiblesPlus 690575 (FGKMRS), by CiCYT through 
the project FPA2012-31880 (GS), and by the Spanish MINECO through the Centro de excelencia
Severo Ochoa Program under grant SEV-2012-0249 (GMS). The work of K.K. is supported by an ESR contract of the European Union network FP7 ITN INVISIBLES mentioned above. The work of S.S. is supported through the grant BES-2013-066480 
of the Spanish MICINN and she also received support during from the Spanish MICINN grant  EEBB-I-15-10242. We would also like
to thank the Kavli Institute for Theoretical Physics - Santa Barbara
(GM) (National Science Foundation grant PHY11-25915), the Mainz
Institute for Theoretical Physics (M), the Physics Department of the University of California San Diego (S), the Aspen Center for
Physics (GM) (National Science Foundation grant PHY-1066293 and the
Simons Foundation) and the Institute for Theoretical Physics IFT-UAM/CSIC (F) for hospitality and/or partial support during the
completion of this work.

\newpage

\appendix

\section{Coleman--Weinberg Potential}
\label{sec-CW}

In Sect.~2.1 it was assumed a specific form for the $SO(5)$ scalar potential broken to $SO(4)$, introducing two 
additional $SO(5)$ breaking parameters $\alpha$ and $\beta$. In this section we will  further motivate 
this assumption. Even assuming that the tree level scalar potential would preserve the global $SO(5)$ 
symmetry, the presence of a $SO(5)$ breaking couplings in the fermionic sector will generate at one-loop level 
$SO(5)$ breaking terms through the Coleman-Weinberg mechanism~\cite{Coleman:1973jx}. The one-loop fermionic contribution 
can be obtained from the field dependent mass matrix $\mathcal{M}$ as 
\bea
V_{\text{loop}} &=&  - \frac{i}{2} \int \frac{d^4 k}{(2\pi)^4} \sum_{n=1}^{\infty} \frac{1}{n} \Tr \l\frac{\Msq}{k^2}\r^n = 
  \frac{i}{2}\int \frac{d^4 k}{(2\pi)^4} \Tr\log\l1-\frac{\Msq}{k^2}\r \nn \\
 &=& -\frac{1}{64\pi^2} \l\Lambda^2\Tr\left[\Msq\right]-\Tr\left[(\Msq)^2 \right]\log\l\frac{\Lambda^2}{\mu^2}\r\right.\,+ \nn \\
 & & \hspace{1.35cm} \left. + \Tr\left[ (\Msq)^2 \log \l \frac{\Msq}{\mu^2} \r \right] - \frac{1}{2} \Tr\left[(\Msq)^2 \right]\r\,,
 \label{Vcwfull}
\eea
where $\Lambda$ is the UV cutoff scale while $\mu$ is a generic renormalization scale. The first two terms on the right-hand side of this equation are divergent, 
respectively quadratically and logarithmically, while the last two terms are finite. For the model under discussion it results:
\bea
\Tr[\Msq]&=& c_1 +  c_2 \, (\phi^T \phi)\,, \label{tracemm}\\
\Tr[(\Msq)^2]&=& d_1 + d_2\,\sigma + d_3\, \s^2 + d_4\,(\phi^T \phi) + d_5 \,(\phi^T \phi)^2 \,,
\label{tracemm2}
\eea
where
\bea
c_1 &=& 2 \Lam_1^2+\Lam_2^2+\Lam_3^2 + M_1^2 + 5\,M_5^2 + (\left\{ \right\} \lraw \left\{\right\}') \hspace{5.35cm}\,, \nn \\
c_2 &=& y_1^2+y_2^2 + (\left\{ \right\} \lraw \left\{\right\}')\,, \nn 
\eea
and
\bea
d_1 &=& M_1^4 + 5 M_5^4 + 2 M_5^2 \l 2 \Lam_1^2 + \Lam_2^2 \r + 2 M_1^2 \Lam_3^2 + 2 \Lam_1^4 + \l\Lam_2^2 + \Lam_3^2\r^2 
     + (\left\{ \right\} \lraw \left\{\right\}')\,,\nn \\
d_2 &=& 4 \l y_1 M_1 + y_2 M_5 \r \Lam_2 \Lam_3 + (\left\{ \right\} \lraw \left\{\right\}')\,, \nn\\
d_3 &=& 2\,y_1^2 \Lam_2^2 - y_2^2 \Lam_1^2 + (\left\{ \right\} \lraw \left\{\right\}') \,, \nn \\
d_4 &=&  4 \,y_1 y_2 M_1 M_5  + 2 \l y_1^2 + y_2^2\r \l M_1^2 + M_5^2\r  + y_2^2 \l \Lam_1^2 + 2\Lam_3^2 \r 
      + (\left\{ \right\} \lraw \left\{\right\}')\,, \nn \\
d_5 &=& y_1^4 + y_2^4 + (\left\{ \right\} \lraw \left\{\right\}')\nn \,.
\eea
In consequence, only the quadratically divergent piece  is seen to remain $SO(5)$ invariant, while the rest 
of the potential introduces an explicit breaking of the $SO(5)$ symmetry to $SO(4)$, see Eq.~(\ref{tracemm2}). The quadratic divergence 
can be thus absorbed in the parameters of the tree-level Lagrangian, and the same holds for the $SO(5)$ invariant component of the 
logarithmically divergent terms ($d_1,d_4$ and $d_5$). However, the presence of the $d_2$ and $d_3$ divergent $SO(5)$-breaking terms 
require to add the two corresponding counterterms in the potential, so as to obtain a renormalizable theory. 
 These two necessary terms are those defined with coefficients $\alpha$ and $\beta$ in the potential definition Eq.~(\ref{Laghs}). 
 The gauge couplings appearing in the covariant derivatives also break explicitly the $SO(5)$ symmetry, but they do not induce extra one-loop divergent contributions to the effective potential, and in consequence only $\alpha$ and $\beta$ are required for consistency.

The computation of the finite part of $V_{\text{loop}}$ should provide the dependence of the parameters on the renormalization scale~\footnote{The gauge bosons induce finite contributions 
which are  usually neglected with respect to the fermionic ones, as their relative ratio is proportional to the ratio between 
the mass scale for the heavy fermions and that for the gauge bosons.}
 and  should thus be equivalent to the computation of their renormalization group equations; this task is beyond the scope of the present paper.

\newpage
\bibliographystyle{JHEP}
\bibliography{UV_A}

\providecommand{\href}[2]{#2}\begingroup\raggedright\begin{thebibliography}{10}

\bibitem{Glashow:1970gm}
S.~L. Glashow, J.~Iliopoulos, and L.~Maiani, {\it {Weak Interactions with
  Lepton-Hadron Symmetry}},  {\em Phys. Rev.} {\bf D2} (1970) 1285--1292.

\bibitem{Gaillard:1974mw}
M.~K. Gaillard, B.~W. Lee, and J.~L. Rosner, {\it {Search for Charm}},  {\em
  Rev. Mod. Phys.} {\bf 47} (1975) 277--310.

\bibitem{Kaplan:1983fs}
D.~B. Kaplan and H.~Georgi, {\it {SU(2) x U(1) Breaking by Vacuum
  Misalignment}},  {\em Phys. Lett.} {\bf B136} (1984) 183.

\bibitem{Georgi:1984af}
H.~Georgi and D.~B. Kaplan, {\it {Composite Higgs and Custodial SU(2)}},  {\em
  Phys. Lett.} {\bf B145} (1984) 216.

\bibitem{Dugan:1984hq}
M.~J. Dugan, H.~Georgi, and D.~B. Kaplan, {\it {Anatomy of a Composite Higgs
  Model}},  {\em Nucl. Phys.} {\bf B254} (1985) 299.

\bibitem{Agashe:2004rs}
K.~Agashe, R.~Contino, and A.~Pomarol, {\it {The Minimal composite Higgs
  model}},  {\em Nucl. Phys.} {\bf B719} (2005) 165--187,
  [\href{http://xxx.lanl.gov/abs/hep-ph/0412089}{{\tt hep-ph/0412089}}].

\bibitem{Contino:2006qr}
R.~Contino, L.~Da~Rold, and A.~Pomarol, {\it {Light custodians in natural
  composite Higgs models}},  {\em Phys. Rev.} {\bf D75} (2007) 055014,
  [\href{http://xxx.lanl.gov/abs/hep-ph/0612048}{{\tt hep-ph/0612048}}].

\bibitem{Manohar:1983md}
A.~Manohar and H.~Georgi, {\it {Chiral Quarks and the Nonrelativistic Quark
  Model}},  {\em Nucl. Phys.} {\bf B234} (1984) 189.

\bibitem{Kaplan:1991dc}
D.~B. Kaplan, {\it {Flavor at SSC energies: A New mechanism for dynamically
  generated fermion masses}},  {\em Nucl. Phys.} {\bf B365} (1991) 259--278.

\bibitem{Aad:2015kqa}
{\bf ATLAS} Collaboration, G.~Aad {\em et.~al.}, {\it {Search for production of
  vector-like quark pairs and of four top quarks in the lepton-plus-jets final
  state in $pp$ collisions at $\sqrt{s}=8$ TeV with the ATLAS detector}},  {\em
  JHEP} {\bf 08} (2015) 105, [\href{http://xxx.lanl.gov/abs/1505.0430}{{\tt
  arXiv:1505.0430}}].

\bibitem{CMS:2015alb}
{\bf CMS} Collaboration, {\it {Search for top quark partners with charge $5/3$
  at $\sqrt{s}=13$ TeV}}, .

\bibitem{Panico:2012uw}
G.~Panico, M.~Redi, A.~Tesi, and A.~Wulzer, {\it {On the Tuning and the Mass of
  the Composite Higgs}},  {\em JHEP} {\bf 03} (2013) 051,
  [\href{http://xxx.lanl.gov/abs/1210.7114}{{\tt arXiv:1210.7114}}].

\bibitem{Carena:2014ria}
M.~Carena, L.~Da~Rold, and E.~Pont\'on, {\it {Minimal Composite Higgs Models at
  the LHC}},  {\em JHEP} {\bf 06} (2014) 159,
  [\href{http://xxx.lanl.gov/abs/1402.2987}{{\tt arXiv:1402.2987}}].

\bibitem{Contino:2011np}
R.~Contino, D.~Marzocca, D.~Pappadopulo, and R.~Rattazzi, {\it {On the effect
  of resonances in composite Higgs phenomenology}},  {\em JHEP} {\bf 10} (2011)
  081, [\href{http://xxx.lanl.gov/abs/1109.1570}{{\tt arXiv:1109.1570}}].

\bibitem{Marzocca:2012zn}
D.~Marzocca, M.~Serone, and J.~Shu, {\it {General Composite Higgs Models}},
  {\em JHEP} {\bf 08} (2012) 013,
  [\href{http://xxx.lanl.gov/abs/1205.0770}{{\tt arXiv:1205.0770}}].

\bibitem{Redi:2012ha}
M.~Redi and A.~Tesi, {\it {Implications of a Light Higgs in Composite Models}},
   {\em JHEP} {\bf 10} (2012) 166,
  [\href{http://xxx.lanl.gov/abs/1205.0232}{{\tt arXiv:1205.0232}}].

\bibitem{Carmona:2014iwa}
A.~Carmona and F.~Goertz, {\it {A naturally light Higgs without light Top
  Partners}},  {\em JHEP} {\bf 05} (2015) 002,
  [\href{http://xxx.lanl.gov/abs/1410.8555}{{\tt arXiv:1410.8555}}].

\bibitem{vonGersdorff:2015fta}
G.~von Gersdorff, E.~Pont\'on, and R.~Rosenfeld, {\it {The Dynamical Composite
  Higgs}},  {\em JHEP} {\bf 06} (2015) 119,
  [\href{http://xxx.lanl.gov/abs/1502.0734}{{\tt arXiv:1502.0734}}].

\bibitem{GellMann:1960np}
M.~Gell-Mann and M.~Levy, {\it {The axial vector current in beta decay}},  {\em
  Nuovo Cim.} {\bf 16} (1960) 705.

\bibitem{Barbieri:2007bh}
R.~Barbieri, B.~Bellazzini, V.~S. Rychkov, and A.~Varagnolo, {\it {The Higgs
  boson from an extended symmetry}},  {\em Phys. Rev.} {\bf D76} (2007) 115008,
  [\href{http://xxx.lanl.gov/abs/0706.0432}{{\tt arXiv:0706.0432}}].

\bibitem{Gertov:2015xma}
H.~Gertov, A.~Meroni, E.~Molinaro, and F.~Sannino, {\it {Theory and
  phenomenology of the elementary Goldstone Higgs boson}},  {\em Phys. Rev.}
  {\bf D92} (2015), no.~9 095003,
  [\href{http://xxx.lanl.gov/abs/1507.0666}{{\tt arXiv:1507.0666}}].

\bibitem{Alonso:2012px}
R.~Alonso, M.~B. Gavela, L.~Merlo, S.~Rigolin, and J.~Yepes, {\it {The
  Effective Chiral Lagrangian for a Light Dynamical "Higgs Particle"}},  {\em
  Phys. Lett.} {\bf B722} (2013) 330--335,
  [\href{http://xxx.lanl.gov/abs/1212.3305}{{\tt arXiv:1212.3305}}]. [Erratum:
  Phys. Lett.B726,926(2013)].

\bibitem{Buchalla:2013rka}
G.~Buchalla, O.~Cata, and C.~Krause, {\it {Complete Electroweak Chiral
  Lagrangian with a Light Higgs at NLO}},  {\em Nucl. Phys.} {\bf B880} (2014)
  552--573, [\href{http://xxx.lanl.gov/abs/1307.5017}{{\tt arXiv:1307.5017}}].

\bibitem{future-paper}
F.~Feruglio~et al. In preparation.

\bibitem{Dixon:2013haa}
L.~J. Dixon and Y.~Li, {\it {Bounding the Higgs Boson Width Through
  Interferometry}},  {\em Phys. Rev. Lett.} {\bf 111} (2013) 111802,
  [\href{http://xxx.lanl.gov/abs/1305.3854}{{\tt arXiv:1305.3854}}].

\bibitem{Agashe:2014kda}
{\bf Particle Data Group} Collaboration, K.~A. Olive {\em et.~al.}, {\it
  {Review of Particle Physics}},  {\em Chin. Phys.} {\bf C38} (2014) 090001.

\bibitem{Gunion:1989we}
J.~F. Gunion, H.~E. Haber, G.~L. Kane, and S.~Dawson, {\it {The Higgs Hunter's
  Guide}},  {\em Front. Phys.} {\bf 80} (2000) 1--448.

\bibitem{Agashe:2006at}
K.~Agashe, R.~Contino, L.~Da~Rold, and A.~Pomarol, {\it {A Custodial symmetry
  for $Zb \bar b$}},  {\em Phys. Lett.} {\bf B641} (2006) 62--66,
  [\href{http://xxx.lanl.gov/abs/hep-ph/0605341}{{\tt hep-ph/0605341}}].

\bibitem{CMS-ATLAS-comb}
{\bf ATLAS, CMS} Collaboration, {\it {Measurements of the Higgs boson
  production and decay rates and constraints on its couplings from a combined
  ATLAS and CMS analysis of the LHC pp collision data at $\sqrt{s}$ = 7 and 8
  TeV}}, .

\bibitem{Panico:2015jxa}
G.~Panico and A.~Wulzer, {\it {The Composite Nambu-Goldstone Higgs}},  {\em
  Lect. Notes Phys.} {\bf 913} (2016) pp.1--316,
  [\href{http://xxx.lanl.gov/abs/1506.0196}{{\tt arXiv:1506.0196}}].

\bibitem{Anastasiou:2009rv}
C.~Anastasiou, E.~Furlan, and J.~Santiago, {\it {Realistic Composite Higgs
  Models}},  {\em Phys. Rev.} {\bf D79} (2009) 075003,
  [\href{http://xxx.lanl.gov/abs/0901.2117}{{\tt arXiv:0901.2117}}].

\bibitem{Ghosh:2015wiz}
D.~Ghosh, M.~Salvarezza, and F.~Senia, {\it {Extending the Analysis of
  Electroweak Precision Constraints in Composite Higgs Models}},
  \href{http://xxx.lanl.gov/abs/1511.0823}{{\tt arXiv:1511.0823}}.

\bibitem{Peskin:1991sw}
M.~E. Peskin and T.~Takeuchi, {\it {Estimation of oblique electroweak
  corrections}},  {\em Phys. Rev.} {\bf D46} (1992) 381--409.

\bibitem{Altarelli:1990zd}
G.~Altarelli and R.~Barbieri, {\it {Vacuum polarization effects of new physics
  on electroweak processes}},  {\em Phys. Lett.} {\bf B253} (1991) 161--167.

\bibitem{Ciuchini:2014dea}
M.~Ciuchini, E.~Franco, S.~Mishima, M.~Pierini, L.~Reina, and L.~Silvestrini,
  {\it {Update of the electroweak precision fit, interplay with Higgs-boson
  signal strengths and model-independent constraints on new physics}},  in {\em
  {International Conference on High Energy Physics 2014 (ICHEP 2014) Valencia,
  Spain, July 2-9, 2014}}, 2014.
\newblock \href{http://xxx.lanl.gov/abs/1410.6940}{{\tt arXiv:1410.6940}}.

\bibitem{Novikov:1992rj}
V.~A. Novikov, L.~B. Okun, and M.~I. Vysotsky, {\it {On the Electroweak one
  loop corrections}},  {\em Nucl. Phys.} {\bf B397} (1993) 35--83.

\bibitem{Orgogozo:2012ct}
A.~Orgogozo and S.~Rychkov, {\it {The S parameter for a Light Composite Higgs:
  a Dispersion Relation Approach}},  {\em JHEP} {\bf 06} (2013) 014,
  [\href{http://xxx.lanl.gov/abs/1211.5543}{{\tt arXiv:1211.5543}}].

\bibitem{Haber:2010bw}
H.~E. Haber and D.~O'Neil, {\it {Basis-independent methods for the
  two-Higgs-doublet model III: The CP-conserving limit, custodial symmetry, and
  the oblique parameters S, T, U}},  {\em Phys. Rev.} {\bf D83} (2011) 055017,
  [\href{http://xxx.lanl.gov/abs/1011.6188}{{\tt arXiv:1011.6188}}].

\bibitem{Lavoura:1992np}
L.~Lavoura and J.~P. Silva, {\it {The Oblique corrections from vector - like
  singlet and doublet quarks}},  {\em Phys. Rev.} {\bf D47} (1993) 2046--2057.

\bibitem{Dawson:2012di}
S.~Dawson and E.~Furlan, {\it {A Higgs Conundrum with Vector Fermions}},  {\em
  Phys. Rev.} {\bf D86} (2012) 015021,
  [\href{http://xxx.lanl.gov/abs/1205.4733}{{\tt arXiv:1205.4733}}].

\bibitem{Heinemeyer:2013tqa}
{\bf LHC Higgs Cross Section Working Group} Collaboration, J.~R. Andersen {\em
  et.~al.}, {\it {Handbook of LHC Higgs Cross Sections: 3. Higgs Properties}},
  \href{http://xxx.lanl.gov/abs/1307.1347}{{\tt arXiv:1307.1347}}.

\bibitem{Aad:2014ioa}
{\bf ATLAS} Collaboration, G.~Aad {\em et.~al.}, {\it {Search for Scalar
  Diphoton Resonances in the Mass Range $65-600$ GeV with the ATLAS Detector in
  $pp$ Collision Data at $\sqrt{s}$ = 8 $TeV$}},  {\em Phys. Rev. Lett.} {\bf
  113} (2014), no.~17 171801, [\href{http://xxx.lanl.gov/abs/1407.6583}{{\tt
  arXiv:1407.6583}}].

\bibitem{Khachatryan:2015qba}
{\bf CMS} Collaboration, V.~Khachatryan {\em et.~al.}, {\it {Search for
  diphoton resonances in the mass range from 150 to 850 GeV in pp collisions at
  $\sqrt{s} =$ 8 TeV}},  {\em Phys. Lett.} {\bf B750} (2015) 494--519,
  [\href{http://xxx.lanl.gov/abs/1506.0230}{{\tt arXiv:1506.0230}}].

\bibitem{Aad:2015agg}
{\bf ATLAS} Collaboration, G.~Aad {\em et.~al.}, {\it {Search for a high-mass
  Higgs boson decaying to a $W$ boson pair in $pp$ collisions at $\sqrt{s} = 8$
  TeV with the ATLAS detector}},  {\em JHEP} {\bf 01} (2016) 032,
  [\href{http://xxx.lanl.gov/abs/1509.0038}{{\tt arXiv:1509.0038}}].

\bibitem{Aad:2015kna}
{\bf ATLAS} Collaboration, G.~Aad {\em et.~al.}, {\it {Search for an
  additional, heavy Higgs boson in the $H\rightarrow ZZ$ decay channel at
  $\sqrt{s} = 8\;\text{ TeV }$ in $pp$ collision data with the ATLAS
  detector}},  {\em Eur. Phys. J.} {\bf C76} (2016), no.~1 45,
  [\href{http://xxx.lanl.gov/abs/1507.0593}{{\tt arXiv:1507.0593}}].

\bibitem{CMS:xwa}
{\bf CMS} Collaboration, {\it {Properties of the Higgs-like boson in the decay
  H to ZZ to 4l in pp collisions at sqrt s =7 and 8 TeV}}, .

\bibitem{CMS:bxa}
{\bf CMS} Collaboration, {\it {Update on the search for the standard model
  Higgs boson in pp collisions at the LHC decaying to W + W in the fully
  leptonic final state}}, .

\bibitem{Khachatryan:2015yea}
{\bf CMS} Collaboration, V.~Khachatryan {\em et.~al.}, {\it {Search for
  resonant pair production of Higgs bosons decaying to two bottom
  quark-antiquark pairs in proton–proton collisions at 8 TeV}},  {\em Phys.
  Lett.} {\bf B749} (2015) 560--582,
  [\href{http://xxx.lanl.gov/abs/1503.0411}{{\tt arXiv:1503.0411}}].

\bibitem{Aad:2015xja}
{\bf ATLAS} Collaboration, G.~Aad {\em et.~al.}, {\it {Searches for Higgs boson
  pair production in the $hh\to bb\tau\tau, \gamma\gamma WW^*, \gamma\gamma bb,
  bbbb$ channels with the ATLAS detector}},  {\em Phys. Rev.} {\bf D92} (2015)
  092004, [\href{http://xxx.lanl.gov/abs/1509.0467}{{\tt arXiv:1509.0467}}].

\bibitem{Holzner:2014qqs}
{\bf ATLAS, CMS} Collaboration, A.~Holzner, {\it {Beyond standard model Higgs
  physics: prospects for the High Luminosity LHC}},
  \href{http://xxx.lanl.gov/abs/1411.0322}{{\tt arXiv:1411.0322}}.

\bibitem{Martin-Lozano:2015dja}
V.~Mart\'in~Lozano, J.~M. Moreno, and C.~B. Park, {\it {Resonant Higgs boson
  pair production in the $ hh\to b\overline{b}\ WW\to
  b\overline{b}{\ell}^{+}\nu {\ell}^{-}\overline{\nu} $ decay channel}},  {\em
  JHEP} {\bf 08} (2015) 004, [\href{http://xxx.lanl.gov/abs/1501.0379}{{\tt
  arXiv:1501.0379}}].

\bibitem{ATLAS-exo}
{\bf ATLAS} Collaboration, {\it {Search for resonances decaying to photon pairs
  in 3.2 fb$^{-1}$ of $pp$ collisions at $\sqrt{s}$ = 13 TeV with the ATLAS
  detector}}, .

\bibitem{CMS:2015dxe}
{\bf CMS} Collaboration, {\it {Search for new physics in high mass diphoton
  events in proton-proton collisions at 13~TeV}}, .

\bibitem{Grzadkowski:2010es}
B.~Grzadkowski, M.~Iskrzynski, M.~Misiak, and J.~Rosiek, {\it {Dimension-Six
  Terms in the Standard Model Lagrangian}},  {\em JHEP} {\bf 10} (2010) 085,
  [\href{http://xxx.lanl.gov/abs/1008.4884}{{\tt arXiv:1008.4884}}].

\bibitem{Coleman:1973jx}
S.~R. Coleman and E.~J. Weinberg, {\it {Radiative Corrections as the Origin of
  Spontaneous Symmetry Breaking}},  {\em Phys. Rev.} {\bf D7} (1973)
  1888--1910.

\end{thebibliography}\endgroup

\end{document}